\documentclass[acmsmall]{acmart}
\AtBeginDocument{%
  }

\setcopyright{acmlicensed}
\copyrightyear{2026}
\acmYear{2026}
\acmDOI{XXXXXXX.XXXXXXX}

\acmJournal{TOSEM}
\acmVolume{1}
\acmNumber{1}
\acmArticle{1}
\acmMonth{1}

\usepackage{amssymb}
\usepackage{algorithm}
\usepackage{algpseudocode}
\usepackage{graphicx}
\usepackage{textcomp}
\usepackage{xcolor}
    
\usepackage{booktabs} 
\usepackage{makecell} 
\usepackage{graphics}

\usepackage{enumitem}
\usepackage{multirow}
\usepackage{comment}

\usepackage{subcaption}
\usepackage{colortbl}

\usepackage{tcolorbox}
\usepackage{fontawesome5} 

\tcbuselibrary{skins, breakable}

\newtcolorbox{summarybox}[2][]{
  enhanced,
  breakable,
  colback=gray!8,
  colframe=black!70,
  fonttitle=\bfseries\small,
  title={#2},
  left=6pt, right=6pt, top=4pt, bottom=4pt,
  #1
}

\definecolor{f1gray}{gray}{0.93}
\definecolor{darkgreen}{RGB}{23,88,3}
\definecolor{methodgray}{gray}{0.96}

\DeclareUnicodeCharacter{2212}{-}

\newcommand \tool{\texttt{AppRay}}{}
\newcommand{\red}[1]{\textcolor{black}{#1}}

\newcommand{\revision}[1]{\textcolor{black}{#1}}
\newcommand{\minor}[1]{\textcolor{black}{#1}}

\begin{document}

\title{From Exploration to Revelation: App-Level Context-Aware Deceptive Pattern Detection for Mobile Applications}

\author{Jieshan Chen}
\email{Jieshan.Chen@data61.csiro.au}
\authornote{Also with TUM-IAS, Garching, Germany.}
\orcid{0000-0002-2700-7478}
\affiliation{%
  \institution{CSIRO’s Data61}
  \city{Sydney}
  \state{NSW}
  \country{Australia}
}

\author{Zhen Wang}
\email{Jeff.Wang@data61.csiro.au}
\orcid{0009-0000-5809-0780}
\affiliation{%
  \institution{CSIRO’s Data61}
  \city{Sydney}
  \state{NSW}
  \country{Australia}
}

\author{Jiamou Sun}
\email{Frank.Sun@data61.csiro.au}
\orcid{0000-0002-5212-7068}
\affiliation{%
  \institution{CSIRO’s Data61}
  \city{Canberra}
  \state{ACT}
  \country{Australia}
}

\author{Zhenchang Xing}
\email{zhenchang.xing@data61.csiro.au}
\authornote{Also with Australian National University, Canberra, ACT, Australia.}
\orcid{0000-0001-7663-1421}
\affiliation{%
  \institution{CSIRO’s Data61}
  \city{Canberra}
  \state{ACT}
  \country{Australia}
}

\author{Qinghua Lu}
\email{Qinghua.Lu@data61.csiro.au}
\orcid{0000-0002-7783-5183}
\affiliation{%
  \institution{CSIRO’s Data61}
  \city{Sydney}
  \state{NSW}
  \country{Australia}
}

\author{Qing Huang}
\orcid{0000-0003-1073-7471}
\email{qh@jxnu.edu.cn}
\affiliation{%
  \institution{Jiangxi Normal University}
  \city{Nanchang}
  \state{Jiangxi}
  \country{China}
}

\author{Xiwei Xu}
\email{Xiwei.Xu@data61.csiro.au}
\orcid{0000-0002-2273-1862}
\affiliation{%
  \institution{CSIRO’s Data61}
  \city{Sydney}
  \state{NSW}
  \country{Australia}
}

\author{Liming Zhu}
\email{Liming.Zhu@data61.csiro.au}
\authornote{Also with University of New South Wales, Sydney, NSW, Australia.}
\orcid{0000-0001-5839-3765}
\affiliation{%
  \institution{CSIRO’s Data61}
  \city{Sydney}
  \state{NSW}
  \country{Australia}
}

\renewcommand{\shortauthors}{Jieshan Chen et al.}

\begin{abstract}

Mobile apps are essential in daily life but frequently employ deceptive patterns, such as visual emphasis or linguistic nudging, to manipulate user behavior. Existing research largely relies on manual detection, which is time-consuming and cannot keep pace with rapidly evolving apps. Although recent work has explored automated approaches, these methods are limited to intra-page patterns, depend on manual app exploration, and lack flexibility. To address these limitations, we present AppRay, a system that integrates task-oriented app exploration with automated deceptive pattern detection to reduce manual effort, expand detection coverage, and improve performance. AppRay operates in two stages. First, it combines large language model–guided task-oriented exploration with random exploration to capture diverse user interface (UI) states. Second, it detects both intra-page and inter-page deceptive patterns using a contrastive learning–based multi-label classifier augmented with a rule-based refiner for context-aware detection. We contribute two datasets, AppRay-Tainted-UIs and AppRay-Benign-UIs, comprising 2,185 deceptive pattern instances, including 149 intra-page cases, spanning 16 types across 876 deceptive and 871 benign UIs, while preserving UI relationships. Experimental results show that AppRay achieves macro/micro averaged precision of 0.92/0.85, recall of 0.86/0.88, and F1 scores of 0.89/0.85, yielding 27.14\% to 1200\% improvements over prior methods and enabling effective detection of previously unexplored deceptive patterns.

\end{abstract}



\maketitle

\section{Introduction}
\label{sec:intro}

People nowadays perform a wide range of daily activities, including shopping, reading, chatting, through laptops or mobile apps, where user interfaces act as an important proxy connecting human with digital functionalities.
However, many apps, unconsciously or intentionally, insert some psychological tricks into their user interfaces to manipulate end-users, either by seducing them to stay longer in their services~\cite{cho2021reflect, monge2023defining}, or by deceiving them to perform some actions that may not be of their best interest~\cite{gray2018dark, gray2021dark, di2020ui}.
For instance, in social media apps, the never-ending auto-play feature continuously streams videos to capture users’ attention, often leading to regret and loss of control upon reflection~\cite{monge2023defining, cho2021reflect}. 
Similarly, native advertisements are visually integrated into normal content, making them indistinguishable and prompting inadvertent clicks. These manipulative design strategies are collectively termed deceptive patterns (or dark patterns, DP)~\cite{brignull2010dark, gray2018dark}.

\begin{figure}
    \centering
    \includegraphics[width=1\textwidth]{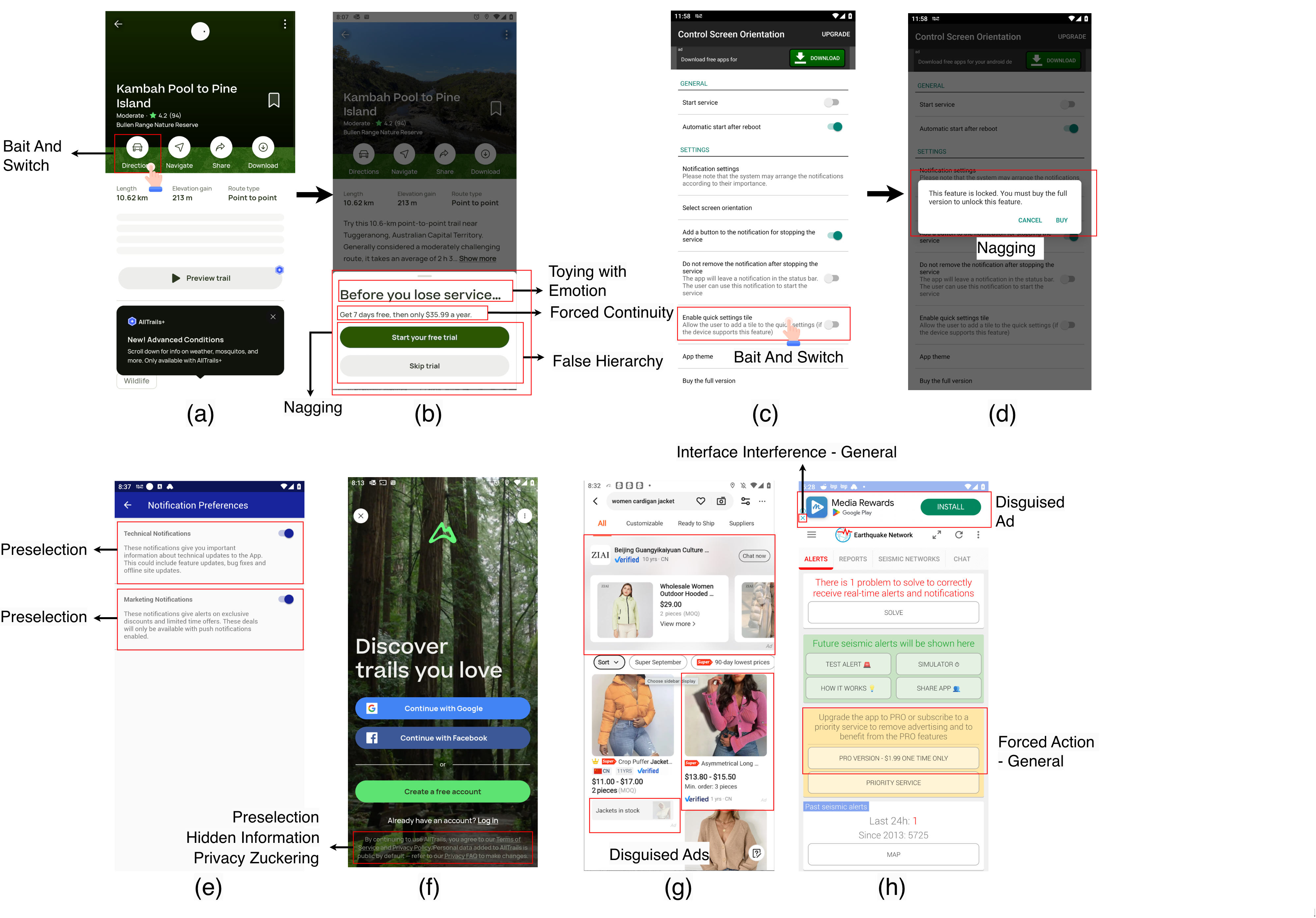}
    \caption{Representative examples of deceptive patterns (DPs) in mobile applications. (a) and (b) illustrate interaction-dependent deceptive patterns, including bait-and-switch, which can only be revealed after the user performs the direction action, exposing a mismatch between the initial affordance and the subsequent system behavior. (b) further exhibits toying with emotion, forced continuity, and false hierarchy. (c) and (d) show another example interaction-dependent deceptive patterns, including bait-and-switch and nagging, where functional options are restricted or repeatedly prompted through system settings. (e) illustrates preselection patterns, while (f) demonstrates that a single UI element can contain multiple deceptive patterns, including preselection, hidden information, and privacy-zuckering. (g) presents disguised advertisements embedded within content feeds, and (h) shows forced action patterns that pressure users into paid upgrades or subscriptions.}
    \label{fig:dp_example}
\end{figure}

While the term may be unfamiliar, deceptive patterns are pervasive (see Figure~\ref{fig:dp_example}).
Mathur et al. \cite{mathur2019dark} found that 11.1\% of 11K shopping websites contained such patterns. Di Geronimo et al.~\cite{di2020ui} revealed that 95\% of 240 popular mobile apps averagely contain at least seven types of DPs.
\revision{These patterns impose harms of escalating severity: they erode user autonomy and privacy through manipulative consent interfaces such as cookie banners that obscure meaningful choices~\cite{bermejo2021website}; exploit attention and reward mechanisms, fostering addictive behaviors  or compulsive interactions \cite{chaudhary2022you}; induce psychological distress, such as anxiety and loss of control \cite{mildner2023engaging, cho2021reflect, monge2022towards}; and cause economic losses through unwanted purchases or installations \cite{mathur2019dark}. 
Companies employing such practices have even faced regulatory sanctions—for instance, the US\$245 million fine imposed on Epic Games for manipulative in-app purchases in Fortnite~\cite{EpicFine}.
Consequently, detecting and mitigating deceptive patterns has become a critical step toward safeguarding users and promoting ethical, trustworthy digital design.}

\revision{Despite its importance, deceptive pattern detection remains challenging.
First, existing approaches primarily rely on rule-based techniques, which often lack robustness~\cite{mansur2023aidui, chen2023unveiling}.
Rule-based methods are inherently unstable when faced with the high diversity of deceptive pattern implementations, where the same pattern can manifest in entirely different visual styles, layouts, or interaction flows.
Enumerating all possible stylistic variants is infeasible, making handcrafted rules brittle and prone to false negatives. A recent work, DPGuard \cite{shi202550}, leverages multi-modal large language models (MLLMs) removing the requirement for rule engineering, but treats detection as a screen-level classification problem, without explicitly modeling interaction-dependent deceptive mechanisms or their localization.
Second, many deceptive patterns are interaction-dependent: they emerge only when users perform specific actions or multi-step tasks, such as in Sneaking into Basket behaviors. These \revision{inter-page}-dependent deceptive patterns cannot be effectively captured through \revision{intra-page} inspection or random testing, yet most existing approaches still rely on manual or random UI collection, which is inefficient and incomplete~\cite{di2020ui, mansur2023aidui, chen2023unveiling}.
Third, current systems typically analyze single UI screens in isolation, neglecting cross-page transitions and application-level context, which are crucial for understanding deceptive behavior~\cite{chen2023unveiling}.
Finally, data scarcity and limited annotation quality remain major bottlenecks, most available datasets cover only a narrow subset of deceptive patterns~\cite{chen2023unveiling, mansur2023aidui} and lack localization support~\cite{di2020ui, shi202550}. These limitations collectively constrain scalability, generalization, and the detection of dynamic, real-world deceptive behaviors.}

Driven by existing research gaps, in this work, we propose a framework, named \tool{}, to automatically collect UIs and detect both \revision{intra-page} and \revision{inter-page} deceptive patterns (DPs) given an app. 
\revision{We adopt a GUI-based detection approach because DPs manifest through what users visually perceive on the rendered interface. The final UI often diverges from its XML or source-code definitions due to runtime styling, device-dependent layout computation, conditional rendering, and occlusion from \revision{inter-page} elements such as pop-ups or scrolling containers. These perceptual and contextual factors—such as weakened contrast, hidden options, or off-screen placement—play a central role in deceptive design and can only be faithfully captured at the UI level. In addition, GUI-based analysis naturally generalizes across heterogeneous mobile frameworks (Android, Flutter, React Native, iOS), offering a unified and robust detection basis independent of underlying implementations.}

Motivated by these challenges, our framework consists of two essential components: (1) an app exploration module that incorporates an LLM-powered app navigator with a classical automated app exploration tool, enhancing the ability to identify and collect UIs containing DPs; (2) a deceptive pattern detector that integrates a contrastive learning-based DP classifier with a rule-based DP refiner, enabling \textbf{context-aware and app-level analysis}. This approach allows for the identification of a broad range of DPs while distinguishing subtle differences between pattern types.

Specifically, given an app under test, our LLM-powered app navigator leverages LLMs to simulate end-users to navigate through the apps to explore and collect diverse UIs. We instructed the LLM to simulate an end-user exploring the app and performing common tasks that are likely to contain DPs. Additionally, we integrated an automated app exploration tool to further increase UI coverage within the app. 
After that, we developed a context-aware DP detector to automatically identify and localise DP instances within and across UIs. We first leveraged the mature property extraction module from UIGuard~\cite{chen2023unveiling} to identify elements and group related elements into different semantic components.
Based on this, we proposed a component-level deceptive pattern detector, powered by a contrastive learning-based multi-label classifier and a rule-based refiner. The multi-label classifier scans each UI page to identify potential DPs, while the rule-based refiner checks preceding and subsequent UIs, modeling the UI relationship, to remove false positives and increase precision.
With these specially designed modules, our detector is able to disambiguate between similar DPs (e.g., distinguishing between nagging and disguised ad patterns) with high precision and recall.

To train and test our proposed techniques, we contribute new datasets, \texttt{AppRay-Tainted-UIs } and \texttt{\revision{AppRay-Benign-UIs}}, that addressed the limitations of existing datasets, covering both \revision{intra-page} and \revision{\revision{inter-page}} deceptive patterns and preserving the relationship between the UIs. We used our collected UIs from our app navigator as a starting point, which captures the action points, screenshots, and relationship among UIs, and conducted three rounds of annotations. 
In total, we annotated 19,722 UIs from 100 apps, and identified 2,185 unique DP instances of 16 types from 876 UIs. Among these instances, 149 are related to multiple UI status (i.e., \revision{\revision{inter-page}} DPs) from 48 apps. We termed it as \texttt{AppRay-Tainted-UIs } dataset. We also collected 871 benign UIs from these collected UIs, termed \texttt{\revision{AppRay-Benign-UIs}} for assessing the likelihood of detecting false positives.

We conducted a comprehensive evaluation to evaluate the performance of the proposed system, including a modular examination,  an ablation study (Section~\ref{sec:rq1.1_exploration} and Section~\ref{sec:rq1.2_detector}), baseline comparison (Section~\ref{sec:rq2_baseline}) and a user study (Section~\ref{sec:rq4_userstudy}). 
The modular examination and ablation study confirm the effectiveness and critical roles of each module of \tool{}.
We found that our \tool{} achieves \red{0.92/0.85} in precision, \red{0.86/0.88} in recall and \red{0.89/0.85} in F1-score for the macro/micro averaged performance respectively, which significantly yields 27.14\% to 1200\% performance gains over prior work in their supported deceptive patterns, and enables effective detections on previously unexplored \revision{intra-page} and \revision{\revision{inter-page}} types.
Finally, our user study confirms the usefulness of the proposed technique.

In summary, the contributions of our work are as follows:
\begin{itemize}
    \item \textbf{End-to-end detection pipeline}: We present, \tool{}, the first systematic approach for end-to-end deceptive pattern detection, covering the entire process from \textit{automated app exploration to pattern revelation.} Unlike prior works that focus solely on \revision{intra-page} screens, \tool{} jointly analyzes \textit{both \revision{intra-page} and \revision{inter-page} deceptive patterns} that emerge through user interactions.
    \item \textbf{Annotated benchmark dataset}: We construct the first benchmark dataset that captures both \revision{intra-page} and \revision{inter-page} deceptive patterns with interaction sequences and element-level annotations.
    The dataset includes 2,185 deceptive instances (149 \revision{inter-page} cases) across 16 types from 876 UIs (AppRay-Tainted-UIs ) and 871 benign UIs (\revision{AppRay-Benign-UIs}), annotated from 19,722 screens across 100 apps, providing the most comprehensive coverage of deceptive pattern behaviors to date. We have released our source code and datasets at \textcolor{blue}{\url{https://github.com/chenjshnn/AppRay}}.
    \item \textbf{Detection of new deceptive pattern types and empirical validation}: Through extensive experiments, \tool{} successfully detected 10 new deceptive pattern types that were previously undetected by AidUI \cite{mansur2023aidui} and 7 new types beyond those covered by UIGuard \cite{chen2023unveiling}. These results not only demonstrate the effectiveness and generality of \tool{} but also expand the frontier of detectable deceptive pattern categories to include multi-page and interaction-dependent cases.
\end{itemize}

\revision{The remainder of this paper is organized as follows. Section 2 introduces the background and taxonomy of deceptive patterns. Section 3 reviews related work and situates our contributions within the existing literature. Section 4 presents the proposed methodology, while Section 5 describes the research questions, datasets, and data collection process. Section 6 reports the experimental setup and results for each research question. Section 7 discusses potential applications of our approach. Section 8 examines threats to validity, and Section 9 outlines limitations, future directions, and broader implications. Finally, Section 10 concludes the paper.}

\section{Background}

\subsection{Deceptive Pattern Taxonomy}

\revision{
Prior work has extensively examined and summarised deceptive pattern taxonomies in the mobile domain. We build upon the integrated taxonomy proposed by Chen et al. \cite{chen2023unveiling} (see their Table 1), which consolidates multiple existing taxonomies \cite{gray2018dark, gunawan2021comparative, di2020ui} into a unified structure. Following this line of work, we adopt their integrated taxonomy as the foundation for our study. 
Specifically, deceptive patterns are commonly organized into five high-level categories. \textbf{Nagging} disrupt user tasks by suddenly displaying irrelevant windows; \textbf{Obstruction} seeks to unnecessarily complicate tasks; \textbf{Sneaking} hides, disguises or delays relevant information to the current user task (e.g., hidden costs); \textbf{Forced Action} forces users to perform some actions to get rewards, unlock features or achieve some tasks; and \textbf{Interface Interference} manipulates the interface to privilege some options over others. Each category contains multiple subtypes, such as Roach Motel, Hidden Costs, Preselection, and Disguised Ads, which our detection model also distinguishes. Given that the taxonomy itself is not our contribution, we refer readers to Table~\ref{tab:full_taxonomy} in the appendix for detailed definitions. Figure~\ref{fig:dp_example} also shows some examples.}

Chen et al.~\cite{chen2023unveiling} classified deceptive patterns into two types, \textbf{static} and \textbf{dynamic}, based on whether contextual information is required for identification. Static patterns can be recognized from a single UI screen, whereas dynamic ones require analyzing multiple screens. They also introduced an ``in-between'' class that spans several pages but can sometimes be detected on a single screen, potentially leading to false positives.

\begin{figure}
    \centering
    \includegraphics[width=0.9\textwidth]{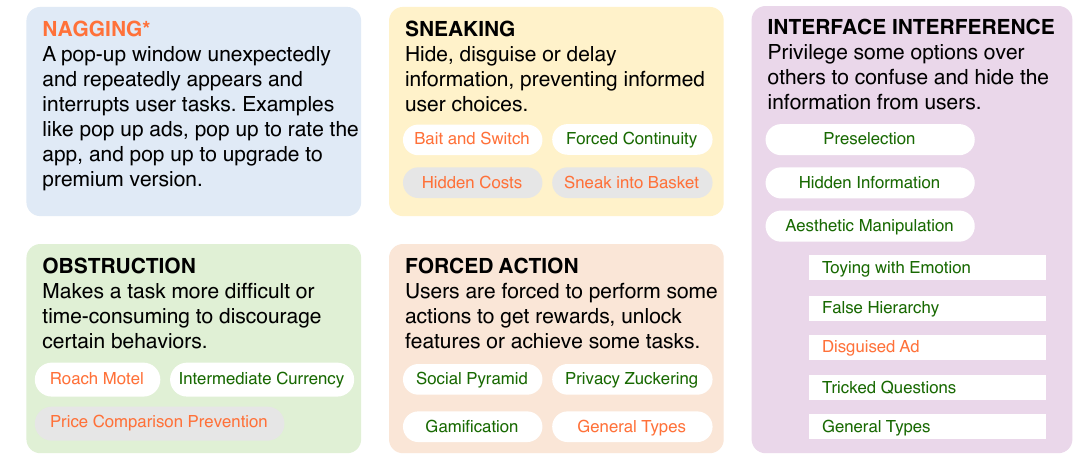}
    \caption{Deceptive pattern taxonomy. It consists of five main strategies. We employ green and orange text colors to indicate the \textcolor{darkgreen}{\revision{intra-page}} and \textcolor{orange}{\revision{inter-page}} deceptive patterns.
    }
    \label{fig:taxonomy}
\end{figure}

\revision{Building on this taxonomy, we follow the same analytical angle but refine it in two ways: (i) we replace the term static and dynamic with the more precise term intra-page and inter-page to avoid confusion with program runtime semantics, following \cite{gray2025getting}, and (ii) we merge the in-between class into the inter-page category, using an asterisk (e.g., Nagging*) to denote such cases directly. Specifically, we categorize deceptive patterns based on whether their deceptive behavior can be observed within a single UI state:
\begin{itemize}
    \item \textbf{Intra-page deceptive patterns}: visible on a single screen without any interaction or state change (e.g., False Hierarchy).
    \item \textbf{Inter-page deceptive patterns}: revealed through UI state changes triggered either by user actions or by system events (e.g., Bait-and-Switch, Nagging).
\end{itemize}
Figure~\ref{fig:taxonomy} presents an overview of the definitions and categories of \textcolor{darkgreen}{intra-page} and \textcolor{orange}{inter-page} deceptive patterns considered in this study. Since the taxonomy itself is not our contribution, we include the comprehensive definitions adapted from \cite{chen2023unveiling} in Table~\ref{tab:full_taxonomy} in the Appendix.
}

\subsection{Positioning Deceptive Patterns within Software Engineering}
\revision{To clarify the use of the term \textit{pattern} in SE, we refer to the software engineering definition in which a pattern is characterised by a stable name, a recurring problem, a reusable solution structure, and its consequences \cite{gamma1995design}. Deceptive UI patterns follow this structure: they recur across applications, employ recognisable manipulation strategies to address a recurring design objective (e.g., steering user choices), and produce predictable consequences such as user harm, loss of trust, or regulatory risk. Given their harmful nature, these strategies align more closely with \textit{anti-patterns}, recurring but undesirable solutions, yet the HCI and SE communities have consistently used the term ``dark/deceptive patterns'' for more than a decade to denote such manipulative interface strategies \cite{gray2018dark,gunawan2021comparative,di2020ui,chen2023unveiling,mansur2023aidui,nie2024shadows}. We follow this established terminology for conceptual continuity and comparability. Moreover, deceptive patterns remain highly relevant to software engineering because they affect key non-functional properties such as usability, transparency, and trust, and their detection aligns with SE concerns including GUI testing, software quality assurance, and the analysis of non-functional requirements.
}

\begin{table*}[t]
\centering
\caption{ \revision{Comparison of representative prior work and AppRay along the deceptive pattern (DP) detection pipeline, covering domain, support for end-to-end pipeline, app exploration, detection capability, support for interaction-aware detection, supported DP types, and dataset characteristics. For prior work, we map the supported deceptive pattern (DP) types to our taxonomy to enable direct comparison. We note that some types considered in prior studies are conceptually distinct from dark patterns and therefore exclude them when calculating the supported DP types (e.g., censored ads, malicious scripts~\cite{liu2020maddroid, liu2025ios}). N/A indicates that the corresponding functionality is not applicable.}
}

\resizebox{1.0\columnwidth}{!}{%
\begin{tabular}{p{2.4cm} p{2.6cm} p{2.2cm} p{2cm} p{3.6cm} p{3.6cm} p{2.0cm} p{3.0cm} p{3.0cm}}
\toprule
\textbf{Type} & \textbf{Example Work} & \textbf{Domain} & \textbf{Support Automatic End-to-End Pipeline?} & \textbf{App Exploration Methods} & \textbf{Detection Methods} & \textbf{Support Interaction-Aware Detection?} & \textbf{Supported DP Types}  & \textbf{DP Dataset} \\

\midrule
\multirow{4}{*}{\begin{minipage}[t]{\linewidth}Dark Pattern \newline Detection \end{minipage}}
& Mathur et al. \cite{mathur2019dark} & Shopping Web & No & Automated rule-based web crawling & Semi-automated detection via HTML-based clustering and manual annotation & Yes & 15 shopping-focused types & Consecutive Web UIs without action semantics \\
& AidUI \cite{mansur2023aidui} & Mobile \& Web & No & Assumes pre-collected UI screens & Automated GUI-based detection within UIs (rule based methods with ML methods) & No  & 5 general types& Isolated UIs \\
& UIGuard \cite{chen2023unveiling} & Mobile & No &  Assumes pre-collected UI screens & Automated GUI-based detection within UIs (rule based methods with ML methods) & No  & 9 general types & Isolated UIs  \\
& DPGuard \cite{shi202550} & Mobile \& Web & Yes & Model-based UI exploration (i.e., Droidbot) & Automated GUI based detection within UIs (binary classifier with MLLM) & No & 20 types (mapped to 14 general types) & Isolated UIs \\
& DarkFleece \cite{yue2024darkfleece} & Mobile Subscriptions & Yes &  Droidbot with rules to locate subscription pages & Automated GUI-based detection within UIs, limited to subscription scenarios & No  & 12 Detailed subscription types (mapped to 5 general types) & Not provided \\

\midrule
\multirow{4}{*}{GUI Testing}
& Monkey \cite{monkey} & Mobile & N/A &  Pseudo-random event generation & N/A & N/A & N/A & N/A \\
& \begin{minipage}[t]{\linewidth} DroidBot \cite{li2017droidbot}, \newline  FastBot2 \cite{lv2022fastbot2} \end{minipage}
& Mobile & N/A & Model-based exploration using UI state graphs & N/A & N/A & N/A & N/A  \\
& \begin{minipage}[t]{\linewidth} GPTDroid \cite{liu2024make},\newline DroidAgent \cite{yoon2024intent} \end{minipage}
& Mobile & N/A & LLM-guided GUI exploration & N/A & N/A & N/A & N/A  \\

\midrule
\multirow{2}{*}{\begin{minipage}[t]{\linewidth}Mobile \newline Advertisements\end{minipage}}
& MadDroid \cite{liu2020maddroid} & Mobile Ads & Yes & Automated rule based collection (i.e., auto-collect main and exit pages with an ad-first exploration strategy) & App network traffic analysis with FastRCNN detection on cross icons & No  & Click-deception Ads (mapped to one general type)& A list of investigated app names \\ 
& Liu et al. \cite{liu2025ios} & Mobile Ads (iOS) & No & Model-Based UI exploration & App network traffic analysis + GPT4o with manual confirmation & No & Deceptive or disruptive ads (mapped to 2 general types) & A list of investigated app names \\ 

\midrule
\textbf{Our Work}
& \textbf{AppRay} & \textbf{Mobile} & \textbf{Yes} & \textbf{Model-based exploration with an LLM-driven explorer tailored for DP discovery}
& \textbf{Automated GUI-based detection across and within UI screens (DL model with rules)} 
& \textbf{Yes} & \textbf{16 general types} & \textbf{Consecutive UIs with action semantics} \\ 

\bottomrule
\end{tabular}
}
\label{tab:priorworks}
\end{table*}

\section{Related Work}
\label{sec:relatedwork}
\revision{
Our work builds upon and intersects three research directions: dark pattern taxonomy and detection, automated GUI testing, and mobile advertisements. Accordingly, we review prior work along these three lines to better position our work. Table \ref{tab:priorworks} summarises the key differences between our work and representative prior works on these three directions.
}

\subsection{Dark Pattern Taxonomy and Detection}
Existing research focuses on different aspects of dark patterns, such as dark pattern taxonomies identification and summarization \cite{gray2018dark, di2020ui, gunawan2021comparative, mathur2019dark, chaudhary2022you, brignull2010dark}, user perception and impacts \cite{cho2021reflect, gak2022distressing, sergeeva2023we, bongard2021definitely}, and efforts toward mitigating or remediating their effects \cite{chen2023unveiling, mansur2023aidui, kollnig2021want, datta2022greasevision}. 

\textbf{Dark Pattern Taxonomies and Summarization}: Early work on dark pattern taxonomies primarily relied on manual or semi-manual exploration of apps and websites to collect interface examples and categorise recurring manipulative design practices \cite{gray2018dark, di2020ui, gunawan2021comparative, mathur2019dark, chaudhary2022you, hidaka2023linguistic, kowalczyk2023understanding, nguyen2022freely}. The concept of dark patterns was initially articulated by Brignull \cite{brignull2010dark} as ``a user interface that has been carefully crafted to trick users into doing things, such as buying insurance with their purchase or signing up for recurring bills.''. Building on this foundation, Gray et al. \cite{gray2018dark} refined the taxonomy into five high-level strategies, Nagging, Interface Interference, Forced Action, Obstruction, and Sneaking, which has since served as a widely adopted conceptual framework. Subsequent studies further enriched these taxonomies across diverse domains, including games \cite{aagaard2022game}, e-commerce websites \cite{mathur2019dark}, and mobile applications \cite{di2020ui}.

\textbf{Understanding User Perception and Impacts}: Another stream of dark pattern research investigates user perceptions of and responses to manipulative interface designs, as well as their broader impacts on end users \cite{cho2021reflect, gak2022distressing, sergeeva2023we, bongard2021definitely}. For instance, Cho et al. \cite{cho2021reflect} analysed feature-level usage logs from multiple social media applications and showed that a substantial proportion of user sessions were associated with regretful experiences, revealing how interface design choices can undermine user autonomy. Sergeeva et al. \cite{sergeeva2023we} examined user reactions to persuasive tactics in permission-based advertising emails and identified several reactance-triggering factors that lead to negative attitudes and reduced trust. Collectively, these studies provide empirical evidence that dark patterns impose measurable cognitive and emotional burdens on users, reinforcing the need for systematic detection and mitigation mechanisms.

\textbf{Mitigation and Remediation}: Recognising the severity of dark patterns and their impacts, researchers have explored various remediation and detection strategies \cite{chen2023unveiling, mansur2023aidui, kollnig2021want, datta2022greasevision}. One line of work focuses on mitigating harm by directly removing or masking deceptive elements in deployed applications. For example, Kollnig et al. \cite{kollnig2021want} introduced GreaseDroid, which allows end users to apply expert-crafted patches to modify application code and eliminate dark patterns. However, such approaches require substantial expertise, are difficult to generalise, and may introduce functional or privacy risks. Datta et al. \cite{datta2022greasevision} later proposed GreaseVision, which lowers the barrier to patch creation by enabling end users to specify modifications via screenshots, though it still relies on manual intervention and post-hoc repair.

\revision{
\textbf{(Semi-)automated Detection}: Complementary to remediation, recent work has begun to explore (semi-)automated detection of dark patterns to proactively alert developers, regulators, and end users. Some approaches focus on web interfaces. Mathur et al. \cite{mathur2019dark} proposed a large-scale crawling framework for e-commerce websites that segments HTML pages into functional components and identifies dark patterns through semi-automated clustering and manual annotation. Raju et al. \cite{raju2021smart} introduced a rule-based notification system that detects predefined dark pattern signatures from webpage source code, though its evaluation primarily relies on illustrative examples rather than systematic benchmarks. Other work targets specific categories of deceptive patterns, such as DarkDialogs \cite{kirkman2023darkdialogs} for consent dialogs and DarkFleece \cite{yue2024darkfleece} for subscription-related fleeceware using feature-driven classifiers.
}

\revision{
The most closely related approaches to our work include AidUI \cite{mansur2023aidui}, UIGuard \cite{chen2023unveiling}, and \revision{DPGuard \cite{shi202550}}, all of which aim to automatically detect general classes of deceptive patterns in application interfaces. AidUI and UIGuard adopt screenshot-based, rule-driven detection pipelines. Both rely on computer vision and optical character recognition (OCR) to extract UI elements and attributes from \revision{intra-page} screenshots, and infer deceptive patterns using handcrafted rules. AidUI \cite{mansur2023aidui} employs UIED \cite{chen2020object} and Google OCR, together with coarse color categorization and spatial heuristics, which yields interpretable but brittle rules and limits support to a small set of predefined deceptive pattern types. UIGuard \cite{chen2023unveiling} extends this paradigm by grounding rule design in a taxonomy-driven analysis and incorporating richer visual cues, such as grouping and layout, via Faster R-CNN and Paddle-OCR. Nevertheless, both approaches depend on manually specified visual cues and fixed logical rules, making them sensitive to UI style variations and limiting their generalization and scalability. DPGuard \cite{shi202550} departs from explicit rule engineering by leveraging multimodal large language models (MLLMs) for deceptive pattern detection across both web and mobile interfaces. It combines a binary classifier with MLLMs and employs prompt mutation to iteratively optimize prompts. While this design reduces reliance on handcrafted rules and enables cross-platform detection, DPGuard formulates detection as a screen-level classification problem, without localizing deceptive mechanisms or modeling how they arise through interaction. Moreover, its performance implicitly depends on the extent to which pretrained MLLMs capture domain knowledge of deceptive patterns. Finally, despite their methodological differences, all these approaches assume pre-collected, \revision{intra-page} UI screens as input. Consequently, they are inherently limited in capturing deceptive behaviors that emerge through multi-step interactions or span multiple pages, and their reliance on external UI collection pipelines further constrains scalability and end-to-end automation.
}

\revision{
Our work addresses these limitations by re-framing deceptive pattern detection as an end-to-end, interaction-aware analysis problem. Instead of treating deception as a \revision{intra-page} property of isolated UI screens, we model it as an interactional phenomenon that emerges through user actions and app-level navigation, enabling the detection of patterns that manifest only across interaction sequences, such as \textit{Sneak into the Basket}.
Beyond interaction awareness, we adopt a principled hybrid reasoning strategy that integrates learning-based and rule-based methods. Learning-based components capture latent and previously unarticulated manifestations of deceptive behavior from data, while rule-based mechanisms encode high-confidence expert knowledge and enforce explicit semantic constraints. This combination balances adaptability and generalization with interpretability and reliability, overcoming the limitations of purely rule-driven or purely data-driven approaches.
}

\subsection{GUI Testing}

\revision{
Automated GUI testing plays a crucial role in ensuring software quality and has been extensively studied in both academia and industry. It typically consists of two core tasks: systematically exploring graphical user interfaces \cite{monkey, li2017droidbot, lv2022fastbot2, lan2024deeply, liu2024make, yoon2024intent, wang2025llmdroid, li2019humanoid}, and leveraging this exploration process to support downstream testing, analysis, and verification objectives \cite{chen2022towards, feng2024prompting, yuan2024designrepair, zhang2023automated, liu2020owl, chen2020unblind}.
}

\revision{
\textbf{GUI Exploration.}
The primary goal of GUI exploration is to achieve high UI coverage, as comprehensive exploration forms the foundation for many downstream tasks. Early approaches relied on non-learning-based techniques. For example, Monkey \cite{monkey} employs pseudo-random event generation to trigger UI actions and explore interface states. To address the inefficiency and limited coverage of random exploration, subsequent work introduced more structured strategies. DroidBot \cite{li2017droidbot} adopts a model-based approach that guides exploration using a state transition graph without requiring instrumentation.
}

\revision{
To better capture realistic interaction behaviors, later work incorporated learning-based methods. Humanoid \cite{li2019humanoid} leverages reinforcement learning to generate human-like interaction sequences, while Fastbot2 \cite{lv2022fastbot2} combines model-based exploration with reinforcement learning and knowledge reuse from prior executions to improve exploration efficiency in large-scale industrial settings. DQT \cite{lan2024deeply} further formulates GUI testing as a deep reinforcement learning problem using graph-based UI representations and curiosity-driven rewards to encourage exploration of unseen states. 
}

\revision{
More recently, research has explored LLM-assisted GUI exploration to better handle complex, multi-step interactions and task-oriented navigation. Approaches such as GPTDroid \cite{liu2024make} and DroidAgent \cite{yoon2024intent} leverage large language models with memory mechanisms and tailored prompt designs to generate intent-driven interaction sequences that simulate realistic user tasks.
}

\revision{
Compared with prior GUI exploration approaches that primarily optimise for coverage or task completion, our work differs at a conceptual level by jointly leveraging automated exploration and task-oriented interaction modeling to explicitly target deceptive behaviors.
}

\revision{
\textbf{Downstream Testing and Analysis.}
Building upon GUI exploration, prior work has leveraged the exploration process and collected UI traces to support both functional and non-functional testing tasks. Functional testing focuses on validating the correctness, robustness, and performance of user interfaces \cite{monkey, liu2024make, wang2025llmdroid, yoon2024intent, li2019humanoid}, as well as reproducing bugs to assist debugging and repair \cite{feng2024prompting}. In contrast, non-functional testing primarily targets UI quality attributes, including the detection of rendering defects, usability issues, and accessibility violations \cite{chen2020unblind, chen2022towards, zhang2023automated, liu2020owl}.
}

\revision{
While these approaches are effective for improving software quality, they are not designed to reason about ethical concerns in interface design, such as deceptive or manipulative practices, which have attracted increasing attention from regulators, industry, and the research community \cite{nouwens2020dark, nguyen2022freely, brignull2010dark, EUAIAct, EUguidelines, CaliforniaLaw}. Our work addresses this orthogonal gap by repurposing automated GUI exploration as an enabling data collection mechanism and extending GUI-based analysis beyond functional and non-functional quality assurance toward ethical risk identification at the application level.
}

\subsection{Mobile Advertisements}

\revision{
Mobile advertisements constitute one of the most prevalent contexts in which deceptive interface patterns arise. Many manipulative design practices are driven by monetization incentives, such as increasing click-through rates, prolonging user engagement, or enforcing subscriptions, making advertisements a natural vehicle for deploying such tactics within \revision{inter-page} user flows.
}

\revision{
Prior research on mobile advertisements has largely focused on system-level risks and ecosystem analysis. A substantial body of work examines the privacy and security implications of mobile advertising infrastructures, including tracking, data leakage, and malicious ad delivery \cite{chen2019revisiting, liu2020maddroid, shao2018understanding}. Other studies have developed automated techniques to identify ad networks, libraries, and fraudulent or abusive advertising behaviours \cite{dong2018frauddroid, liu2020maddroid, liu2025ios}. While a subset of this work touches upon deceptive advertisements, the primary emphasis remains on measuring prevalence, security threats, or policy violations rather than characterising manipulative interface behaviours. For instance, MadDroid \cite{liu2020maddroid} conducted a large-scale analysis of Android applications and identified multiple categories of devious advertisement content, such as click-deceptive images, censored images, and malicious redirection behaviours, revealing that a non-trivial fraction of apps embeds problematic ads. Similarly, Liu et al. \cite{liu2025ios} audited iOS applications and reported several classes of abusive advertisements, including deceptive or disruptive ads and external security threats, using LLM-assisted detection followed by manual verification.
}

\revision{
Complementary to system-level analyses, several studies investigate user perceptions of mobile advertisements. Analyses of large-scale app reviews reveal that users frequently report frustration with intrusive or excessive ads, particularly in terms of timing, frequency, and placement, as well as indirect burdens such as performance degradation and battery drain \cite{gui2017aspects, gao2021users, gao2022understanding}. These findings suggest that user harm often arises not only from malicious intent but also from interface design choices that disrupt task flow or undermine user autonomy.
}

In contrast to prior work that centres on ad content, delivery mechanisms, or subjective user complaints, our work adopts a human-computer interaction perspective to operationalise deceptive pattern taxonomies for identifying manipulative interface behaviours within advertisement-driven interaction flows. Rather than analysing advertisements as isolated content, we focus on how ad-related interfaces embed concrete design strategies that steer user behaviour across interactions.

\revision{
Finally, it is important to distinguish advertisements from dark patterns. While advertisements frequently provide a context in which dark patterns emerge, largely due to monetization incentives or, in some cases, security concerns, the two are not synonymous. Advertisements serve as a content delivery and monetization mechanism, whereas dark patterns constitute a broader class of manipulative interface designs that erode user autonomy across diverse contexts, including consent dialogs, subscription management, and multi-step interaction flows. In practice, advertisements and dark patterns partially overlap, but they differ fundamentally in scope and intent.
}

\section{Methodology}

\begin{figure}[]
\centering
  \includegraphics[width=\linewidth]{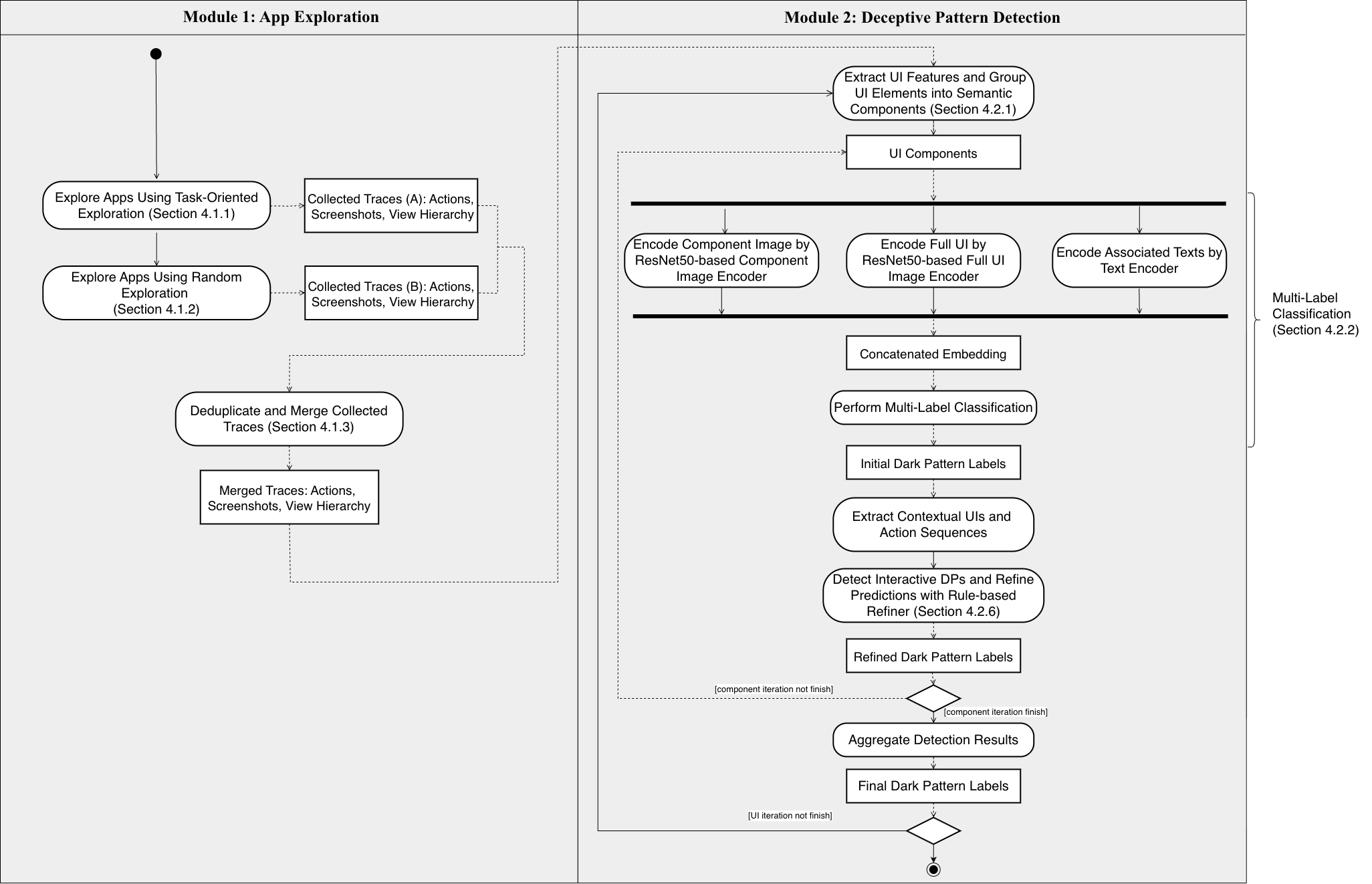}
  \caption{\revision{Overview of the end-to-end pipeline of \tool{}. \tool{} consists of two modules. (1) \textit{App Exploration} collects UI screenshots, view hierarchies, and action sequences via task-oriented and random exploration, followed by trace deduplication and merging. (2) \textit{Deceptive Pattern Detection} groups UI elements into semantic components, encodes component images, full UI images, and associated texts using multi-modal encoders, and predicts initial labels via multi-label classification. A rule-based refiner then leverages UI context and interaction sequences to detect \revision{inter-page} deceptive patterns and refine predictions, producing the final dark pattern labels.}
  }
  \label{fig:appray_uml_activity}
\end{figure}

\revision{Figure~\ref{fig:appray_uml_activity} shows the end-to-end workflow of our \tool{}. Our system consists of an app exploration module and a deceptive pattern detection module. The exploration module installs the app, executes it in real time, explores its UI, captures rendered screens and their view hierarchies, and records all interacted elements and actions. The deceptive pattern detection module then leverages the collected information to identify deceptive patterns across UIs.}

\subsection{App Exploration}
\label{sec:appexploration}
The app exploration module has two phases. 
We first conduct task-oriented exploration based on large language model (LLM) to perform tasks that are prone to have potential deceptive patterns.
We then employ existing app exploration tool to collect as many UI status as possible. 
\revision{Unlike prior GUI exploration approaches that optimise coverage or task completion, our method synergistically combines learning-based exploration with LLM-driven, task-oriented interaction modeling to induce deceptive behaviors. This design balances exploration efficiency and cost while substantially increasing the likelihood of discovering interaction-dependent dark patterns.}

\subsubsection{Task-Oriented App Exploration}
\label{sec:taskAppExploration}

While random exploration can capture many UI states, it overlooks UI semantics, often getting stuck on certain pages and failing to perform meaningful actions.
Unfortunately, certain DPs only come to light during logical, sequential exploration of the app. 
Failing to do any of the steps can not lead to the final page that contains DPs.
Furthermore, some routine tasks can be hotspots for some DPs. 
Simply navigating to the notification settings page might unveil the ``Preselection'' tactic, where choices are made on behalf of the user without clear consent.
Thus, task-oriented exploration, which mirrors human interaction, is essential.
To this end, we harness the commonsense knowledge of large language models, i.e. GPT4~\cite{openai2023gpt4}, for targeted app exploration.

We first define a set of common tasks and a feature-specific task.
Then, given an app under test and the task, we 
(1) obtain and process the current UI's view hierarchy into text; (2) feed this to the LLM and predict the next action for the current task; (3) perform the action on behalf of LLM and repeat Steps 1-3 until reaching the max step, or until the LLM determines that the task is complete.

\textbf{Task Definitions.}
We consider general tasks and feature-based tasks.
The general tasks are common tasks that are prone to contain deceptive patterns~\cite{di2020ui}.
For example, most apps should have notification setting page to control the way the app inform users. Therefore, by setting the task ``go to the setting page and turn off all notification'', we can check whether these notifications are enabled by default.
Feature-based tasks focus on app-specific functionalities, initially targeting shopping features to reveal potential DPs. The primary goal of task-oriented exploration is to delve into more UI states coherently, complementing the limitations of random exploration. Completing these tasks is secondary and inconsequential to the exploration process.
\revision{Specifically, Table \ref{tab:tasks} summarizes the LLM-based task-oriented app tasks, their selection rationale in terms of potential deceptive patterns, and the corresponding task completion rates.}

\begin{table*}[]
    \caption{Task Definitions, Potential Associated deceptive Patterns, and Results}
    \resizebox{1.0\textwidth}{!}{
    \begin{tabular}{l p{0.55\linewidth} p{0.35\linewidth} p{0.15\linewidth}}
    \hline
    \textbf{ID} & \textbf{Task} & \textbf{Potential deceptive Patterns} & \textbf{Success Rate}\\ \hline
    \textbf{T1} & Register an account  & Preselection, Nagging, Privacy Zuckering & 56.41\% \\ 
    \textbf{T2} & Sign in  & Preselection, Nagging, False Hierarchy & 78.21\% \\ 
    \textbf{T3} & Go to setting page, go through all notification related pages & Preselection & 82.83\% \\ 
    \textbf{T4} & Go to setting page, go through all privacy related setting & Preselection, Privacy Zuckering & 84.0\%  \\ 
    \textbf{T5} & Check if we can subscribe to premium account, if so, read  through all contents on the subscription page  & False Hierarchy, Preselection, Forced Continuity & 83.33\%   \\ 
    \textbf{T6} & (Optional) Select any product you like with proper attributes  (like size), add to cart, proceed to checkout & Sneak into the Basket, Hidden Costs, Preselection & 30.0\% \\
    \textbf{T7} & Sign out the app & Roach Motel, False Hierarchy & 58.11\%\\
    \hline
    \end{tabular}
    }
    \label{tab:tasks}
\end{table*}

\textbf{UI Information Extraction.}
\label{subsubsec:info_ext}
To feed data to GPT4, we first convert the UI information into textual format.
Each user interface is represented as a tree structure with leaf UI elements like Buttons for interaction, TextViews for conveying textual information, ImageViews for displaying images, and layout elements organising the leaf elements into a structure way like LinearLayout for placing the elements in a linear order.
Each element has attributes such as text, clickable, and classname, indicating its properties. We use the Android Debugging Bridge (ADB) to obtain the current UI's view hierarchy, extract key information, and feed it into the LLM to understand the UI and make decisions.
In detail, we extract the following information:
\begin{itemize}[leftmargin=17pt]
    \item \textbf{Classname} of UI elements indicate their functionalities, such as a Button that allows user interaction.
    \item \textbf{Resource ID} is used as a unique identifier for UI elements, indicating their semantic meaning. 
    \item \textbf{Text Content}: Text and Content-description contains the text on elements or the accessibility label for image-based buttons.
    \item \textbf{Action-Related Attributes} including clickable, scrollable, checked, indicates possible actions and the element's current status. For example, a clickable and checked CheckBox element means it can be clicked and is currently checked.
    \item \textbf{Bounds} specifies the element position on the UI.
\end{itemize}

\textbf{Prompt Engineering and Action Space.}
We adopted the in-context learning with few shot examples to optimise the prompt and facilitate the task-oriented app exploration.  
\revision{The initial prompt followed the best practices from existing works~\cite{vu2023voicify, wang2023enabling}.}
We provide overall instructions to GPT, including the expected output, and one example for each action to clarify the task. 
We consider five common actions, namely tap, scroll, type, back and stop. Stop is specially designed when the LLM determines that the task is finished.
Specifically, our prompt consists of three main parts, the system prompt that define the role, basic actions, and some instructions, the few shot examples, and the user prompt for prompting the LLM to predict the next action. 
\revision{We prepared a separate set of 10 apps outside our test set, specifically for iterative prompt refinement. For each iteration, we ran the prompt across these apps, identified systematic error types, e.g., premature stopping in multi-scroll scenarios, and not terminating after completing the task, and consolidated these into 10 corrective instructions. This refinement process required approximately five iterations before the prompt consistently performed well across all apps in the evaluation subset. The final prompt was then applied to the full set of 100 apps.}
An example of our prompt can be found in the appendix - Figure~\ref{fig:prompt}.

\textbf{Overall Interaction between Device and LLM.} 
For each round, we first obtain the UI information from the device using \revision{UI Automator 2, an official Android automation framework that enables programmatic interaction with mobile applications and provides access to runtime UI hierarchies and widget attributes,} and convert it into text format. We then fill the prompt template with the current task, history actions, UI information, and query GPT for the next actions. Once we obtain the actions, we extract the action point using the bounds of the target element, and perform the action on the device using \revision{UI Automator 2}. This process is iterative until the LLM provides a stop action or the number of steps reaches the predefined threshold.

\subsubsection{Random Exploration}

We use the state-of-the-art automated app exploration tool, FastBot2~\cite{lv2022fastbot2}, developed by ByteDance, to capture as many UI states as possible. FastBot2 employs a probabilistic model and a reinforcement learning model to collaboratively determine the best strategy for uncovering unique UI states. It identifies the optimal action based on the current UI information, available UI elements, and historical actions, leading to an effective exploration strategy. 
Note that this automated tool can be replaced by any other app exploration tool, and it is not our primary contribution. 

\subsubsection{UI Deduplication and Merging} 
After we obtain UIs from both automatic and targeted exploration, we conduct post-processing to merge these obtained UIs by identifying duplicate UIs and UIs that are not related to the apps.
For deduplication, we consider two measures, i.e., UI screenshot matching and view hierarchy matching. 
For UI screenshot matching, we check whether their images are the same by simple pixel matching method. 
For view hierarchy matching, we extract all the leaf elements of the screen. 
As some elements are invisible, we check their ``visible-to-users'' attribute and only keep ones that are visible. 
Some elements are too small and meaningless, like the divider, so we remove elements that has a width or height less than 5.
We only keep the class, resource-id and the ``checked'' attributes of the elements to avoid saving many UIs with subtle or meaningless differences.
For example, some UIs may be dynamic but show a similar meaning, like the music app will recommend different music every time we refresh the list.
After filtering, if two view hierarchies are the same, we consider them as duplicates.

\subsection{Deceptive Pattern Detection (DPD)}
\label{sec:detector}
Exiting work mainly focus on single UI detection, which are prone to have false negatives and can not deal with deceptive patterns that are related to several UIs.
As we incorporate the exploration in our work, our method has the capability to deal with \revision{\revision{inter-page}} DPs.

Given a UI under test, we initiate the detection process by extracting properties of the UI elements, grouping them into semantically distinct clusters/components using heuristic rules. Subsequently, we deploy our proposed component-based deceptive pattern detector on each identified element or component.
Once we obtain the initial detection results, we employ our rule refiner to check surrounding UIs and make final decisions.
An illustration example can be found at Appendix~\ref{sec:inference_process}.

\subsubsection{UI Information Extraction and Grouping}
\label{sec:ui_extraction_grouping}
\begin{figure}[t]
    \centering

    \begin{subfigure}{0.19\linewidth}
        \includegraphics[width=\linewidth]{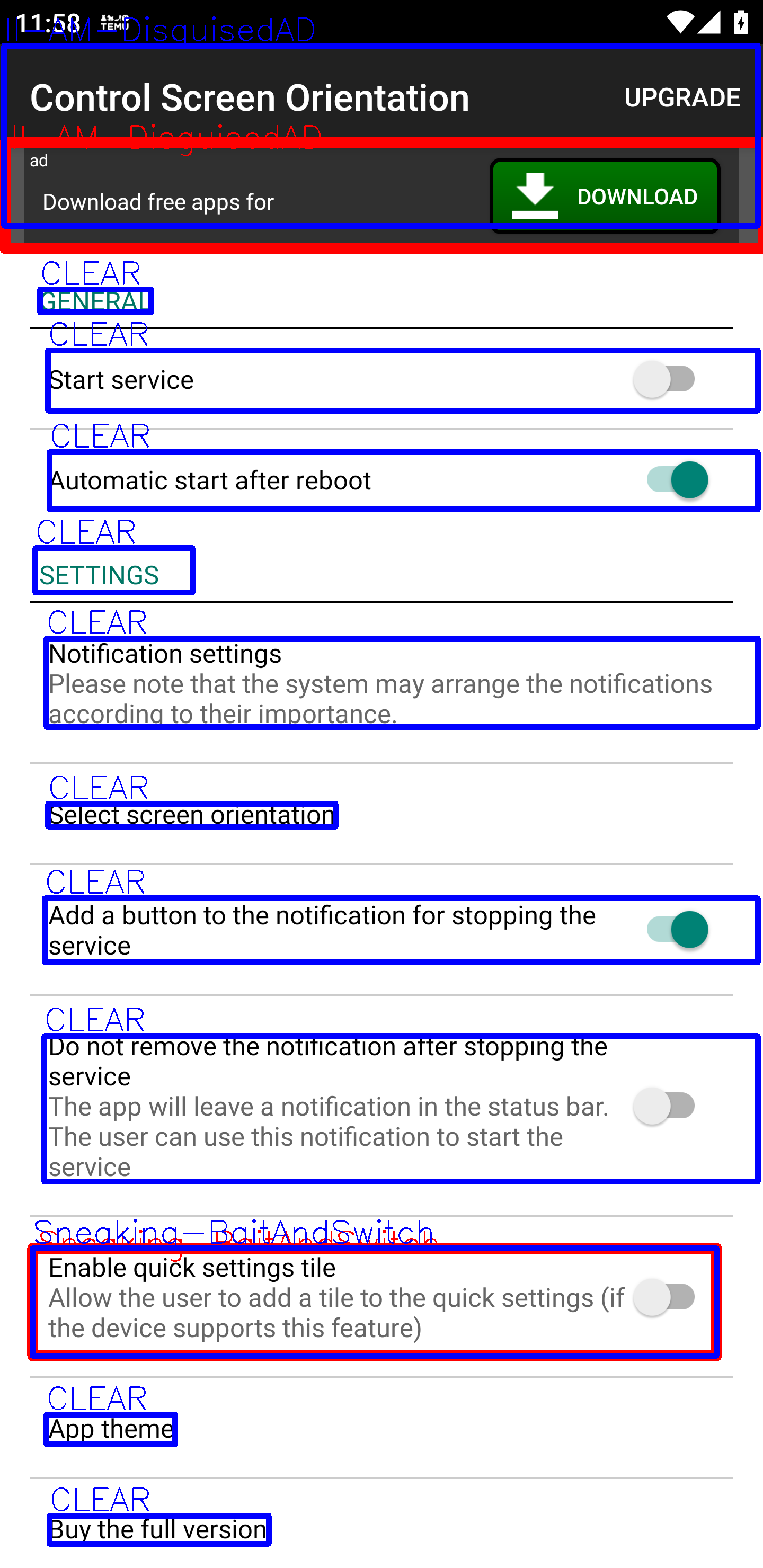}
        \caption{Disguised Ads \& Bait and Switch}
    \end{subfigure}
    \hfill
    \begin{subfigure}{0.19\linewidth}
        \includegraphics[width=\linewidth]{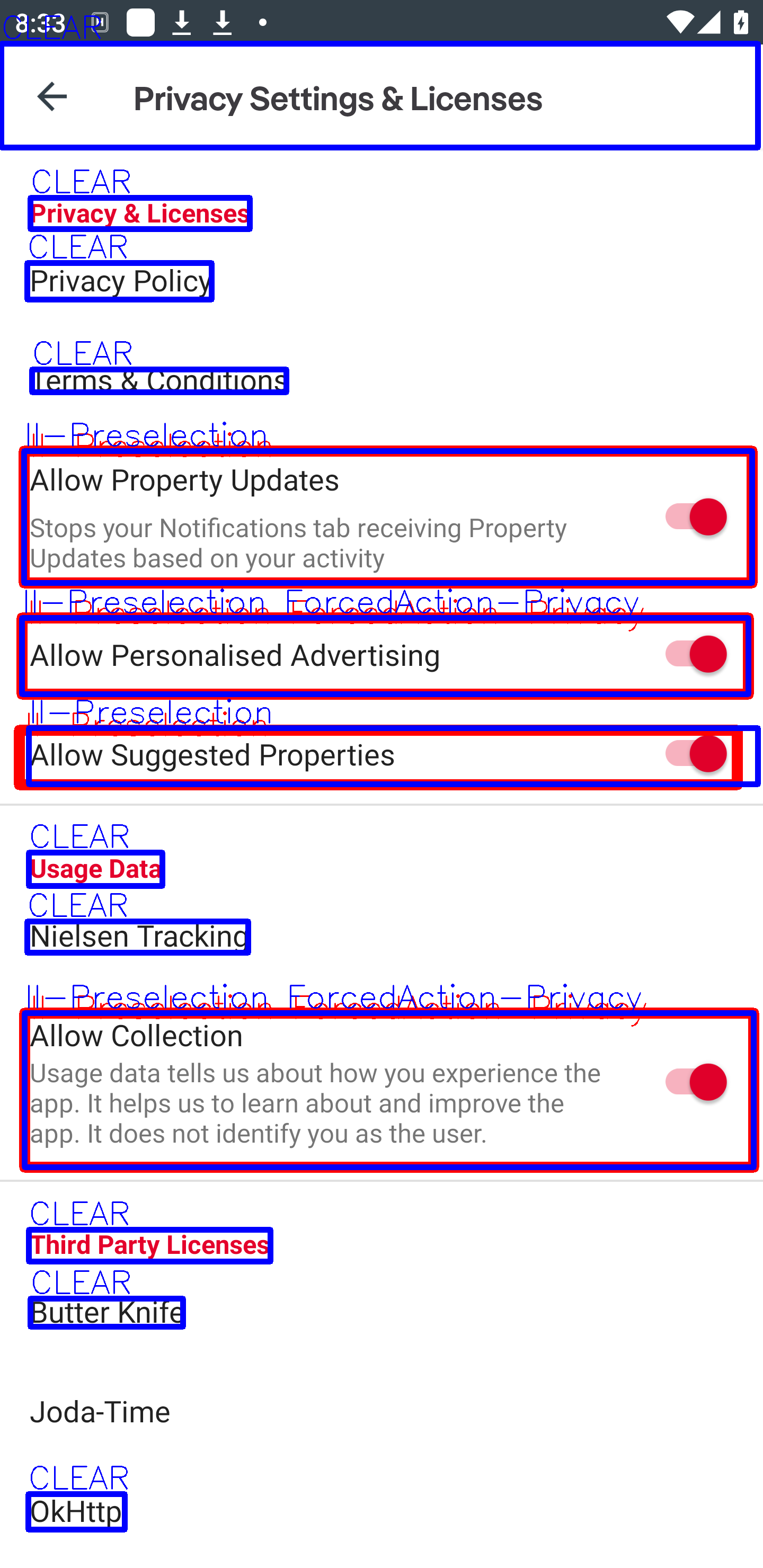}
        \caption{Preselection}
    \end{subfigure}
    \hfill
    \begin{subfigure}{0.19\linewidth}
        \includegraphics[width=\linewidth]{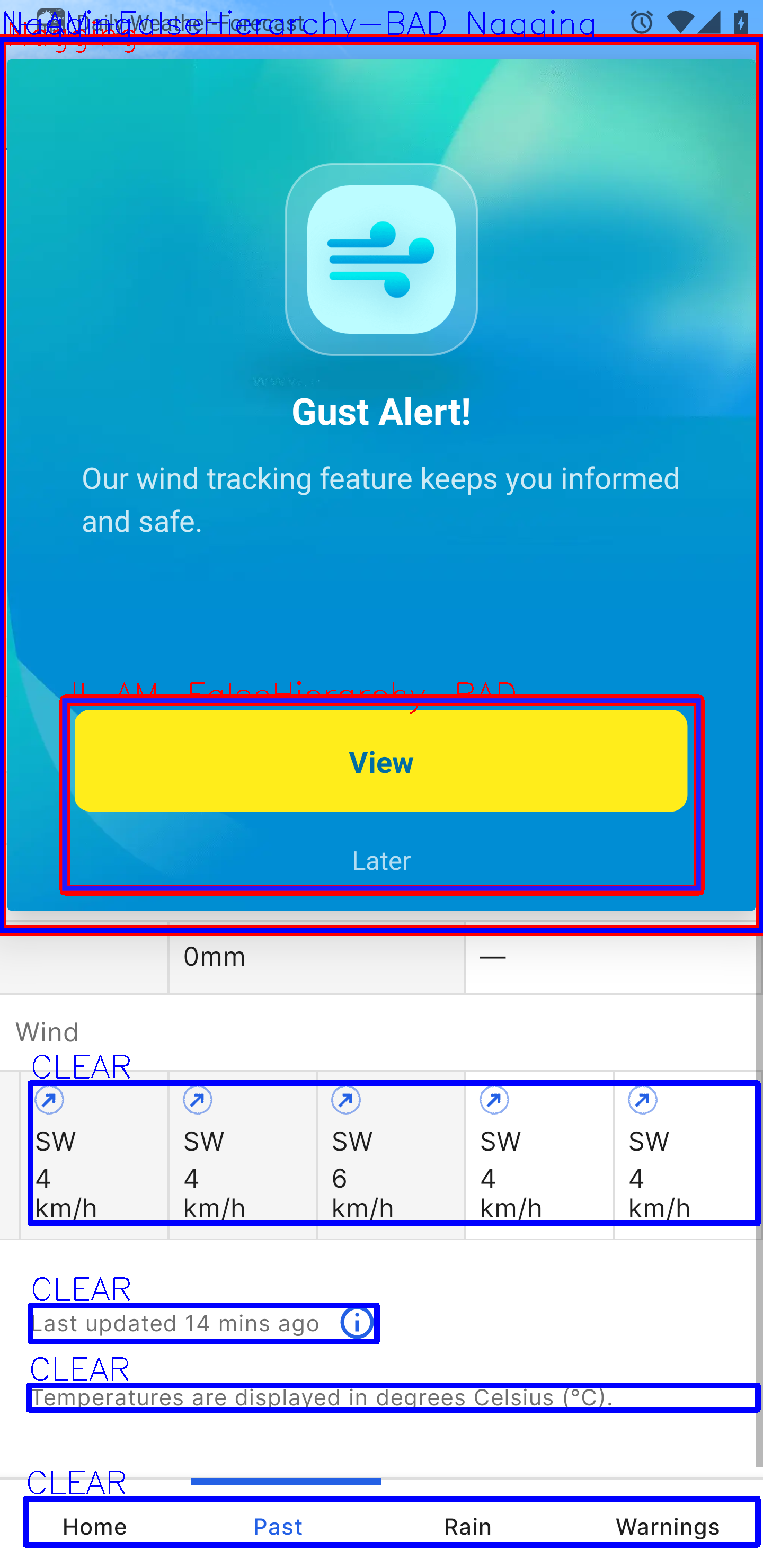}
        \caption{Nagging \& False Hierarchy}
    \end{subfigure}
    \hfill
    \begin{subfigure}{0.19\linewidth}
        \includegraphics[width=\linewidth]{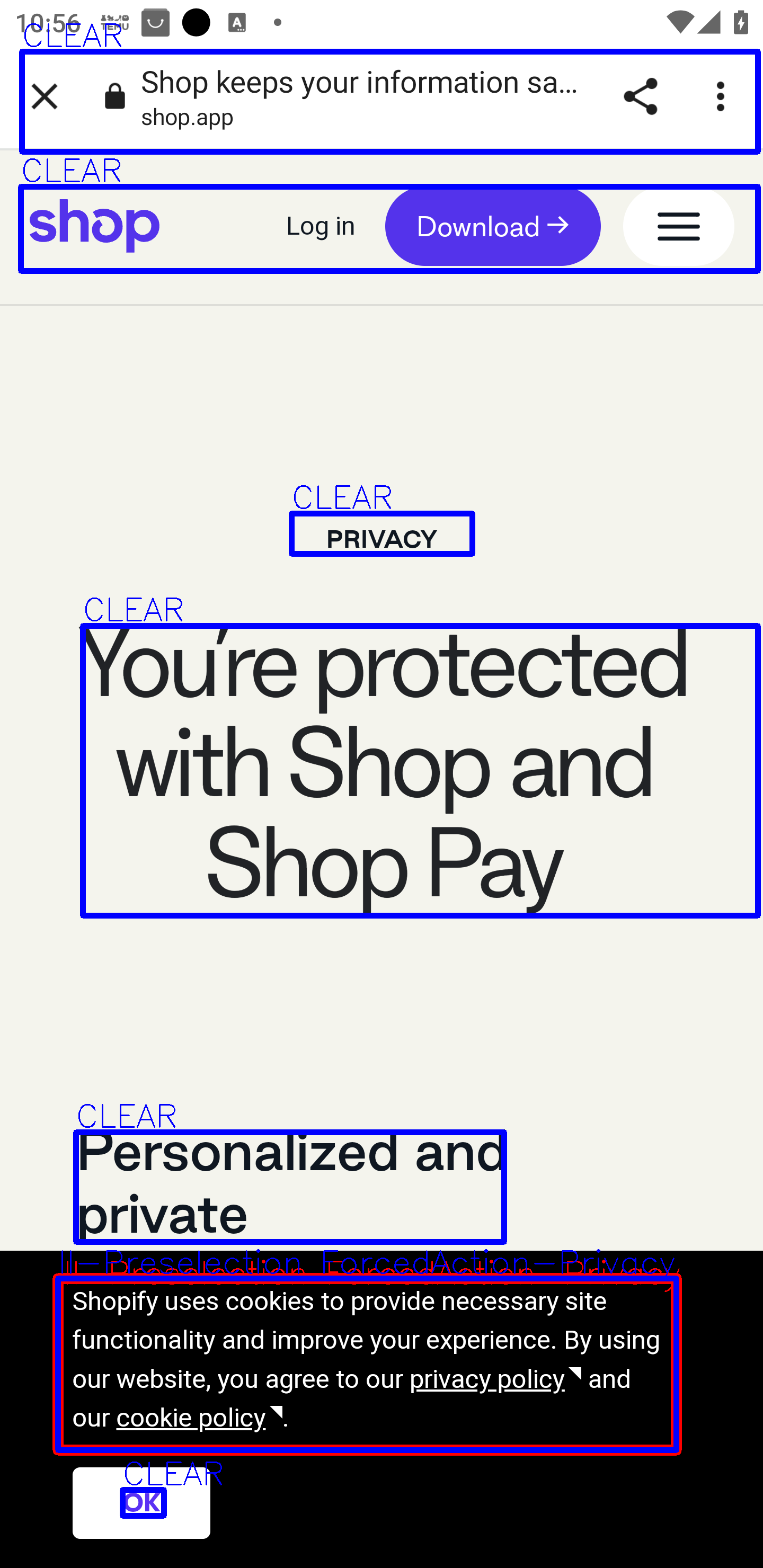}
        \caption{Privacy Zuckering}
    \end{subfigure}
    \hfill
    \begin{subfigure}{0.19\linewidth}
        \includegraphics[width=\linewidth]{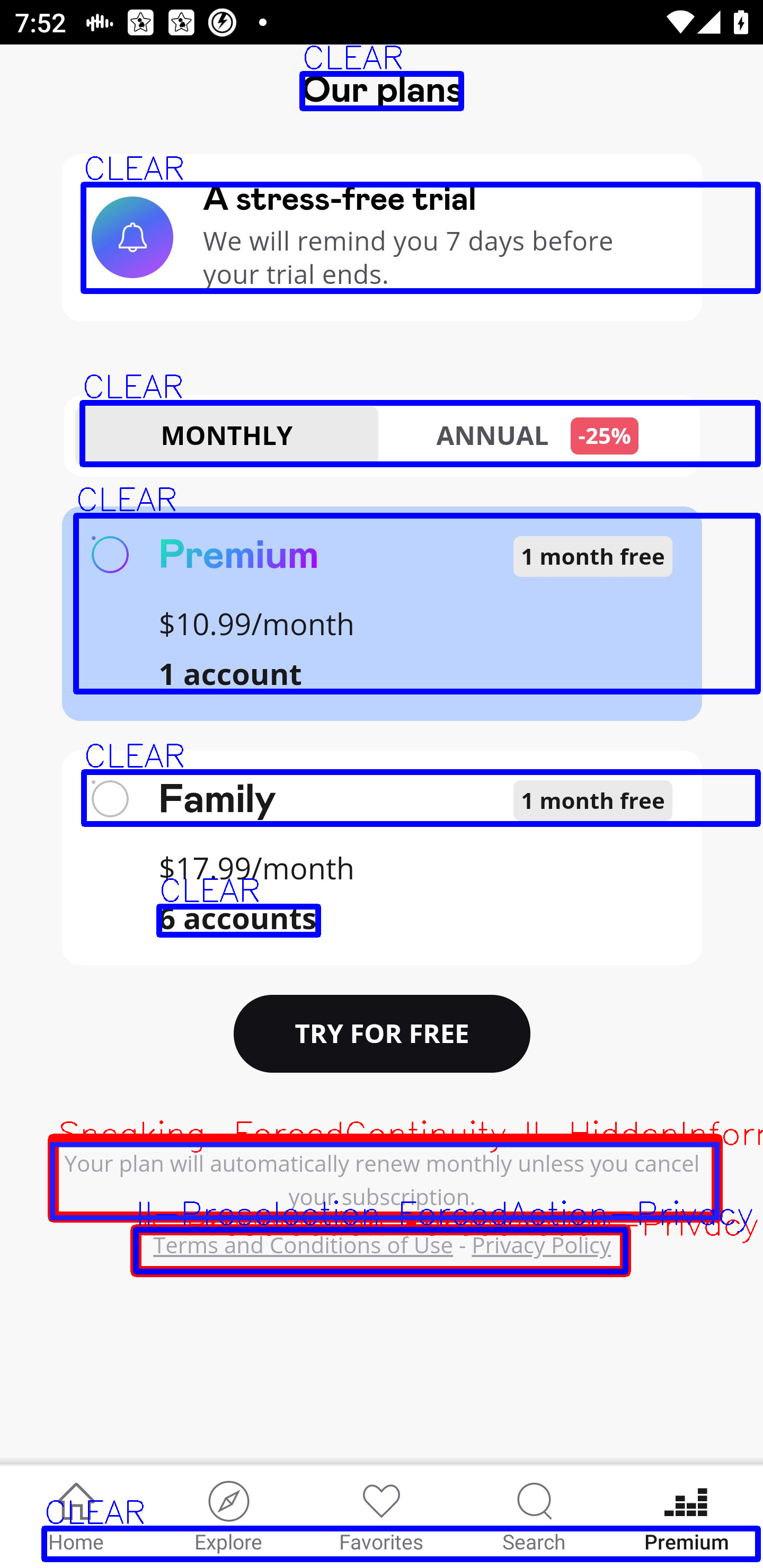}
        \caption{Hidden Information}
    \end{subfigure}

    \caption{
    \revision{
    Examples of UI component grouping produced by our grouping approach.
    Horizontally aligned elements (e.g., label–toggle pairs) are grouped into rows,
    vertically adjacent elements form stacked components (e.g. False Hierarchy),
    and elements fully enclosed by a container (e.g., Nagging) are merged into a single semantic component.
    }
    }
    \label{fig:ui_group_examples}
\end{figure}

We directly leverage UIGuard~\cite{chen2023unveiling} to extract low-level UI information, including element bounding boxes, element types, textual content, and interaction states. Building upon these atomic UI elements, we then group semantically related elements into higher-level UI components (e.g., all elements within a confirmation pop-up are grouped into a single semantic cluster). Our grouping strategy is three heuristics rules and \revision{non-iterative. Given a set of detected UI elements, we first generate candidate groups using three complementary heuristic rules. }

\begin{itemize}[leftmargin=17pt]
    \item \textbf{Horizontal Alignment:} UI elements that are horizontally aligned within a fixed pixel tolerance are grouped together, capturing common row-based layouts such as icon–text–toggle patterns.
    \item \textbf{Vertical Adjacency:} Elements that are vertically adjacent within a predefined threshold are clustered to model stacked structures such as lists, menus, or multi-line components. 
    \item \textbf{Containment:} If an element is fully contained within the bounding box of a larger element, it is merged with that element, which is particularly effective for dialogs, banners, and card-based layouts. If multiple larger elements overlap, the method creates separate components for each layer.
\end{itemize}

\revision{
All thresholds are defined in pixel space and tuned on a held-out development set of mobile screenshots (e.g. 1440×2960 resolution). A tolerance of 50 pixels balances robustness to detector jitter with preservation of semantic boundaries, while the 70\% overlap criterion reliably merges duplicate group proposals without collapsing distinct UI regions. }

\revision{
Figure~\ref{fig:ui_group_examples} shows typical grouping results produced by our method. The red bounding boxes indicate the ground-truth annotations with their corresponding labels, while the blue bounding boxes are produced by our grouping method. Elements that do not correspond to any DP are labeled as \texttt{CLEAR}.
For advertisement banners and privacy consent notices (a)(d), text and action buttons are grouped into a single component using the containment rule.
In the settings pages (a)(b), each option row, which usually contains a text label and a toggle switch, is grouped through horizontal alignment, while different rows remain separate.
For full-screen pop-up dialogs (c), all internal elements such as titles, descriptions, and buttons are merged into one component because they are fully contained within the dialog.
In subscription and pricing pages (e), each plan card is grouped as an individual component using containment, while renewal notices and legal text are separated based on vertical spacing.  }

\subsubsection{Component-Based Deceptive Pattern Detector}

After identifying a candidate component (a group of UI elements), we use three inputs for detection: the cropped image of the component, the full UI image in which it appears, and the associated UI text. These inputs are processed by our component-based deceptive pattern (DP) detector. 
The detector consists of (1) a contrastive learning based multi-label classifier that predicts a shortlist of possible DP types, and (2) a lightweight post-processing rule-based refiner (see Section~\ref{subsec:rule_refiner}) that produces the final detection results.

The classifier adopts a multi-modal Siamese architecture to jointly model visual and textual information. It includes a BERT-based text encoder for UI text and a ResNet50-based image encoder for UI images. As shown in Figure~\ref{fig:appray_uml_activity},
\revision{the visual branch contains \textbf{two image encoders with shared weights}. One encoder processes the cropped image of the candidate component, and the other processes the full UI image. The shared-weight ResNet50 encoders have the following benefits. First, the cropped component image and the full UI image belong to the same visual domain. Sharing weights ensures that both inputs are mapped into a common feature space, which makes their representations directly comparable and easier to combine. Second, weight sharing encourages the model to focus on meaningful visual differences caused by context rather than learning separate feature extractors for each input type. This is important for distinguishing deceptive patterns that have similar visual appearance but differ in how they are presented within the UI. Third, using a single set of weights significantly reduces the number of parameters, lowering training overfit and inference latency, which is important for large-scale analysis and mobile deployment.
}

The embeddings from the two image branches and the text encoder are concatenated and passed to a fully connected classifier to predict DP labels. The rule-based refiner then adjusts these predictions using predefined rules based on UI structure, semantics, and interaction behavior. Because deceptive pattern instances are limited and unevenly distributed across classes, we apply data augmentation to underrepresented classes during training to improve robustness and overall detection performance.

\subsubsection{Data Augmentation}

Data augmentation is used to increase training diversity and to mitigate data scarcity and class imbalance across different deceptive pattern (DP) categories. Following prior work, we apply both image-level transformations and text-level perturbations to improve model robustness and generalization~\cite{DBLP:journals/corr/abs-2105-03075,yang2023imagedataaugmentationdeep}. All augmentation operations are applied only to the training set. No augmentation is used for validation or testing.

\revision{For \textbf{image augmentation}, we apply a sequence of transformation functions $ T = \left [ T_1, ... , T_n \right ]$ to UI images and DP component images. Given the input image $I$, the augmented images are obtained as $ I' = T_n(T_{n-1}(…T_1(I)))$. The image augmentation pipeline consists of the following steps:}

\revision{\begin{itemize}[leftmargin=17pt]
\item Resize the input image so that the shorter side is 256 pixels.
\item Randomly crop approximately 1\% of the image area to introduce spatial variation.
\item Apply random horizontal flip with probability 0.5.
\item Apply random brightness and contrast adjustment with probability 0.2, where the adjustment values are sampled from a small range around the original values.
\item Add small Gaussian noise with probability 0.1 to improve robustness to minor visual perturbations.
\item Normalize the image using the dataset mean and standard deviation.
\end{itemize}
}

\revision{For \textbf{text augmentation}, we apply lightweight perturbations to the UI text and textual descriptions associated with DP components. The text augmentation pipeline includes the following steps:}
\revision{\begin{itemize}[leftmargin=17pt]
\item Tokenize the input text and convert it to lowercase.
\item Perform synonym replacement at the word level with probability 0.1, while avoiding stopwords and DP-specific keywords.
\item Apply random word deletion with probability 0.05.
\item Apply random word swapping between nearby words with probability 0.05.
\end{itemize}
}

\revision{This augmentation process generates a larger and more diverse set of training samples. The model can learn more robust features for different types of deceptive patterns and improve generalization.}

\subsubsection{Contrastive Learning and Loss Functions}
Our dataset includes 16 classes of deceptive patterns, where presenting various ambiguities that can confuse the classifier. Contrastive learning is a technique used to learn representations by comparing and contrasting samples in a high-dimensional space~\cite{DBLP:journals/corr/abs-2011-00362}. This approach is particularly effective in unsupervised and semi-supervised learning scenarios, where labeled data is scarce. The main idea is to bring \revision{representations of samples from the same class} closer together \revision{while pushing representations from different classes} further apart in the feature space. For instance, the DPs ``Disguised AD'' and ``Nagging'' can both be located at the bottom of a UI with similar attributes. However, their main contexts are quite different for the two DPs. ``Disguised AD'' fakes as a normal feature, appearing similar to surrounding components, while ``Nagging'' repeatedly pops up, obstructing the user's normal experience and urging them to agree to something. Without contrastive learning, the model would struggle to distinguish between these two patterns. We will provide detailed comparisons how contrastive learning helps address this issue in Section~\ref{sec:rq1.2_detector}.

\textbf{Contrastive Loss:} 
Given \revision{an anchor sample} $x_i$ and \revision{a positive sample} $x_j$ \revision{from the same class}, and \revision{a set of of negative samples $x_k$ from different classes, let $f_w(\cdot )$ denote the learned embedding function. The constrastive loss is defined as:}
\begin{equation}
L_{D}=−log\frac{exp(D\left( x_i, x_j \right)/\tau)} {\sum_{k=1}^{N}exp(D\left( x_i, x_k \right)/\tau)}
\end{equation}
where $D\left( x_i, x_j \right) = \left\| f_{w}(x_i), f_{w}(x_j) \right\|_2$ \revision{denotes} the similarity between \revision{two embeddings }$f_w(x_i)$ and $f_w(x_j)$. $\left\| \cdot  \right\|$ is a similarity metrics (e.g., Cosine similarity or Euclidean distance), $\tau$ is a temperature parameter and $N$ is the number of negative samples. It encourages the representations of positive pairs to be similar and those of negative pairs to be dissimilar. 

The \revision{overall }contrastive loss can be expressed as:
\begin{equation}
L_{contrastive} = (1-y)L^{-}_{D}+yL^{+}_{D} 
\end{equation}
where the term $y$ specifies whether the two given data inputs ($x_i$ and $x_j$) are similar ($y=1$) or dissimilar ($y=0$). The $L^{-}_{D}$ term stands for the contrastive distance $L_{D}$ when the data inputs are from the different classes. The $L^{+}_{D}$ term stands for the contrastive distance $L_{D}$ when the data inputs are from the same classes. 

\textbf{Cross-entropy loss} is used to measure the performance of the final classifier and can be defined as:
\begin{equation}
L_{crossentropy} = - \sum_{i=1}^{C} y_i \log(\hat{y}_i)
\end{equation}
\revision{where $C$ is the number of DP classes, $y_i$ is the ground-truth label, and $\hat{y}_i$ is is the predicted label.}

\textbf{Class Weight and Overall Loss:} \revision{Due to class imbalance in the dataset, training without adjustment may bias the model toward majority classes. To address this issue, we introduce class weights that assign larger penalties to minority classes so that the model will pay more attention to these minority classes during training. The overall loss function is defined as:
\begin{equation}
\mathbb{L} = L_{crossentropy} + L_{contrastive}
\end{equation}
and the final weighted loss is:
\begin{equation}
\mathcal{L} = \sum_{c=1}^{\left\| C \right\|}w_c\mathbb{L}
\label{eq:final_loss}
\end{equation}
}

\textbf{Negative Samples:} 
Negative samples help the model learn to differentiate between distinct classes by providing examples that are not similar to the anchor sample. By learning to correctly identify these dissimilar examples, the model develops a clearer understanding of the boundaries between different classes in the embedding space. \revision{Negative sampling is performed during contrastive learning in each training iteration}. In the deceptive pattern detection problem, \revision{for each anchor sample $x_i$, we select negative samples $x_k$ from different DP classes to construct contrastive pairs. These negative samples are used in the contrastive loss $L_D$ to push embeddings of different classes apart.} 

\revision{We adopt the following task-oriented negative sampling strategies}: 
\begin{enumerate}[leftmargin=15pt]
    \item \textbf{Random negative sampling}. Negative samples are randomly chosen from other classes. \revision{This strategy is mainly used in the early stage of training to stabilize optimization.}
    \item \textbf{Hard negative sampling}~\cite{DBLP:journals/corr/abs-2010-04592}. Hard negatives are selected based on their similarity to the anchor samples. \revision{A negative sample $x_k$ is considered hard if:
    \begin{equation}
        D^+\left( x_i, x_j \right) - D^-\left( x_i, x_k \right) > \xi
    \end{equation}
    where $\xi$ is a hyperparameter controlling the hardness level of negative samples relative to the anchor samples. This strategy helps the model learn finer distinctions between visually similar DP classes.
    }
    \item \textbf{Balanced negative sampling}. The number of negative samples is balanced with \revision{the number of} positive samples \revision{to prevent biased towards dominant classes during contrastive learning.} 
\end{enumerate}

\subsubsection{Model Training}
Our multi-label classifier is a Siamese model that consists of two pipelines: an image embedding pipeline and a text embedding pipeline, followed by a final MLP classifier with a sigmoid activation function for the final predictions. Each pipeline processes its respective inputs to generate meaningful embeddings. These embeddings are then combined and fed into the final classifier. \revision{Each pipeline extracts modality-specific embeddings, which are later combined for DP prediction. We do not train the entire model in an end-to-end manner. Instead, we use a \textbf{two-stage training strategy}. This design choice is motivated by the learning imbalance between image and text modalities observed in our experiments:
\begin{itemize}
\item \textbf{Modality dominance.} In our dataset, text features often provide stronger and clearer signals than visual features (e.g., explicit consent-related words). During end-to-end training, the model quickly minimizes the loss by relying on the text pipeline, which reduces the gradient contribution to the image encoder. As a result, the image encoder remains undertrained and fails to learn meaningful visual representations.
\item \textbf{Gradient interference.}  
The image encoder (ResNet50) and the text encoder (BERT) require very different learning rates to train effectively. ResNet50 typically uses a learning rate around 0.003, while BERT requires a much smaller rate such as 3e-5. When trained jointly, large parameter updates from the image pipeline can overpower the smaller updates from the text pipeline, leading to unstable optimization and unbalanced learning.
\item \textbf{Inferior multimodal performance.}  
Due to modality dominance and gradient imbalance, end-to-end training leads to poorly balanced multimodal representations. The final classifier relies mainly on one modality, which reduces the contribution of the other modality and results in lower overall detection accuracy, especially for minority DP classes that depend on visual context.
\end{itemize}
To address these issues, we adopt the following two-stage training process, as summarized in Algorithm~\ref{alg:model_training}:}

\begin{itemize}[leftmargin=15pt]
    \item \revision{The training dataset consists of images, including both UI component images and their corresponding full UI images, and associated textual content extracted via Optical Character Recognition (OCR). Each training instance is annotated with multiple labels indicating the deceptive patterns present in the component. During training, data augmentation is applied to both image and text modalities to mitigate class imbalance and enrich underrepresented deceptive pattern classes.}
    \item \revision{\textbf{Training the image encoder:} The image embedding pipeline is built on ResNet-50 and processes both the cropped UI component images and their corresponding full UI images using shared weights. During this stage, a temporary fully connected classification head is attached to the encoder to provide supervised signals for image feature learning.}
    \revision{\item \textbf{Training the text encoder}: The text embedding pipeline is based on BERT and encodes the associated UI text. During this stage, a temporary fully connected classification head is attached to the encoder to provide supervised guidance for learning text representations independently.}
    \item \revision{\textbf{Training the final MLP classifier}: After both embedding pipelines are trained, the temporary classification heads are removed and the encoder weights are frozen. The resulting image and text embeddings are then concatenated and used to train the final MLP classifier.}
    \item The entire training process is optimized using the Adam optimizer, with the overall loss function $\mathcal{L}$  \revision{in equation~\ref{eq:final_loss}.}
\end{itemize}

\begin{algorithm}[!htbp]
\caption{DP Detector Training Procedure (DA + NS)}
\label{alg:model_training}
\small
\begin{algorithmic}[1]
\Require Training set $\mathcal{D}=\{(I_c, I_f, T, y)\}$ where $I_c$ is component image, $I_f$ is full UI image, $T$ is OCR text, and $y$ is multi-label DP labels
\Ensure Image encoder $f_{\text{img}}$, Text encoder $f_{\text{text}}$, and Final MLP classifier $h$

\State \textbf{Init:} image encoder $f_{\text{img}}$ (ResNet50) and text encoder $f_{\text{text}}$ (BERT)

\Statex
\State \textbf{Stage 1: Train image encoder}
\For{each epoch}
  \For{each mini-batch $(I_c, I_f, T, y)$ in $\mathcal{D}$}
    \State \textbf{Image Aug.:} $(\tilde{I}_c,\tilde{I}_f)\gets \text{AugmentImg}(I_c,I_f)$
    \State \textbf{Neg. Sampling:} $(\tilde{I}_c^{-},\tilde{I}_f^{-})\gets \text{NegSampleImg}(I_c,I_f)$
    \State $z_c \gets f_{\text{img}}(\tilde{I}_c)$, $z_f \gets f_{\text{img}}(\tilde{I}_f)$, $z_c^{-} \gets f_{\text{img}}(\tilde{I}_c^{-})$, $z_f^{-} \gets f_{\text{img}}(\tilde{I}_f^{-})$ \Comment{shared weights}
    \State $z_{\text{img}} \gets [z_c; z_f]$, $z_{\text{img}}^{-} \gets [z_c^{-}; z_f^{-}]$ \Comment{concatenate crop + full UI}
    \State $\hat{y} \gets g_{\text{img}}(z_{\text{img}}, z_{\text{img}}^{-})$
    \State Update $f_{\text{img}}, g_{\text{img}}$ using Adam and loss $\mathcal{L}$ (Eq.~\ref{eq:final_loss})
  \EndFor
\EndFor

\Statex
\State \textbf{Stage 2: Train text encoder}
\State Attach temporary classifier $g_{\text{text}}$ to $f_{\text{text}}$
\For{each epoch}
  \For{each mini-batch $(I_c, I_f, T, y)$ in $\mathcal{D}$}
    \State \textbf{Text Aug.:} $\tilde{T} \gets \text{AugmentText}(T)$
    \State \textbf{Neg. Sampling:} $\tilde{T}^{-} \gets \text{NegSampleText}(T)$
    \State $z_{\text{text}} \gets f_{\text{text}}(\tilde{T})$, $z_{\text{text}}^{-} \gets f_{\text{text}}(\tilde{T}^{-})$
    \State $\hat{y} \gets g_{\text{text}}(z_{\text{text}}, z_{\text{text}}^{-})$
    \State Update $f_{\text{text}}, g_{\text{text}}$ using Adam and loss $\mathcal{L}$ (Eq.~\ref{eq:final_loss})
  \EndFor
\EndFor

\Statex
\State \textbf{Stage 3: Train final multi-modal classifier}
\State Freeze weights $f_{\text{img}}$ and $f_{\text{text}}$
\State Initialize final MLP classifier $h$
\For{each epoch}
  \For{each mini-batch $(I_c, I_f, T, y)$ in $\mathcal{D}$}
    \State \textbf{Image Aug.:} $(\tilde{I}_c,\tilde{I}_f)\gets \text{AugmentImg}(I_c,I_f)$
    \State \textbf{Text Aug.:} $\tilde{T} \gets \text{AugmentText}(T)$
    \State \textbf{Neg. Sampling:} $\tilde{T}^{-} \gets \text{NegSampleText}(T)$, $(\tilde{I}_c^{-},\tilde{I}_f^{-})\gets \text{NegSampleImg}$
    \State $z_c \gets f_{\text{img}}(\tilde{I}_c)$, $z_f \gets f_{\text{img}}(\tilde{I}_f)$, $z_c^{-} \gets f_{\text{img}}(\tilde{I}_c^{-})$, $z_f^{-} \gets f_{\text{img}}(\tilde{I}_f^{-})$
    \State $z_{\text{img}} \gets [z_c; z_f]$, $z_{\text{img}}^{-} \gets [z_c^{-}; z_f^{-}]$
    \State $z_{\text{text}} \gets f_{\text{text}}(\tilde{T})$, $z_{\text{text}}^{-} \gets f_{\text{text}}(\tilde{T}^{-})$
    \State $z \gets [z_{\text{img}}; z_{\text{text}}]$, $z^{-} \gets [z_{\text{img}}^{-}; z_{\text{text}}^{-}]$
    \State $\hat{y} \gets h(z, z^{-})$
    \State Update $h$ using Adam and loss $\mathcal{L}$ (Eq.~\ref{eq:final_loss})
  \EndFor
\EndFor

\State \Return $f_{\text{img}}, f_{\text{text}}, h$
\end{algorithmic}

\end{algorithm}

\subsubsection{Rule-Based Refiner}
\label{subsec:rule_refiner}
Deceptive patterns often incorporate domain knowledge and interactions across multiple pages. However, DPD method faces limitations when it analyses pages independently. The DPs appear as standard buttons or text on an individual page, making detection difficult. For instance, in Figure~\ref{fig:dp_example}(a), the \textbf{Direction} button in the car icon has no difference subtly from the other four buttons. The only way to confirm a \textbf{Bait and Switch} pattern is by examining the next page. As shown in Figure 1(b), the following page is categorized as \textbf{Nagging}, indicating that the previous action led to an unrelated page. Similarly, detecting \textbf{Roach Motel} requires analyzing multiple pages to determine whether users can easily enter a flow but have difficulty leaving it. 

By incorporating predefined domain knowledge and \revision{page-to-page} relationships, the rule refiner enhances the detection of \revision{\revision{inter-page}} DPs. \revision{The rule-based refiner is applied as a post-processing step after the DPD method generates initial multi-label predictions. It uses two sources of information: 
\begin{itemize}
    \item Navigation relationships between pages.
    \item Domain-specific rules describing known dependencies and conflicts between deceptive patterns.
\end{itemize}
The refiner checks whether the predicted labels are consistent with these rules. If a label is missing but required by a rule, it is added. If a label violates a rule, it is removed.} This process also improves precision. The DPD method is designed as a high-recall multi-label classifier and may assign extra labels.
For example, in Figure~\ref{fig:dp_example}(h), the upper red bounding box is initially labeled as \textbf{Interface Interference – General}, \textbf{Disguised Ad}, and \textbf{Bait and Switch} by the DPD method. However, if the destination page is not labeled as \textbf{Nagging}, the rule-based refiner removes the \textbf{Bait and Switch} label, since this pattern requires a misleading click followed by an unrelated page.

\begin{algorithm}[!t]
\caption{Rule-Based Refiner}
\label{alg:refiner}
\small
\begin{algorithmic}[1]
\Require Pages $P$, navigation edges $E$, initial predictions $Y$, domain rules $K$
\Ensure Refined predictions $\hat{Y}$
\State $\hat{Y} \gets Y$

\For{each navigation edge $(p_{\text{src}}, e, p_{\text{dst}})$ in $E$}
    \State $L_{\text{src}} \gets \hat{Y}[e]$
    \State $L_{\text{dst}} \gets$ all labels on page $p_{\text{dst}}$
    \If{\text{"Nagging"} $\in L_{\text{dst}}$ and $e$ is clickable}
        \State add \text{"Bait and Switch"} to $\hat{Y}[e]$
    \EndIf
    \If{\text{"Bait and Switch"} $\in L_{\text{src}}$ and \text{"Nagging"} $\notin L_{\text{dst}}$}
        \State remove \text{"Bait and Switch"} from $\hat{Y}[e]$
    \EndIf
\EndFor

\For{each user session $S = \langle p_1, p_2, \dots, p_n \rangle$}
    \State $enter\_easy \gets$ entry requires few clicks
    \State $exit\_hard \gets \textbf{true}$, $obstruct \gets 0$
    \For{each page $p_i$ in $S$}
        \If{no clear exit option on $p_i$}
            \State $obstruct \gets obstruct + 1$
        \Else
            \State $exit\_hard \gets$ exit is hidden or requires many steps
        \EndIf
    \EndFor
    \If{$enter\_easy$ and $exit\_hard$ and $obstruct$ is high}
        \State add \text{"Roach Motel"} to all pages in $S$
    \EndIf
\EndFor

\For{each element $e$}
    \State apply conflict rules in $K$ to $\hat{Y}[e]$
\EndFor

\State \dots
\State \Return $\hat{Y}$
\end{algorithmic}
\end{algorithm}

\revision{Algorithm~\ref{alg:refiner} shows the simplified procedure of the rule-based refiner. The algorithm first applies cross-page rules based on navigation edges, then applies session-level rules for multi-page patterns, and finally resolves label conflicts using predefined constraints.}


\section{Research Questions and Datasets}

\subsection{Research Questions}
In this section, we evaluate our proposed system by investigating the following research questions (RQs): 

\begin{itemize}[leftmargin=13pt]
    \item \textbf{RQ1}: How effective is each module (i.e., app exploration and detector modules) in our proposed system?  
    \revision{We decompose this question into two parts:
    \begin{itemize}
        \item \textbf{RQ1.1 App Exploration Module}: How effective is our technique at navigating apps and collecting UIs?
        \begin{itemize}
            \item \textbf{RQ1.1.1 Task Completion}: How well does the LLM-based app navigator complete the tasks?
            \item \textbf{RQ1.1.2 UI Collection}: How effective is the overall exploration strategy in gathering diverse UI states from apps? 
        \end{itemize}
        \item \textbf{RQ1.2 Dark Pattern Detector}: How well does \tool{} perform in detecting dark patterns? We conducted an ablation study to examine the contributions of each sub-module.
    \end{itemize}
    }

    \item \textbf{RQ2}: How effective is the proposed method in detecting deceptive patterns in UIs compared to existing techniques?

    \item \textbf{RQ3 Usefulness}: To what extent can \tool{} assist humans in manual exploration and identification for an app under test?
\end{itemize}

\subsection{Datasets}
\label{sec:dataset}
As mentioned in Section~\ref{sec:intro}, existing datasets fall short in providing a dataset that covers both \revision{intra-page} and \revision{\revision{inter-page}} deceptive patterns with localisation labels. Therefore, in our work, we leverage the data collected through our app exploration module to fill the gap.

\subsubsection{App Collection}
\label{sec:app_collection}
\begin{figure}[]
\centering
  \includegraphics[width=0.8\linewidth]{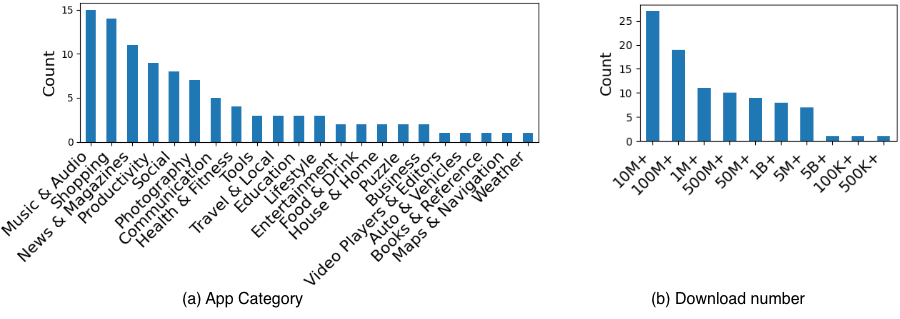}
  \caption{Distributions of our final app set by application category and download count. \revision{The distribution demonstrates that the final app set spans diverse application categories and covers a wide range of download numbers.}}
  \label{fig:app_dist}
\end{figure}

We collected 350 top apps from Google App Store across 22 app categories, and crawled their metadata including ratings, number of downloads, app description, app category, app name, app package name. 
We then use app package name to obtain their latest app package from third-party websites such as AppCombo and AppMirror. Meantime, we saved the app version and app files for future replication. 
After that, we manually installed and tested each app to check if they are usable in emulator. We excluded game apps that leverage game engine like unity. We stopped until we obtained 105 usable apps.
We randomly sampled 1 app from 5 different categories as a \textit{withhold subset for RQ3}, and the rest 100 apps are used in RQ1 and RQ2.
We notice that most apps require email or phone number verification to register accounts, which are currently not supported by our system and are out of our scope. Therefore, for each app, we first register an account and save the account information for further usage. Such information can be easily provided by the app provider.
Figure~\ref{fig:app_dist} shows the distributions of app categories and downloads for our final app set, with ratings ranging from 2.7 to 4.9 (mean: 4.37, variance: 0.15, median: 4.45). 

\subsubsection{UI Collection and Annotation}
We ran our app exploration module on the 100 collected apps and obtained 45,292 UIs. After removing duplicates, we were left with 8,408 unique UIs. To maintain the sequence of explored UIs during annotation, we included some duplicate UIs, resulting in a total of 19,722 UIs that need to be annotated.
Given the large scale of data, we only annotate UIs that contain unique DP instances, i.e., if the same instance appear before, we would not annotate it.

\textbf{AppRay-Tainted-UIs }:
\revision{Three authors contributed to the dataset annotation. Two authors independently annotated all 19,722 UIs, and a third author was involved to adjudicate cases where discrepancies could not be resolved by the first two annotators.}
To conduct annotation, two of the authors first discussed and went through the taxonomy and examples~\cite{chen2023unveiling, mansur2023aidui, di2020ui} to get a commonsense of the definition of deceptive patterns. The two authors then annotated all 19,722 UIs separately by \revision{drawing boxing boxes and adding deceptive pattern types} of any identified instances in these UIs using a open-source tool LabelImg~\cite{labelImg}. We modified LabelImg to show the action on the UI to assist annotation. For \textit{\revision{\revision{inter-page}} deceptive patterns}, we additionally added a linker (a digit) to indicate connections between pages. For example, if UI1, UI2, and UI3 contribute to an \revision{\revision{inter-page}} DP, we include a small bounding box with the same digit label on each, signifying their relation to the same \revision{\revision{inter-page}} DP. 
\revision{After the independent round, we merged the two annotation files for each UI, marking the source (i.e., which annotator identified each instance), and visualised them using the same tool, LabelImg. Conflicts were identified based on whether annotations on the same UI referred to the same instance and type, using overlap-based bounding-box matching and type consistency. For discrepant instances, the two annotators first discussed them to reach consensus; if disagreement remained, a third author was involved to adjudicate. }

As a result, the \revision{first annotator} identified 2,223 instances across 873 UIs, while the \revision{second annotator} identified 2,183 instances from 701 UIs. Automated matching methods found 1,914 matching annotations (76.8\%). \revision{We calculated the inter-rater agreement between the two annotators using Cohen’s Kappa, which yielded a value of 0.688, indicating substantial agreement.} 
Discrepancies mainly arose from missed instances, due to the large volume of 19,722 UIs, and from challenges in distinguishing between Nagging Ads and Disguised Ads. After observing frequent disagreements between these two types, we conducted an additional focused round to ensure rigor. Specifically, we sampled 30 instances per type and independently annotated them, followed by an in-depth discussion of these 60 examples to clarify the distinctions. After reaching consensus on the criteria, the two annotators independently re-annotated the same 60 instances using the established guidelines, achieving 100\% agreement. Finally, one annotator re-annotated all identified Disguised Ads and Nagging Ads instances to ensure consistency with the agreed guidelines.

As a result, we obtained 2,185 \texttt{unique} deceptive pattern instances of 16 types from 876 UIs across 97 apps. Among these instances, 149 of them are related to multiple UIs (i.e., \revision{\revision{inter-page}} DPs) from 48 apps. We denoted this dataset as \texttt{AppRay-Tainted-UIs }. This dataset will be used to answer RQ1 and RQ2. We reported the distribution of each DP type in Table~\ref{tab:rq3_detector_baselines_results}.

\textbf{\revision{AppRay-Benign-UIs}}:
Apart from that, we randomly sampled five consecutive sequences of UIs (length=3) that do not have annotated DP instances from each app. The aim of this dataset is to assess the performance of our tool in reporting false positives while preserving the sequential information. Two of the authors went through these samples, and annotate whether they contain DPs or not. In total, we sampled 1,469 UIs, and ended with 871 benign UIs. We denoted this dataset as \texttt{\revision{AppRay-Benign-UIs}}. This dataset will be used to answer RQ2.

\textbf{Data Splitting}:
Given that our detector requires training, \revision{we performed five-fold cross-validation with application-level partitioning}, i.e., the instances from the same app will be allocated to a same split/fold. 
\revision{Due to the imbalanced structure of the data, where each app contain varying numbers of UIs and instances, enforcing an exact 20\% instance split per fold was infeasible. Instead, applications were assigned to one of five folds such that each fold covered approximately 20\% of the total instances, while ensuring that all instances from the same application remained within the same fold. In each run, four folds were used for training and the remaining fold was used for testing. Consequently, each cross-validation run resulted in an approximate 80/20 split between training and testing data, with folds reasonably balanced in terms of instance counts. We perform 5-fold cross-validation in all the experiments.}

\begin{figure}
    \centering
        \includegraphics[width=0.4\textwidth]{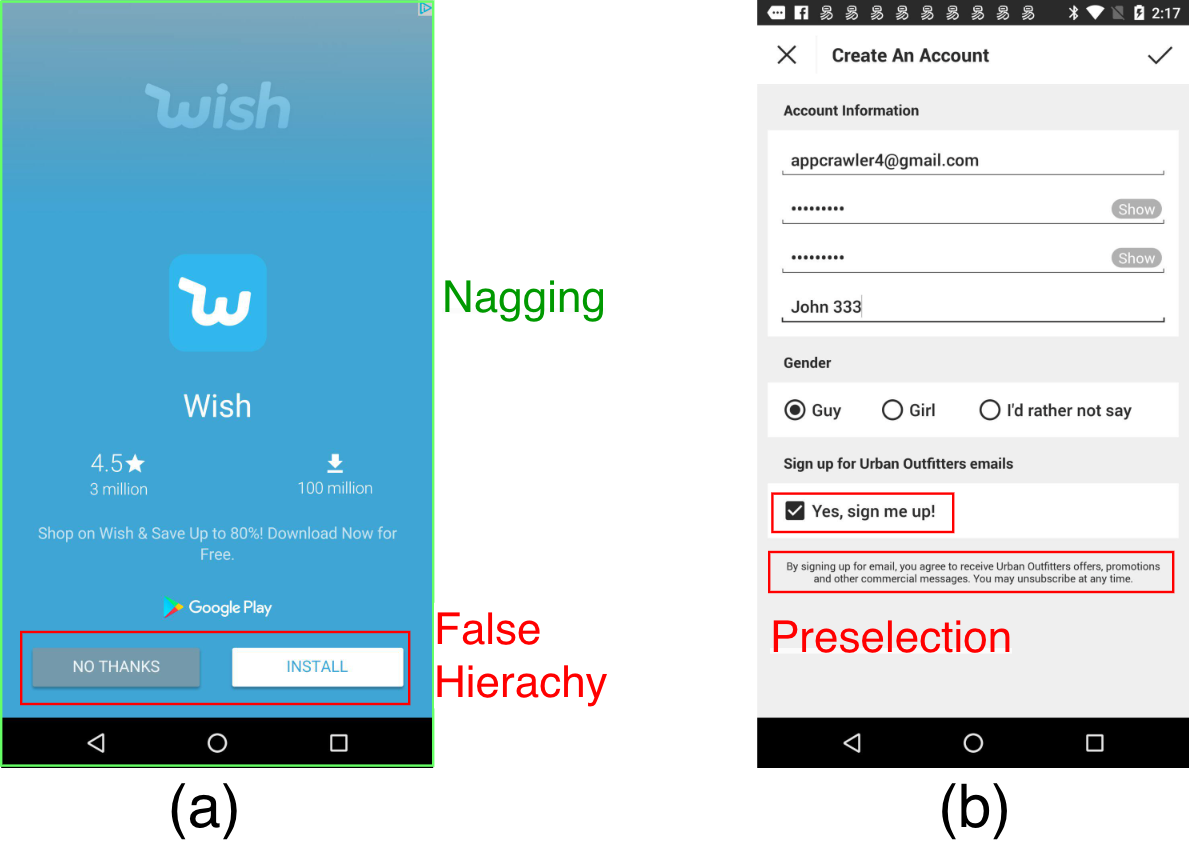}
        \caption{Missing annotations in ContextDP-mobile dataset.}
        \label{fig:contextdp-example}
\end{figure}

\subsection{Existing Single-UI DP Datasets}
In addition to our newly collected datasets, we utilized datasets from previous works, including Chen et al.'s dataset (UIGuard-Rico) \cite{chen2023unveiling} and ContextDP \cite{mansur2023aidui}, to mitigate the potential threat regarding to datasets. \textbf{UIGuard-Rico} comprises 4,999 benign UIs and 1,353 malicious UIs with 1,660 deceptive pattern instances from 1,023 mobile apps. To ensure consistency, we mapped their taxonomy to ours. For ContextDP, we extracted 175 malicious UIs with 197 DP instances and 164 benign UIs from mobile UI datasets, applying a similar taxonomy mapping. We termed it as \textbf{ContextDP-mobile}. For ContextDP-mobile, our analysis of the detection results revealed that both UIGuard and AppRay had low precision. Upon closer examination, we found that many DPs were missed. For example, in Figure~\ref{fig:contextdp-example}, the original annotations miss three deceptive pattern instances (in red) defined in their taxonomy.
To address this, we re-annotated the dataset based on the original annotations, following the same procedure in Section~\ref{sec:dataset}. 
After re-annotations, we have 212 malicious UIs with 320 deceptive patterns, and 127 benign UIs.
Details are provided in the supplementary materials\cite{appray_material}.
Since these datasets lack contextual information about preceding and subsequent UIs, they can only evaluate AppRay's capability in detecting \revision{intra-page} DP types, not the \revision{\revision{inter-page}} types.
\revision{The results on these datasets are largely consistent with our findings on the AppRay datasets. To avoid redundancy, we therefore present the detailed analysis and results in the Appendix Section \ref{sec:baseline_datasets_analysis}.}

\begin{figure}
    \centering
    \includegraphics[width=0.8\textwidth]{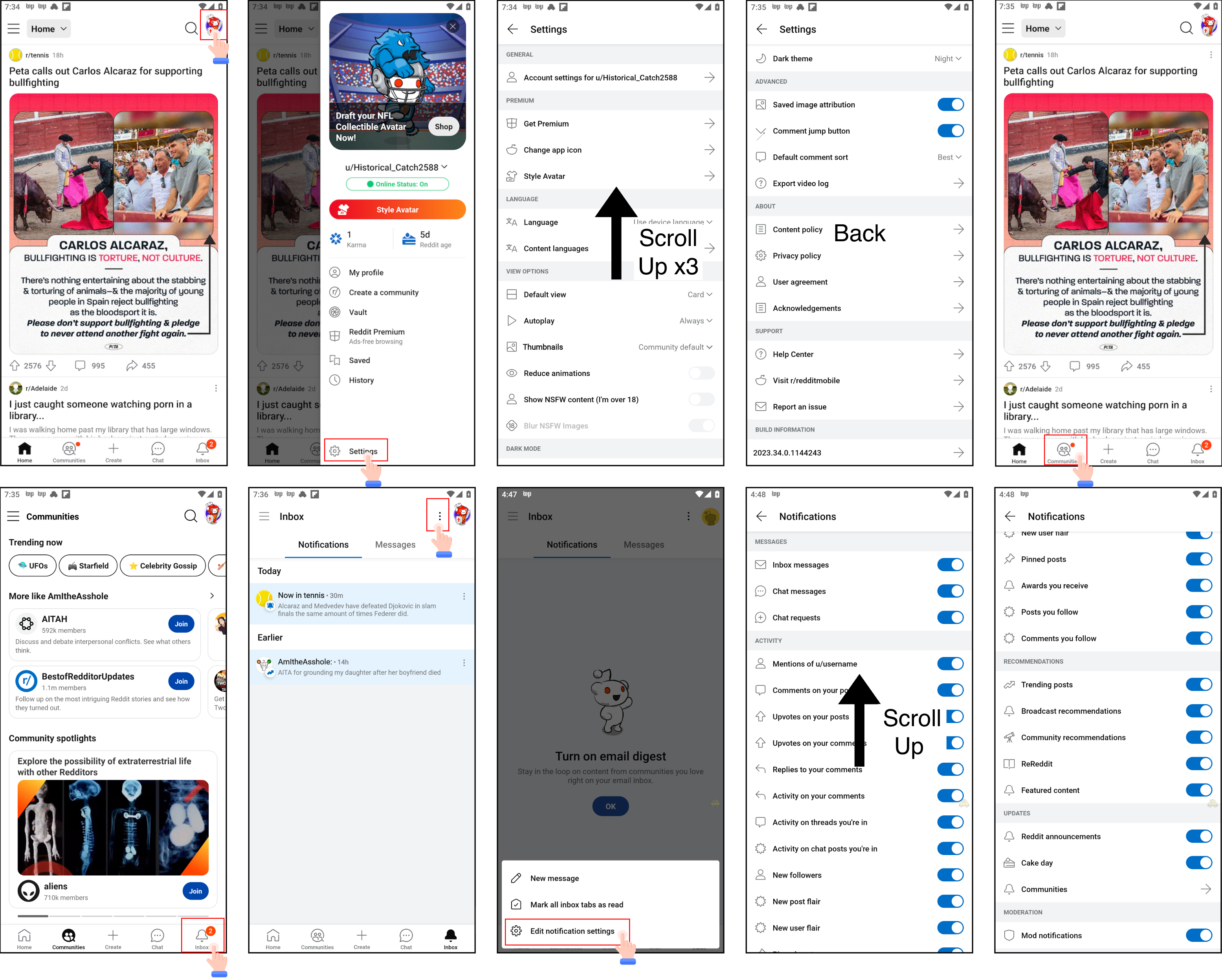}
    \caption{\revision{Execution trace of Task 3 (Go to setting page, go through all notification relate pages) in the Reddit app by our LLM-based task-oriented app explorer. The explorer initially locates the settings page by clicking the top-right profile icon; however, this page does not contain notification settings. It then returns to the main interface, explores alternative tabs, and subsequently discovers the notification settings page, illustrating a trial-and-error navigation strategy.}}
    \label{fig:example-reddit}
\end{figure}

\section{Experiments}

\minor{All experiments were conducted on a workstation running Ubuntu 20.04.3 LTS,  with NVIDIA RTX A6000 GPU.}

\subsection{RQ1: Effectiveness of each module in \tool{}}
\label{sec:rq1.1_exploration}

This section aims to measure the contribution and effectiveness of each module in our \tool{}. We consider the following two sub-questions regarding the effectiveness of the app exploration module (RQ1.1) and deceptive pattern detector module (RQ1.2).
\begin{itemize}[leftmargin=13pt]
    \item \textbf{RQ1.1 App Exploration Module}: How effective is our technique at navigating apps and collecting UIs?
    \item \textbf{RQ1.2 Dark Pattern Detector}: How well does \tool{} perform in detecting dark patterns? We conducted an ablation study to examine the contributions of each sub-module.
\end{itemize}

\subsubsection{RQ1.1: Effectiveness of App Exploration Module}

This section assessed the performance of our app exploration module introduced in Section~\ref{sec:appexploration}. We considered two sub-questions:
\begin{itemize}[leftmargin=13pt]
    \item \textbf{RQ1.1.1}: How well does the LLM-based app navigator complete the tasks?
    \item \textbf{RQ1.1.2}: How effective is the overall exploration strategy in gathering diverse UI states from apps?
\end{itemize}

\textbf{Setup}:
We used the 100 apps collected in Section~\ref{sec:app_collection} in this RQ. 
For \textit{RQ1.1.1}, We ran all tasks in Table~\ref{tab:tasks} in these apps using the LLM task-oriented app exploration module. We adopted \textit{Task Completion Rate} to evaluate whether our approach can successfully finish a task. We manually examined each task to see if it is completed or not. Note that whether the task is completed or not is not a key factor in our proposed system. The key goal here is to explore more UI status by logically using the app.
For \textit{RQ1.1.2}, We further ran Fastbot on these apps to collect more UIs. Two ablations are considered in this RQ, i.e., Fastbot only and LLM only. We adopt number of unique UI status and UI activity as the metrics.

\textbf{Results for RQ1.1.1}:
Among 100 apps, we in total ran 556 tasks, with 11.5 steps on average for each task. The task completion rate is 75.14\%, which shows that our approach has certain capabilities to perform dark pattern sensitive tasks.
As seen in Table~\ref{tab:tasks}, the detailed completion rate for \revision{T1 (Registration), T2 (Sign In), T3 (Notification Setting), T4 (Privacy Setting), T5 (Subscription), T6 (Shopping) and T7 (Sign Out)} are 56.41\%, 78.21\%, 82.83\%, 84.0\%, 83.33\%, 30.0\% and 58.11\% respectively.
We observed that our tool performs tasks logically, identifying relevant UI elements instantly in some cases. For more complex interactions, it employs a trial-and-error strategy, resembling human behavior when navigating unfamiliar apps. 
For some tasks, it can immediately figure out the relevant UI elements to interact~\footnote{ More examples can be found in supplementary material\cite{appray_material}}.
Apart from these straight-forward interaction traces, for apps that have different or more complex interaction ways, our tool can also locate the target page via a trial-and-error strategy. This behaviour is similar to human when users are not familiar with the apps.
For example, in Figure~\ref{fig:example-reddit}, our tool first goes to the setting page and fails to find the notification setting. Then it goes back, and navigates to other tabs, and finally finds the notification setting under the Inbox tab. 
The main failure reasons are due to missing information in view hierarchy~\cite{chen2020unblind}, email/phone verification code requirement. Our tool also identified a usability issue during the exploration. 
Examples can be found in the supplementary materials\cite{appray_material}.

\textbf{Results for RQ1.1.2}:
Among these 100 apps, GPT4 performed 6,342 actions and collected 2,575 unique UIs (Mean=25.75, Std=9.73) from 668 activities (Mean=6.68, Std=4.05) and FastBot2 performed in total 33,649 actions and collected 8,680 unique UIs (Mean=59.86, Std=41.51) from 733 activities (Mean=7.33, Std=6.4), and 
we merged the collected UIs from both modules, and obtained 8,680 unique UIs (Mean=86.8, Std=46.76) from 1,135 activities (Mean=11.35, Std=8.33). 
Among the 8,680 unique UIs, 1.82\% are visited by both modules, 69.37\% are unique from FastBot2, and 28.81\% are unique from GPT4.
Among the 1,135 activities, 23.44\% are visited by both modules, 41.15\% are unique from FastBot2, and 35.42\% are unique from GPT4.
From these statistics, we can see that these two modules can complement each other.
Fastbot2 is fast and effective, good at collecting many UIs in a short time in an efficient and low-cost way, while GPT4 can go deeper with logical exploration solution and explore unique UI pages.

\begin{summarybox}{Summary: RQ1.1 — Effectiveness of App Exploration Module}
\minor{
The LLM-based navigator successfully completed \textbf{75.14\%} of dark-pattern-sensitive tasks across 100 apps. The overall exploration strategy, combining LLM-guided and random exploration, is effective at gathering diverse UI states: the two modules are \textbf{complementary} rather than redundant, with FastBot2 excelling at broad, efficient coverage and GPT-4 enabling deeper, semantically coherent navigation, together yielding \textbf{8,680 unique UIs} from \textbf{1,135 activities}.
}
\end{summarybox}

\subsubsection{RQ1.2: Effectiveness of the Detector Module}
\label{sec:rq1.2_detector}

To understand the effectiveness of each sub-module in our component-based detector (Section~\ref{sec:detector}), we conducted an ablation study which aims to evaluate the contributions of different components in our multi-label classification dark pattern detector. It includes an image embedding pipeline, a text embedding pipeline, and a final MLP classifier. Additionally, the study will also assess different optimization methods such as data augmentation, contrastive learning, and class weights. By systematically removing or modifying one factor at a time, we will observe its impact on the overall model performance.
Note that we will report the overall results of our detector module in Section~\ref{sec:rq2_baseline} comparing with baselines.

\textbf{Setup:} 
We utilized the AppRay-Tainted-UIs  dataset, as described in Section~\ref{sec:dataset}, to evaluate the ablation variants. The dataset includes full UI images, sub-images of individual elements, and the text contained within those elements. If an element does not contain any text, it is assigned the placeholder ``This element does not contain texts.'' The data inputs consist of a list of a combination of sub-images, full UI images, and the corresponding text for examination. 

\revision{As described in Section~\ref{sec:detector}, our AppRay detector adopts a dual-encoder architecture, comprising a \textbf{text encoder} (BERT) for processing textual UI content and a \textbf{visual encoder} (ResNet50) for processing UI images. For training, we employed \textbf{three training strategies}, namely data augmentation (DA), negative sampling (NS) and class weighting (CW). To examine the contribution of different architectural components and training strategies, we conduct ablation studies by incrementally enabling the two encoder branches and three training strategies. All ablation variants operate on the same inputs and differ only in the enabled encoders and training strategies. All models are trained and evaluated under the same five-fold cross-validation setting with application-level partitioning, as described earlier. The evaluated ablation variants and their configurations are summarized in Table~\ref{tab:rq1.2_ablation}.}
Detailed performance for each ablation by type can be found in the Appendix - Table~\ref{tab:detailed_ablation}.

\begin{table}[]
    \centering
    \caption{\revision{Ablation results of the AppRay's detector. All methods take the same input, while different rows enable different encoders, training strategies and usage of the rule refiner. Specifically, the text encoder (Text-only), the visual encoder (Image-only), and training strategies including data augmentation (DA), negative sampling (NS), and class weighting (CW) are incrementally enabled, and a rule refiner. The last row corresponds to the full AppRay model.} 
    \minor{Improvements of the full model over the text-only and image-only variants are statistically significant at the 99\% confidence level (all p<0.001), as assessed by paired t-tests across five folds.}
    } 
    \resizebox{\columnwidth}{!}{%
    \begin{tabular}{l c c c c c c | c c c | c c c}
        \hline
        \multirow{2}{*}{\textbf{Method}} 
        & \multicolumn{2}{c}{\textbf{Encoders Used}} 
        & \multicolumn{3}{c}{\textbf{Training Strategy}} 
        & \multicolumn{1}{c}{\textbf{Refiner}} 
        & \multicolumn{3}{c|}{\textbf{Micro Avg.}} 
        & \multicolumn{3}{c}{\textbf{Macro Avg.}} \\
        \cline{2-6} \cline{7-13}
        & \textbf{Text} & \textbf{Image} 
        & \textbf{DA} & \textbf{NS} & \textbf{CW}  & \textbf{Rules} 
        & \textbf{Pre.} & \textbf{Rec.} & \textbf{F1} 
        & \textbf{Pre.} & \textbf{Rec.} & \textbf{F1}  \\
        \hline
        AppRay (Text + DA + NS + CW) 
        & \checkmark & -- 
        & \checkmark & \checkmark & \checkmark & -- 
        & 0.837 & 0.805 & 0.821 
        & \textbf{0.879} & 0.729 & 0.758 \\

        AppRay (Image + DA + NS + CW)
        & -- & \checkmark 
        & \checkmark & \checkmark & \checkmark & -- 
        & 0.773 & 0.845 & 0.800 
        & 0.840 & 0.649 & 0.675 \\
        \hline
        AppRay (Text + Image) 
        & \checkmark & \checkmark 
        & -- & -- & -- & -- 
        & 0.867 & 0.835 & 0.851 
        & 0.797 & 0.707 & 0.722 \\

        AppRay (Text + Image + DA)
        & \checkmark & \checkmark 
        & \checkmark & -- & -- & -- 
        & 0.897  & 0.859  & 0.877  
        & 0.813  & 0.752  & 0.767 \\

        AppRay (Text + Image + DA + NS) 
        & \checkmark & \checkmark 
        & \checkmark & \checkmark & -- & -- 
        & \textbf{0.906} & \textbf{0.881} & \textbf{0.893}
        & 0.874 & 0.734 & 0.770 \\

        AppRay (Text + Image + DA + NS + CW) 
        & \checkmark & \checkmark 
        & \checkmark & \checkmark & \checkmark & -- 
        & \textbf{0.903}  & 0.855  & 0.878  
        & 0.841  &  0.812 & 0.820 \\

        \textbf{AppRay (Full Model)}
        & \checkmark & \checkmark 
        & \checkmark & \checkmark & \checkmark & \checkmark 
        & 0.846 & \textbf{0.879} & 0.852 
        & 0.869 & \textbf{0.844} & \textbf{0.848}  \\
        \hline
    \end{tabular}
    }
    \label{tab:rq1.2_ablation}
\end{table}

\textbf{Evaluation metrics:} 
We use precision, recall and F1-score as metrics to evaluate the performance. Specifically, \textit{Precision} is calculated by $Precision = TP/(TP+FP)$, where true positive (TP) indicates that a detected instance is correctly classified as the intended deceptive pattern type, while false positive (FP) refers to a detection that is incorrectly classified.
Recall is calculated by $Recall = TP/(TP+FN)$, where FN indicates the ground truth elements that should be predicted but wrongly classified by our model. Finally, F1-score is computed by the harmonic mean of the precision and recall, i.e., $F1-Score = 2 \times Precision \times Recall/(Precision+Recall)$. We calculate the micro average and macro average for all types to illustrate the overall performance.

\textbf{Results}:
As shown in Table~\ref{tab:rq1.2_ablation}, we compare the performance of different model variants using micro- and macro-averaged Precision, Recall, and F1-score. \revision{When using a single modality, AppRay (Text+DA+NS+CW)} achieves strong micro-average performance, with an F1-score of 0.821, while \revision{AppRay (Image+DA+NS+CW)} also demonstrates the effectiveness of image-based features. Combining both modalities, the \revision{AppRay (Text+Image+DA+NS+CW) further improves both micro-averaged and macro-averaged performance.}

The addition of each training strategy gradually boosts the performance of the DP detector, with 
\revision{AppRay(Text + Image + DA + NS)}  achieves the highest scores across all three micro-average metrics: Precision (0.906), Recall (0.881), and F1-score (0.893). This indicates that the combination of data augmentation and negative sampling with contrastive learning significantly improves the model capability. 
\revision{AppRay (Full Model)} achieves the highest macro-average F1-score (0.848) and Recall (0.844), which indicates its consistent performance across all classes, even those with fewer instances. This suggests that Class Weight helps solve the problem of class imbalance and the highest F1 score in macro-average metric shows that it performs well on minority DP classes.

\revision{To examine whether the performance improvement is statistically significant, we repeated all experiments using 5-fold cross-validation to ensure stable results. We then applied paired t-tests on the per–deceptive-pattern category performance to compare the full model with the Text-only model and the Image-only model. For the comparison between the full model and the AppRay (Text+DA+NS+CW) model, the p-values across the five folds are 0.000132, 0.000214, 0.000256, 0.000192, and 0.000298. For the comparison between the full model and the image-only model, the p-values are 0.000558, 0.000613, 0.000360, 0.000791, and 0.000611. All p-values are below 0.01, indicating that the improvements are statistically significant at the 99\% confidence level.}

\begin{table}[t]
\centering
\caption{Per–deceptive-pattern false positive rate (FPR) analysis using one-vs-rest confusion matrices.}
\label{tab:fpr_analysis}
\resizebox{0.5\columnwidth}{!}{%
\begin{tabular}{lccccc}
\toprule
\textbf{Deceptive Pattern} & \textbf{FPR} & \textbf{FP} & \textbf{TP} & \textbf{FN} & \textbf{TN} \\
\midrule
Nagging                 & 0.0040 & 8 & 146 & 44 & 1987 \\
Bait and Switch         & 0.0094 & 20 & 67 & 0 & 2098 \\
Forced Continuity       & 0.0180 & 37 & 124 & 24 & 2000 \\
Roach Motel             & 0.0014 & 3 & 4 & 2 & 2176 \\
Intermediate Currency   & 0.0093 & 20 & 41 & 0 & 2124 \\
Social Pyramid          & 0.0000 & 0 & 15 & 0 & 2170 \\
Privacy Zuckering       & 0.0046 & 9 & 164 & 84 & 1928 \\
Gamification            & 0.0014 & 3 & 10 & 0 & 2172 \\
Forced Action -- General & 0.0084 & 17 & 139 & 21 & 2008 \\
Preselection            & 0.0130 & 20 & 647 & 34 & 1484 \\
Hidden Information      & 0.0032 & 6 & 304 & 3 & 1872 \\
Toying with Emotion     & 0.0076 & 16 & 63 & 5 & 2101 \\
False Hierarchy         & 0.0057 & 12 & 81 & 2 & 2090 \\
Disguised Ads           & 0.0106 & 22 & 107 & 0 & 2056 \\
Tricked Questions       & 0.0000 & 0 & 4 & 3 & 2178 \\
Interface Interference -- General & 0.0061 & 13 & 43 & 4 & 2125 \\
\bottomrule
\end{tabular}
}
\end{table}

\revision{
To further analyze model errors, we report per–deceptive-pattern false positive rates (FPR) using one-vs-rest confusion matrices in the multi-label setting. Table~\ref{tab:fpr_analysis} summarizes the FPR together with the numbers of false positives (FP), true positives (TP), false negatives (FN), and true negatives (TN) for each deceptive pattern. Overall, the model shows low false positive rates across most deceptive patterns. No class exceeds an FPR of 0.02, indicating that the model adopts a generally conservative prediction strategy and rarely assigns deceptive pattern labels incorrectly.
}
\revision{Among all patterns, \textbf{Forced Continuity} has the highest false positive rate (FPR = 0.018), followed by \textbf{Preselection} (0.013) and \textbf{Disguised Ads} (0.0106). These patterns have broad and overlapping definitions and often co-occur with other interface interference or forced-action strategies. As a result, the model may over-predict these labels in ambiguous cases, leading to higher false positive counts, even though true positive detection remains strong (e.g., TP = 647 for Preselection).}

\revision{In contrast, several deceptive patterns, including \textbf{Social Pyramid}, \textbf{Tricked Questions}, and \textbf{Roach Motel}, exhibit near-zero false positive rates. This indicates high specificity, where the model predicts these patterns only when strong evidence is present. While this behavior effectively reduces false alarms, it may also reflect limited training samples or narrowly defined pattern characteristics, which can constrain recall for these classes.}

\revision{Overall, the confusion matrix–based analysis confirms that the proposed model maintains a low false positive rate across deceptive pattern types, supporting its reliability in practical detection scenarios.}

\begin{summarybox}{Summary: RQ1.2 — Effectiveness of the Detector Module}
\minor{
AppRay's detector performs well in detecting deceptive patterns, achieving a micro/macro F1 of \textbf{0.852/0.848}. The ablation study confirms that each sub-module contributes positively: multimodal fusion, data augmentation, contrastive learning with hard negative sampling, class weighting, and rule-based refinement all improve performance incrementally, with the full model yielding the best overall results.
}
\end{summarybox}

\subsection{RQ2: Effectiveness in Detecting Deceptive Patterns}
\label{sec:rq2_baseline}

In this RQ, we used \textbf{AppRay-Tainted-UIs } and existing datasets to evaluate the performance of our deceptive pattern detection module, and then used \textbf{\revision{AppRay-Benign-UIs}} to examine the false positive rates.

\textbf{Baselines}: 
We considered three methods as our baselines, namely AidUI \cite{mansur2023aidui}, UIGuard \cite{chen2023unveiling} \revision{and DPGuard \cite{shi202550}}. 
\revision{AidUI and UIGuard are rule-based systems with general UI information extraction techniques, while DPGuard is an MLLM based method with a binary classifier and prompt engineering. Detailed introduction of these three methods can be found in Section~\ref{sec:relatedwork}.
All these methods released their implementations and pretrained models, and we directly used their public repositories \cite{AidUICode, UIGuardCode, DPGuardCode} in the experiments.
Given that DPGuard relies on multi-modal large language models, we considered three  backbone models, i.e., their originally used GPT-4o, the most recent GPT-5.2 and Gemini3-Pro, denoted as DPGuard (GPT-4o), DPGuard (GPT-5.2) and DPGuard (Gemini3-Pro). }

\textbf{Metrics:} We adopted the same metrics, \textit{precision, recall and F1-score} as mentioned in Section~\ref{sec:rq1.2_detector}. \revision{As DPGuard only predicts the presence of deceptive patterns at the screen level without localizing individual instances, we adopt a conservative mapping strategy for comparison. Specifically, when DPGuard reports a particular deceptive pattern type for a UI, all ground-truth instances of that type within the UI are counted as true positives; otherwise, all such instances are counted as false negatives. This evaluation protocol favors DPGuard and may therefore overestimate its instance-level detection performance.
In addition, we also record the \textit{time consumption} for each method. 
}

\subsubsection{Results for AppRay-Tainted-UIs }
\label{sec:resultsAppRayDark}
\revision{
Table~\ref{tab:rq3_detector_baselines_results} summaries the performance of all baselines and our \tool{} on the AppRay-Tainted-UIs  dataset. 
Overall, our \tool{} achieves the best performance across both micro-averaged and macro-averaged metrics, substantially outperforming all baselines. Specifically, AppRay obtains a micro-averaged F1 score of 0.89 and a macro-averaged F1 score of 0.85, indicating strong performance both on frequent and rare deceptive pattern categories.
\revision{Among the baselines, DPGuard is the strongest performer. When restricting the comparison to the deceptive pattern types supported by DPGuard (Gemini3-Pro), AppRay achieves micro-averaged precision, recall, and F1 scores of 0.92, 0.86, and 0.89, respectively, and macro-averaged precision, recall, and F1 scores of 0.84, 0.90, and 0.86. This corresponds to improvements of 27.14\% in micro-averaged F1 and 38.71\% in macro-averaged F1 over DPGuard (Gemini3-Pro). 
Compared with the weakest baseline, AidUI, on the deceptive pattern types it supports, our method achieves micro- and macro-averaged F1 scores of 0.91 and 0.90, corresponding to improvements of 1200\% and 650\%, respectively.}
Next, we identified three key aspects that differentiate the performance of these methods by analysing their detailed performance compared  to our \tool{}.
}

\begin{table*}
    \centering
    \caption{Performance of baselines and \tool{} in the AppRay-Tainted-UI dataset.} 
    \resizebox{1.0\columnwidth}{!}{%
        \begin{tabular}{l c
                | c c >{\columncolor{f1gray}}c 
                | c c >{\columncolor{f1gray}}c
                |c c >{\columncolor{f1gray}}c
                |c c >{\columncolor{f1gray}}c
                |c c >{\columncolor{f1gray}}c
                |c c >{\columncolor{f1gray}}c
                }
        \toprule
        
         & &  \multicolumn{3}{c|}{\textbf{AidUI}} & \multicolumn{3}{c|}{\textbf{UIGuard}} & \multicolumn{3}{c|}{\textbf{DPGuard (GPT-4o)}} & \multicolumn{3}{c|}{\textbf{DPGuard (GPT-5.2)}} & \multicolumn{3}{c|}{\textbf{DPGuard (Gemini3-pro)}} & \multicolumn{3}{c}{\texttt{\textbf{\tool{}}}}\\
         \midrule
         \textbf{DP Category} & \textbf{\# of Instance} 
         & \textbf{Pre.} & \textbf{Rec.} & \textbf{F1} 
         & \textbf{Pre.} & \textbf{Rec.} & \textbf{F1}  
         & \textbf{Pre.} & \textbf{Rec.} & \textbf{F1}  
         & \textbf{Pre.} & \textbf{Rec.} & \textbf{F1} 
         & \textbf{Pre.} & \textbf{Rec.} & \textbf{F1} 
         & \textbf{Pre.} & \textbf{Rec.} & \textbf{F1} \\
         \midrule
         Nagging               & 190 & 0.14 & 0.03 & 0.05 & 0.22 & 0.06 & 0.10 & 0.75 & 0.18 & 0.29 & 0.73 & 0.24 & 0.36 & 0.61 & 0.52 & 0.56 & 0.95 & 0.77 & \textbf{0.85} \\
         Bait And Switch       & 67  & -    & -    & -    & -    & -    & -    & -    & -    & -    & -    & -    & -    & -    & -    & - & 0.77 & 1.00 & \textbf{0.87} \\
         Forced Continuity     & 148 & -    & -    & -    & 0.00 & 0.00 & 0.00 & 0.83 & 0.94 & 0.88 & 0.86 & 1.00 & \textbf{0.93} & 0.89 & 1.00 & \textbf{0.94} & 0.77 & 0.84 & 0.81 \\
         Roach Motel           & 6   & -    & -    & -    & -    & -    & -    & 0.50 & 1.00 & 0.67 & 1.00 & 0.50 & 0.67 & 0.67 & 1.00 & \textbf{0.80} & 0.57 & 0.67 & 0.62 \\
         Intermediate Currency & 41  & -    & -    & -    & -    & -    & -    & 0.93 & 0.87 & 0.90 & 0.88 & 1.00 & \textbf{0.94} & 0.83 & 1.00 & 0.91 & 0.67 & 1.00 & 0.80 \\
         Social Pyramid        & 15  & -    & -    & -    & 1.00 & 0.25 & 0.40 & 1.00 & 1.00 & \textbf{1.00} & 0.80 & 1.00 & 0.89 & 1.00 & 1.00 & \textbf{1.00} & 1.00 & 1.00 & \textbf{1.00} \\
         Privacy Zuckering     & 248 & -    & -    & -    & 1.00 & 0.15 & 0.27 & 0.88 & 0.39 & 0.54 & 0.75 & 0.39 & 0.51 & 0.91 & 0.51 & 0.66 & 0.95 & 0.66 & \textbf{0.78} \\
         Gamification          & 10  & -    & -    & -    & -    & -    & -    & 0.50 & 0.50 & 0.50 & 0.50 & 0.50 & 0.50 & 0.20 & 0.50 & 0.29 & 0.80 & 1.00 & \textbf{0.89} \\
         ForcedAction-General  & 160 & 0.00 & 0.00 & 0.00 & 1.00 & 0.53 & 0.69 & 0.80 & 0.42 & 0.55 & 1.00 & 0.47 & 0.64 & 0.86 & 0.95 & \textbf{0.90} & 0.89 & 0.87 & \textbf{0.88} \\
         Preselection          & 681 & 0.00 & 0.00 & 0.00 & 0.84 & 0.21 & 0.33 & 0.87 & 0.66 & 0.75 & 0.93 & 0.68 & 0.79 & 0.96 & 0.78 & 0.86 & 0.97 & 0.95 & \textbf{0.96} \\
         Hidden Information    & 307 & -    & -    & -    & -    & -    & -    & 0.85 & 0.27 & 0.41 & 0.88 & 0.44 & 0.58 & 0.59 & 0.24 & 0.34 & 0.98 & 0.99 & \textbf{0.98} \\
         Toying with Emotion   & 68  & 0.42 & 0.33 & 0.37 & -    & -    & -    & 0.41 & 0.93 & 0.57 & 0.32 & 0.93 & 0.48 & 0.60 & 1.00 & 0.75 & 0.80 & 0.92 & \textbf{0.86} \\
         False Hierarchy       & 83  & 1.00 & 0.05 & 0.09 & 0.25 & 0.10 & 0.14 & 0.64 & 0.67 & 0.65 & 0.38 & 1.00 & 0.55 & 0.37 & 0.90 & 0.53 & 0.87 & 0.98 & \textbf{0.92} \\
         Disguised Ads         & 107 & 0.22 & 0.15 & 0.18 & 0.00 & 0.00 & 0.00 & 0.60 & 0.46 & 0.52 & 0.62 & 0.62 & 0.62 & 0.29 & 0.15 & 0.20 & 0.83 & 1.00 & \textbf{0.91} \\
         Tricked Questions     & 7   & -    & -    & -    & -    & -    & -    & -    & -    & -    & -    & -    & -    & -    & -    & - & 1.00 & 0.50 & \textbf{0.67} \\
         Interface Interference - General 
                               & 47  & -    & -    & -    & 0.00 & 0.00 & 0.00 & 0.00 & 0.00 & 0.00 & 0.33 & 0.29 & 0.31 & 0.00 & 0.00 & 0.00 & 0.77 & 0.91 & \textbf{0.83} \\
         \midrule
         \textbf{Micro Avg.} & 2,185 
         & 0.16 & 0.04 & 0.07 
         & 0.66 & 0.16 & 0.25 
         & 0.76 & 0.55 & 0.64 
         & 0.69 & 0.63 & 0.66 
         & 0.73 & 0.68 & 0.70
         & 0.92 & 0.86 & \textbf{0.89} \\
         \textbf{Macro Avg.} & 2,185 
         & 0.30 & 0.09 & 0.12 
         & 0.48 & 0.14 & 0.21 
         & 0.68 & 0.59 & 0.59 
         & 0.71 & 0.65 & 0.63 
         & 0.63 & 0.68 & 0.62
         & 0.85 & 0.88 & \textbf{0.85} \\
         \bottomrule
        \end{tabular}
    }
    \label{tab:rq3_detector_baselines_results}
\end{table*}

\revision{
\textbf{Perceptual Sensitivity to Visual Cues.}
We first observe clear differences across methods in their sensitivity to visual cues of varying scale and salience. Rule-based methods (AidUI and UIGuard) consistently perform worst across metrics, as they rely on explicit visual or textual indicators and fail when deceptive content lacks predefined cues. For example, in cases where advertisements omit explicit ad icons or keywords (Figure~\ref{fig:disguisedAds}(a)), both rule-based methods miss the instances entirely.
Among MLLM-based methods, performance varies by model. GPT-based variants, particularly DPGuard (GPT-5.2), achieve stronger results on categories requiring fine-grained visual recognition, such as \textit{Interface Interference – General} and \textit{Hidden Information}. These categories often involve small or visually subtle elements, including tiny close buttons or low-contrast text, which DPGuard (Gemini-3-Pro) frequently fail to detect. In contrast, DPGuard (Gemini-3-Pro) shows a bias toward detecting large and visually salient elements, which limits its coverage of smaller or less prominent UI components but enhance its performance in Nagging patterns, which normally involves larger UI components.
AppRay demonstrates more stable performance across both large and small UI elements by systematically analyzing UI components at multiple granularities, leading to improved coverage across visually diverse deceptive patterns.
}

\revision{
\textbf{Normative Ambiguity in Deceptive Pattern Definitions.}
Performance differences are particularly evident in deceptive pattern categories with ambiguous or normatively defined boundaries, such as \textit{Gamification} and \textit{Toying With Emotion}. For \textit{Gamification} patterns, DPGuard (Gemini-3-Pro) exhibits lower precision, frequently classifying the presence of points, coupons, or reward mechanisms in shopping apps as deceptive. While such interpretations are semantically plausible, these design elements are not inherently deceptive under our taxonomy, leading to increased false positives.
A similar effect is observed for Toying With Emotion, where ambiguity in normative boundaries leads MLLM-based methods to adopt aggressive detection strategies. As a result, all MLLM variants achieve very high recall for this category (above 0.93). However, this aggressive strategy introduces a clear precision trade-off by overextending the definition of emotional manipulation. Specifically, DPGuard (Gemini-3-Pro) classifies phrases such as “do you love AllTrails?” and “best value” labels as instances of emotional manipulation, while DPGuard (GPT-4o) and DPGuard (GPT-5.2) adopt an even more aggressive strategy, further treating discount-related cues (e.g., “−48\% off” and “buy one get 1 30\% off”) as deceptive. In both cases, these interpretations result in false positives under our taxonomy, illustrating how normative ambiguity can lead to over-detection despite high recall. In contrast, AppRay achieves more balanced performance across these categories by leveraging deceptive-pattern-specific supervision, enabling it to better align detection decisions with the normative boundaries defined in our taxonomy.
}

\revision{
\textbf{Context and Interaction Awareness.}
Finally, interaction-dependent deceptive patterns reveal substantial limitations in methods that analyze UI screens in isolation. For light interaction-dependent patterns such as Nagging, although DPGuard (Gemini-3-Pro) outperforms other baseline methods, it still struggles to distinguish whether an upgrade or advertisement page constitutes a deceptive pattern or a legitimate UI triggered intentionally by the user. For example, when a user explicitly navigates to a subscription page, the resulting subscription UI is benign; however, when the same page is automatically triggered by the app without user intent, it becomes a deceptive pattern.
Moreover, both rule-based and MLLM-based baselines frequently fail to detect patterns such as Bait-and-Switch, which only emerge through multi-step interactions or across multiple screens. AppRay addresses this limitation by explicitly modeling logical connections across pages via its rule refiner, enabling it to capture deceptive behaviors that depend on user actions or system-triggered flows. As a result, AppRay consistently identifies interaction-dependent deceptive patterns that are missed by other methods.
}

\revision{
\textbf{Failure Analysis for \tool{}.}
While our \tool{} shows a promising capability for detecting deceptive patterns through cross- and within-UI analysis, our methods still make mistakes. We observe two main failure issues: grouping and model errors. Over-reliance on GT groupings during training can cause incorrect hierarchy detection when visual differences exist within semantic groups. Model errors occur when elements visually resemble dark patterns, like mistaking a user profile for pre-selection due to color changes.
}

\subsubsection{Results for \revision{AppRay-Benign-UIs}}
We then measure the false positive rates for these methods. 
The average rates at which benign UIs are misidentified as malicious UIs with DPs by AidUI, UIGuard, \revision{DPGuard (GPT-4o), DPGuard (GPT-5.2), DPGuard (Gemini3-Pro)} and \tool{} are 14\%, 6\%, \revision{19\%, 26\%, and 19\%} and 7.6\%, respectively.
\revision{Overall, false positive rates remain moderate across all methods, with \tool{} achieving a comparatively low rate of 7.6\%. False positives from rule-based baselines mainly stem from rigid heuristics that over-interpret surface-level visual or textual cues, leading to benign design choices being flagged as deceptive. In contrast, MLLM-based methods tend to produce higher false positives due to overly aggressive semantic inference, occasionally misinterpreting standard interface elements or benign user information requests as deceptive patterns. AppRay exhibits a small number of false positives when encountering visual motifs commonly associated with specific deceptive pattern categories, reflecting residual noise learned from the training data rather than systematic modeling errors.}

\subsubsection{Time Consumption}
We also measured the time consumption and computational cost of each method \minor{for deceptive pattern detection at the UI level}. On average, AidUI requires 10.7 s per UI, while UIGuard takes 12.9 s per UI. Among MLLM-based baselines, DPGuard (GPT-4o) processes each UI in 2.60 s, DPGuard (GPT-5.2) in 3.60 s, and DPGuard (Gemini-3-Pro) in 35.04 s. Our \tool{} processes each UI in 0.658 s. Notably, for DPGuard, the dominant source of latency arises from waiting for responses from external MLLM services, whereas its internal binary classifier incurs negligible overhead (approximately 0.1 s per UI).
In contrast, \tool{} does not rely on external network calls and therefore avoids network latency and service waiting time, resulting in the most efficient end-to-end inference among all evaluated methods.

\minor{Given that the baseline methods already demonstrate limited effectiveness at the UI level, and that they do not support the app exploration phase necessary for app-level analysis, we report the end-to-end time consumption of \tool{} only.
The exploration phase requires an average of 21 minutes 33 seconds per application. Based on the 86.8 UIs collected on average per application (Section 6.1.1), the detection phase requires an additional 57 seconds (86.8 × 0.658s). In total, AppRay takes an average of approximately 22.5 minutes to fully analyse an application. 
}

\begin{summarybox}{Summary: RQ2 — Comparison with Existing Methods}
\minor{
AppRay significantly outperforms all baselines, achieving \textbf{micro/macro F1 of 0.89/0.85} — up to \textbf{1200\%} over the weakest and \textbf{27.1\%} over the strongest baseline. As the only method covering all \textbf{16 pattern types}, AppRay overcomes the key limitations of prior work: rule brittleness, normative over-detection in MLLMs, and inability to capture inter-page patterns. It also achieves a low false-positive rate of \textbf{7.6\%} and the fastest inference at \textbf{0.658s} per UI.
}
\end{summarybox}

\subsection{RQ3: Usefulness}
\label{sec:rq4_userstudy}
This research question aims to evaluate the usefulness of the proposed technique.

\textbf{Setup}:
We randomly selected 5 apps from different categories as mentioned in Section~\ref{sec:dataset}. First, we deployed our \tool{} to gather UIs and perform detection on the collected UIs. We also annotated these UIs to evaluate the effectiveness of our UI exploration module.
We then recruited two experts, each with over two years of experience in deceptive pattern research, to conduct this study. 
Both experts were asked to perform an inspection walk-through of these popular apps to obtain as many UIs as possible while exploring the apps' main features. They were also instructed to document any DPs identified during this process.
Specifically, each app was explored by one researcher for up to 15 minutes, during which they performed the tasks listed in Table~\ref{tab:tasks}. They were encouraged to use the apps for their intended purposes as well.
After manually obtaining the UIs from apps, one of the authors manually labeled the potential DPs in the collected UIs to ensure annotation consistency as in Section~\ref{sec:dataset}.

\begin{figure}
    \centering
    \includegraphics[width=0.6\textwidth]{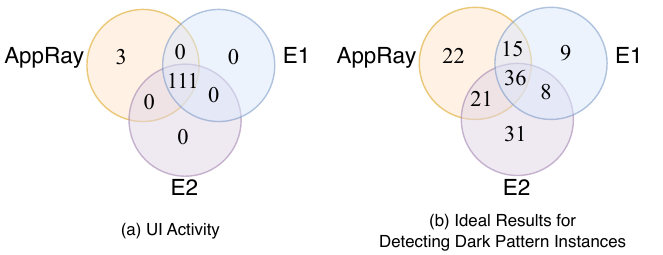}
    \caption{The Venn diagrams for Expert 1, Expert 2 and \tool{} in collected UI activities and identified instances.
    }
    \label{fig:venn_diagrams}
\end{figure}

\textbf{Results}:
In total, \tool{} collected in total 2,726 UIs from these 5 apps. After removing duplicates, we obtained 522 unique UIs from 114 Activities, and detected 90 unique DP instances. Upon annotation on these collected UIs, we obtained 87 DP instances from 11 types.
For the human experts, E1 collected in total 369 unique UIs from 111 unique activity, and identified 68 unique DP instances of 11 DP types. E2 collected in total 404 unique UIs from 111 unique activity, and identified 103 unique DP instances of 11 DP types. These results are directly recorded in their initial reports.
As illustrated in Figure~\ref{fig:venn_diagrams}, AppRay, E1, and E2 exhibit significant overlaps in visited UI activity.  

\revision{Upon closer examination, we identify three underlying reasons for the differences in deceptive pattern detection across the experts and AppRay, which can be grouped into two broader categories: UI exploration coverage and detection during review.}

\revision{Within UI exploration coverage, two distinct factors account for most of the discrepancies.
The first factor concerns differences in \textit{exploration depth and the ability to handle complex interaction logic.} Expert 2 demonstrated stronger persistence and adaptability in navigating multi-step or nested UI flows, exploring every accessible page, tab, and link, and effectively returning to previous states when trapped in deep navigation structures. This allowed Expert 2 to uncover both more and rarer deceptive patterns. Expert 1 also performed well but occasionally skipped certain \revision{inter-page} elements when they seemed unlikely to yield new insights. In comparison, AppRay, despite being prompted to click through all \revision{inter-page} elements, sometimes ignored them or became stuck in complex navigation paths without the ability to backtrack or recover, highlighting its current limitations in reasoning about interaction context.
The second factor involves \textit{randomly triggered events} that are independent of any explorer’s ability. Some deceptive patterns appeared only under stochastic system conditions, for example, a feedback pop-up encountered by Expert 2 during the exploration of Google Translate but unseen by Expert 1 or AppRay.}

\revision{The second category, \textit{detection during review}, reflects human limitations rather than exploration differences. Even when the same UIs were captured, some deceptive patterns were overlooked because the experts had to review a large number of interfaces, and certain patterns were subtle or time-consuming to annotate. In contrast, AppRay does not suffer from fatigue or attentional decline, allowing it to maintain consistent detection performance across large-scale datasets. Together, these observations suggest that human experts excel at contextual reasoning and adaptive exploration, while AppRay provides scalability and reliability.}

\revision{
To further contextualize these findings, we compared AppRay with the two experts across 33 tasks from five applications. AppRay successfully completed 24 tasks (72.7\%) and detected 65 deceptive patterns, while Expert~1 and Expert~2 identified 56 and 86, respectively. Although Expert~1 completed all tasks, their total count was lower than AppRay’s, likely due to fatigue during prolonged reviews (about 75 minutes), which led to missing subtle patterns such as small or greyed text. Expert~2’s higher count reflected their more thorough post-task exploration, in which they continued to investigate each application beyond the task boundaries, consistent with the earlier observation on exploration depth. These results suggest that detection performance depends not only on task completion but also on the depth and consistency of exploration, implying that more advanced UI agents could improve performance by enabling deeper and more adaptive interaction rather than merely completing more tasks.
}

\begin{summarybox}{Summary: RQ3 — Usefulness}
\minor{
AppRay detects a comparable number of deceptive patterns to human experts (65 vs.\ 56--103) without fatigue-induced misses, confirming its practical value as a scalable pre-screening tool. Human experts remain superior in adaptive, deep exploration; together, they are more effective than either alone.
}
\end{summarybox}

\section{Applications}

\revision{The detection results can be further used to benefit different stakeholders, including the end-users, app-providers and the legal regulators. }

\subsection{Stakeholder Benefits and Use Cases}
\revision{For end-users, we can highlight the position of the existing deceptive patterns to warn them, avoiding them been tricked without notice. For example, for the “Forced Continuity” pattern, which auto-continue the app subscription after offering the end-users a free trial. By highlighting this issue on this app, the users can get noticed; they can either give up using this service; or setting up a calendar event to remind themselves when the trial is about to end to avoid potential financial loss. Through this process, the end-users can gradually accumulate the knowledge of deceptive patterns. Future research works on using the detection results for repairing these patterns can further benefit end-users.}

\revision{For \textbf{app providers}, we find that online design sharing platforms always contains embedded deceptive patterns. When designing the prototype for their apps, the app designers may unconsciously inherit these manipulative designs. Our tool can warn the potential reuse of them and avoid using them. On the other hand, some patterns are consciously inserted into the apps, and our work can be further developed to connect with existing regulations. When detecting a deceptive pattern, we can link to the existing regulations, showing the consequences and linking the real lawsuit cases to highlight the consequences. However, it is also potential that the app providers use our tool to avoid been detected, that means, when our tool can detect their deceptive patterns, they then further develop the patterns to make it un-detectable by our tools. 
On the other hand, we acknowledge that many applications rely on in-app advertisements as a source of revenue. Our concern is not advertising itself, but deceptive implementations of it. Developers can still monetize effectively by following responsible practices, such as clear labeling and transparency about sponsored content, which are also increasingly mandated by regulations (e.g., GDPR, EU Digital Services Act). For instance, in Figure 1(g), although ads were technically labeled, the labels were barely visible, leading users to click without realizing the content was sponsored. Making such disclosures more prominent is both a more ethical practice and a more sustainable business strategy.}

\revision{Lastly, for \textbf{regulators}, the detection algorithms from our work can be further engineered into an app, that supervise the existence of dark patterns in people’s daily usage of apps. The real statistics can ground regulators with the evidence to implement their policy. For app market, the detected patterns can be added to the app install page to warm the end-users of the existence of dark patterns, and as a way to motivate the app to reduce the usage. Ultimately, effective constraints on the use of deceptive patterns must come from regulatory frameworks. In recent years, substantial progress has been made, with new regulations, including EU’s Digital Services Act, Digital Markets Act, and Data Act proposal, and the U.S. California Privacy Rights Act (CPRA), emerging in response to sustained advocacy from the research community and broader societal pressures.}

\subsection{Deployment Modes}

\minor{
AppRay supports two potential and complementary deployment modes: an offline, pre-deployment mode suitable for software testing and auditing workflows, and an online, real-time mode for runtime detection during active application use.
}

\minor{
\textbf{Offline, Pre-Deployment Analysis}. In the offline mode, AppRay operates as a passive analysis tool similar to conventional software testing. Given an application package (e.g., an APK file) as input, AppRay automatically collects UI states by executing the application in a mobile simulator (e.g., Genymotion), performs deceptive pattern detection over the collected UIs, and produces a structured detection report. This mode is particularly suited for developers, auditors, and app marketplace operators seeking to assess applications prior to release. In practice, this mode could be realized as a web-based submission service, like VirusTotal~\cite{virusTotal}, through which stakeholders upload APK files, app metadata, or UI screenshots to obtain detection reports, or alternatively by running the AppRay source code directly. A GPU based backbone could enhance the efficiency. The resulting reports can be consumed by diverse stakeholders: developers may use them to identify and remediate deceptive patterns prior to release; app markets like Google Play may incorporate findings into app review pipelines or require developer remediation before publication; regulatory bodies may leverage detection evidence to support compliance enforcement; and end users may be presented with visual overlays highlighting detected issues within the application UI, in a manner analogous to prior work such as GPTVoiceTasker~\cite{vu2024gptvoicetasker} and UIGuard~\cite{chen2023unveiling}.
}

\minor{
\textbf{Online, Real-Time Detection.} In the online mode, AppRay performs deceptive pattern detection during live user interaction with the target application. Since the Deceptive Pattern Detection module operates directly on UI states, it can be applied incrementally to UI screens encountered during runtime without requiring prior app-level analysis. For deployment, the detection service may be instantiated either on-device or as a remote cloud service that receives the current UI state, executes the detection pipeline, and returns results to a client-side rendering layer to produce UI overlays, leveraging the accessibility service similar to prior works~\cite{vu2024gptvoicetasker, chen2023unveiling}. To further reduce computational overhead, the system may maintain a cache of historical detection results, whereby incoming UI states are first matched against previously processed entries and the full detection pipeline is invoked only upon a cache miss.
}

\minor{
Note that full production deployment of either mode would require additional engineering effort beyond the scope of this work, such as the development of submission interfaces or client applications, persistent storage and indexing infrastructure, or real-time UI overlay rendering pipelines. Our current work focuses on the core detection methodology; the deployment pathways described above are intended to illustrate practical realisation scenarios.
}
\section{Threat to Validity}

\textbf{Internal Validity.}
\revision{Internal validity concerns whether the observed results are attributable to the proposed techniques rather than confounding factors. A potential internal threat arises from the dataset construction process, which relies on large-scale human annotation. Given the scale of the dataset (19,722 UIs), annotators may be more likely to identify clearer or more salient deceptive patterns, while subtler or harder-to-detect instances may be missed, particularly due to fatigue effects. This annotation-driven selection bias may skew the data distribution toward easier cases, potentially inflating the observed detection performance. We mitigate this risk by having two authors independently annotate the dataset and cross-validate their results. In addition, to control annotation quality and reduce fatigue effects, we limited each annotator to at most five hours of annotation per day (approximately 1,500 UIs per day). The inter-rater agreement between the two annotators is 0.688, indicating a substantial agreement. Moreover, we also leveraged two datasets from prior works with different researchers annotating the data~\cite{chen2023unveiling, mansur2023aidui}.}

\revision{Another potential internal threat arises from the implementation of baseline models, as differences in implementation details could affect baseline performance and confound the comparison. We mitigate this threat by directly adopting the official source code and released pretrained models provided by the original authors \cite{AidUICode, UIGuardCode, DPGuardCode}, without any modification, thereby ensuring faithful and consistent baseline execution.  
}

\textbf{Construct Validity.}
\revision{Construct validity concerns whether the measurements used in this study accurately capture the intended constructs. A key construct in our work is the presence and type of deceptive patterns in UIs. Identifying dark patterns inherently involves subjective judgment, particularly for visually or semantically similar patterns such as \emph{Nagging Ads} and \emph{Disguised Ads}. To reduce construct ambiguity, two authors independently annotated all UIs following a shared taxonomy and guidelines, with disagreements resolved through discussion and adjudication by a third author. We further conducted an additional focused annotation round on frequently confused pattern types to refine annotation criteria and improve consistency.}

\textbf{External Validity:} 
First, our selection of top trending apps introduces potential external threat regarding sampling bias. To mitigate this, we ensured diversity across app categories, download and ratings (see Figure~\ref{fig:app_dist}), and incorporated baseline datasets from previous works~\cite{chen2023unveiling, mansur2023aidui}, though these only support evaluation of specific DP types.
Second, the task selection in Table~\ref{tab:tasks} may introduce external threat due to limited comprehensiveness. We mitigated this by selecting common app functionalities, and our user study confirms that AppRay matches or exceeds human experts' detection capabilities (Figure~\ref{fig:venn_diagrams}). Future work can be extended to broader task types to enhance this work.

Lastly, our hybrid methodology may introduce external validity threats regarding generalisability. To understand this, we evaluated AppRay on historical datasets\cite{mansur2023aidui, chen2023unveiling} from 2017 and 2020, whereas AppRay's training data was collected in 2024. 
\minor{Despite this temporal gap and the differences in app corpora and implementation contexts across these datasets, AppRay achieves micro-averaged F1 scores of 0.92 and 0.90 respectively (Tables 7–9), consistently outperforming all baselines. 
This demonstrates that AppRay captures fundamental semantic and visual concepts of manipulative design rather than surface-level stylistic features tied to a specific period or app corpus, and aligns with the relative stability of deceptive pattern taxonomies over time. }

\minor{
Another related threat concerns newly emerging deceptive patterns that fall outside the 16 types defined in our adopted taxonomy. As mobile app design practices evolve, new forms of manipulative design may arise that require additional annotation effort and model retraining to support. This limitation is inherent to any supervised detection approach in this domain; however, the cross-temporal generalization results above suggest that AppRay is likely to remain effective as long as the underlying manipulative intent and its manifestation in UI design remain consistent with established taxonomic definitions.
}

\section{Limitation, Future Work and Implications}
We present \tool, an end-to-end system that makes deceptive-pattern detection scalable by coupling automated app exploration with a multi-modal detector that captures both \revision{intra-page} and interaction-dependent patterns. Beyond improved metrics, our results indicate a practical path from ad-hoc, labor-intensive audits toward repeatable, ecosystem-scale screening.

\revision{
\textbf{Exploration coverage is the primary technical bottleneck.} While coordinating a task-oriented LLM navigator with random exploration substantially reduces manual effort, many missed patterns stem from incomplete UI coverage rather than misclassification. This reflects limitations of both LLM-based and random explorers, as well as a finite task set. Future work should pursue two complementary directions: (1) stronger UI agents with higher task completion and robustness (e.g., persistent memory, error prevention, rollback, and retrieval-augmented prompting) \cite{ran2024guardian, wen2024autodroid, zhang2025appagent, lee2024mobilegpt, agrawal2025uinavbench, vu2024gptvoicetasker}; and (2) a broader yet curated set of representative tasks and routines to improve coverage while controlling time and cost.
}

\revision{
\textbf{Scope is currently mobile-first.} Our GUI-level approach is largely framework-agnostic within mobile ecosystems because it operates on rendered screens rather than platform-specific code. We focus on mobile due to its high real-world impact (e.g., over 20 million active users in Australia and 4.5 hours/day spent in apps) \cite{AustraliaAppMarket}. However, extending \tool{} to web, desktop, and smart-TV interfaces will require adapting to different interaction paradigms and layout constraints, and is an important direction for broadening regulatory and industry applicability.
}

\revision{
\textbf{Bridging detection to remediation requires new signals.} GUI-level detection captures manipulation as users experience it, but developers and auditors often need actionable traces to locate root causes. Integrating program-analysis artifacts (e.g., layout code, event handlers, server-driven configuration) could connect observed UI behaviors to implementation-level evidence and support debugging and remediation. This remains challenging due to heterogeneous frameworks, dynamic UI generation, and feature flags, suggesting that code-level analysis should augment, not replace, GUI-level detection.
}

\revision{
\textbf{Dataset scaling and long-tail coverage remain open challenges.} Constructing AppRay's datasets required intensive manual exploration and annotation with around two-month annotations for 100 apps, and rare or highly \revision{inter-page} patterns are underrepresented relative to their diversity in the wild. Future work should reduce annotation cost and expand coverage via active learning, semi-automatic mining from large app corpora, and systematic perturbations that generate diverse variants of known patterns, especially for interaction-dependent cases.
}

\revision{
\textbf{Human-in-the-loop use is the realistic deployment model.} Expert studies suggest automated detectors are best positioned as high-throughput assistants rather than replacements. Experts excel at deep, persistent exploration and rare-case reasoning but are vulnerable to fatigue and scale. \tool{} offers consistent screening and prioritization, yet still struggles with complex navigation and borderline cases. These findings motivate workflows where \tool{} pre-screens large app ecosystems, surfaces evidence-backed candidates with explanations and visualizations, and enables more targeted, higher-quality audits by regulators and practitioners.
}

\revision{
\textbf{Ambiguity in definitions limits both annotation and automation.} Disagreements among \tool{}, baselines, and human annotators often arise from normative uncertainty, particularly for categories involving emotional framing or gamified incentives, where the boundary between persuasive design and dark patterns is contested. While resolving these boundaries is beyond this paper’s scope, our results highlight the need for closer collaboration between HCI, legal scholarship, and regulators to refine operational guidelines that can be consistently annotated and reliably detected at scale.
}

\minor{
\textbf{Relationship to Security Attacks.} While deceptive patterns are primarily an ethical and regulatory concern, they are not entirely unrelated to certain security threats. Phishing attacks, in particular, often employ UI-level deception techniques that overlap with deceptive patterns, for instance, using false visual hierarchy to make malicious links appear more prominent, or employing misleading consent dialogs to coerce credential disclosure. In this sense, deceptive pattern detection could serve as a complementary signal in phishing identification pipelines. However, the two should not be conflated: deceptive patterns are employed by otherwise legitimate applications for commercial gain, whereas phishing involves deliberate impersonation with malicious intent. AppRay is designed for the former and cannot determine malicious intent or detect network-level attacks such as DoS. Exploring the intersection of deceptive pattern analysis and security attack detection remains an open and potentially fruitful direction for future work.
}

\section{Conclusion}

In this paper, we take a first step toward addressing deceptive pattern detection at the app level, covering both \revision{intra-page} and \revision{\revision{inter-page}} deceptive patterns. We propose \tool{}, a fully automated system that combines LLM-based, task-oriented app exploration with automated testing to detect deceptive patterns across multiple related UIs. We further contribute two large-scale datasets, \revision{AppRay-Tainted-UIs} and \revision{AppRay-Benign-UIs}, which span a broad range of deceptive pattern types while preserving UI-level sequential information.

Experimental results demonstrate that \tool{} consistently outperforms existing methods and enables effective detection of previously unexplored \revision{intra-page} and \revision{\revision{inter-page}} deceptive patterns, \revision{improving micro- and macro-averaged F1 scores by 27.14\% and 38.71\% over the strongest baseline and by 1200\% and 650\% over the weakest baseline on their supported types in the \revision{AppRay-Tainted-UIs} dataset.} In addition, our user study confirms the practical value of \tool{} in assisting human experts with automated deceptive pattern identification.

\revision{Looking forward, we plan to enhance UI coverage through more advanced UI agents and expanded task designs, and to further improve dataset completeness and diversity to support the discovery of more complex interaction-dependent deceptive patterns.}

\begin{acks}
Jieshan Chen is partially funded by the Dieter Schwarz Foundation and the Technical University of Munich – Institute for Advanced Study, Germany.
\end{acks}

\bibliographystyle{ACM-Reference-Format}
\bibliography{reference}


\begin{thebibliography}{78}


\ifx \showCODEN    \undefined \def \showCODEN     #1{\unskip}     \fi
\ifx \showISBNx    \undefined \def \showISBNx     #1{\unskip}     \fi
\ifx \showISBNxiii \undefined \def \showISBNxiii  #1{\unskip}     \fi
\ifx \showISSN     \undefined \def \showISSN      #1{\unskip}     \fi
\ifx \showLCCN     \undefined \def \showLCCN      #1{\unskip}     \fi
\ifx \shownote     \undefined \def \shownote      #1{#1}          \fi
\ifx \showarticletitle \undefined \def \showarticletitle #1{#1}   \fi
\ifx \showURL      \undefined \def \showURL       {\relax}        \fi
\providecommand\bibfield[2]{#2}
\providecommand\bibinfo[2]{#2}
\providecommand\natexlab[1]{#1}
\providecommand\showeprint[2][]{arXiv:#2}

\bibitem[app(2026)]%
        {appray_material}
 \bibinfo{year}{2026}\natexlab{}.
\newblock \bibinfo{booktitle}{\emph{AppRay's source code and datasets}}.
\newblock
\urldef\tempurl%
\url{https://github.com/chenjshnn/AppRay}
\showURL{%
Retrieved 18 Jan 2026 from \tempurl}


\bibitem[Aagaard et~al\mbox{.}(2022)]%
        {aagaard2022game}
\bibfield{author}{\bibinfo{person}{Jacob Aagaard}, \bibinfo{person}{Miria Emma~Clausen Knudsen}, \bibinfo{person}{Per B{\ae}kgaard}, {and} \bibinfo{person}{Kevin Doherty}.} \bibinfo{year}{2022}\natexlab{}.
\newblock \showarticletitle{A Game of Dark Patterns: Designing Healthy, Highly-Engaging Mobile Games}. In \bibinfo{booktitle}{\emph{CHI Conference on Human Factors in Computing Systems Extended Abstracts}}. \bibinfo{pages}{1--8}.
\newblock


\bibitem[Agency(2024)]%
        {CaliforniaLaw}
\bibfield{author}{\bibinfo{person}{California Privacy~Protection Agency}.} \bibinfo{year}{2024}\natexlab{}.
\newblock \bibinfo{booktitle}{\emph{Avoiding dark patterns: clear and understandable language, symmetry in choice}}.
\newblock
\urldef\tempurl%
\url{https://cppa.ca.gov/pdf/enfadvisory202402.pdf}
\showURL{%
Retrieved 18 December 2025 from \tempurl}


\bibitem[Agrawal et~al\mbox{.}(2025)]%
        {agrawal2025uinavbench}
\bibfield{author}{\bibinfo{person}{Harsh Agrawal}, \bibinfo{person}{Eldon Schoop}, \bibinfo{person}{Xinlei Pan}, \bibinfo{person}{Anuj Mahajan}, \bibinfo{person}{Ari Seff}, \bibinfo{person}{Di Feng}, \bibinfo{person}{Ruijia Cheng}, \bibinfo{person}{Andres Romero Mier~Y Teran}, \bibinfo{person}{Esteban Gomez}, \bibinfo{person}{Abhishek Sundararajan}, {et~al\mbox{.}}} \bibinfo{year}{2025}\natexlab{}.
\newblock \showarticletitle{UINavBench: A Framework for Comprehensive Evaluation of Interactive Digital Agents}. In \bibinfo{booktitle}{\emph{Proceedings of the IEEE/CVF International Conference on Computer Vision}}. \bibinfo{pages}{23353--23363}.
\newblock


\bibitem[Bermejo~Fernandez et~al\mbox{.}(2021)]%
        {bermejo2021website}
\bibfield{author}{\bibinfo{person}{Carlos Bermejo~Fernandez}, \bibinfo{person}{Dimitris Chatzopoulos}, \bibinfo{person}{Dimitrios Papadopoulos}, {and} \bibinfo{person}{Pan Hui}.} \bibinfo{year}{2021}\natexlab{}.
\newblock \showarticletitle{This website uses nudging: Mturk workers' behaviour on cookie consent notices}.
\newblock \bibinfo{journal}{\emph{Proceedings of the ACM on human-computer interaction}} \bibinfo{volume}{5}, \bibinfo{number}{CSCW2} (\bibinfo{year}{2021}), \bibinfo{pages}{1--22}.
\newblock


\bibitem[Bongard-Blanchy et~al\mbox{.}(2021)]%
        {bongard2021definitely}
\bibfield{author}{\bibinfo{person}{Kerstin Bongard-Blanchy}, \bibinfo{person}{Arianna Rossi}, \bibinfo{person}{Salvador Rivas}, \bibinfo{person}{Sophie Doublet}, \bibinfo{person}{Vincent Koenig}, {and} \bibinfo{person}{Gabriele Lenzini}.} \bibinfo{year}{2021}\natexlab{}.
\newblock \showarticletitle{” I am Definitely Manipulated, Even When I am Aware of it. It’s Ridiculous!”-Dark Patterns from the End-User Perspective}. In \bibinfo{booktitle}{\emph{Designing Interactive Systems Conference 2021}}. \bibinfo{pages}{763--776}.
\newblock


\bibitem[Brignull(2010)]%
        {brignull2010dark}
\bibfield{author}{\bibinfo{person}{Harry Brignull}.} \bibinfo{year}{2010}\natexlab{}.
\newblock \bibinfo{booktitle}{\emph{Deceptive Design}}.
\newblock
\urldef\tempurl%
\url{https://www.darkpatterns.org/}
\showURL{%
Retrieved 17 August 2022 from \tempurl}


\bibitem[Chaudhary et~al\mbox{.}(2022)]%
        {chaudhary2022you}
\bibfield{author}{\bibinfo{person}{Akash Chaudhary}, \bibinfo{person}{Jaivrat Saroha}, \bibinfo{person}{Kyzyl Monteiro}, \bibinfo{person}{Angus~G Forbes}, {and} \bibinfo{person}{Aman Parnami}.} \bibinfo{year}{2022}\natexlab{}.
\newblock \showarticletitle{“Are You Still Watching?”: Exploring Unintended User Behaviors and Dark Patterns on Video Streaming Platforms}. In \bibinfo{booktitle}{\emph{Designing Interactive Systems Conference}}. \bibinfo{pages}{776--791}.
\newblock


\bibitem[Chen et~al\mbox{.}(2019)]%
        {chen2019revisiting}
\bibfield{author}{\bibinfo{person}{Gong Chen}, \bibinfo{person}{Wei Meng}, {and} \bibinfo{person}{John Copeland}.} \bibinfo{year}{2019}\natexlab{}.
\newblock \showarticletitle{Revisiting mobile advertising threats with madlife}. In \bibinfo{booktitle}{\emph{The World Wide Web Conference}}. \bibinfo{pages}{207--217}.
\newblock


\bibitem[Chen et~al\mbox{.}(2020a)]%
        {chen2020unblind}
\bibfield{author}{\bibinfo{person}{Jieshan Chen}, \bibinfo{person}{Chunyang Chen}, \bibinfo{person}{Zhenchang Xing}, \bibinfo{person}{Xiwei Xu}, \bibinfo{person}{Liming Zhu}, \bibinfo{person}{Guoqiang Li}, {and} \bibinfo{person}{Jinshui Wang}.} \bibinfo{year}{2020}\natexlab{a}.
\newblock \showarticletitle{Unblind your apps: Predicting natural-language labels for mobile gui components by deep learning}. In \bibinfo{booktitle}{\emph{Proceedings of the ACM/IEEE 42nd International Conference on Software Engineering}}. \bibinfo{pages}{322--334}.
\newblock


\bibitem[Chen et~al\mbox{.}(2023a)]%
        {UIGuardCode}
\bibfield{author}{\bibinfo{person}{Jieshan Chen}, \bibinfo{person}{Jiamou Sun}, \bibinfo{person}{Sidong Feng}, \bibinfo{person}{Zhenchang Xing}, \bibinfo{person}{Qinghua Lu}, \bibinfo{person}{Xiwei Xu}, {and} \bibinfo{person}{Chunyang Chen}.} \bibinfo{year}{2023}\natexlab{a}.
\newblock \bibinfo{booktitle}{\emph{Replication Package for UIGuard}}.
\newblock
\urldef\tempurl%
\url{https://github.com/chenjshnn/UIST23-UIGuard}
\showURL{%
Retrieved 18 December 2025 from \tempurl}


\bibitem[Chen et~al\mbox{.}(2023b)]%
        {chen2023unveiling}
\bibfield{author}{\bibinfo{person}{Jieshan Chen}, \bibinfo{person}{Jiamou Sun}, \bibinfo{person}{Sidong Feng}, \bibinfo{person}{Zhenchang Xing}, \bibinfo{person}{Qinghua Lu}, \bibinfo{person}{Xiwei Xu}, {and} \bibinfo{person}{Chunyang Chen}.} \bibinfo{year}{2023}\natexlab{b}.
\newblock \showarticletitle{Unveiling the Tricks: Automated Detection of Dark Patterns in Mobile Applications}.
\newblock \bibinfo{journal}{\emph{arXiv preprint arXiv:2308.05898}} (\bibinfo{year}{2023}).
\newblock


\bibitem[Chen et~al\mbox{.}(2022)]%
        {chen2022towards}
\bibfield{author}{\bibinfo{person}{Jieshan Chen}, \bibinfo{person}{Amanda Swearngin}, \bibinfo{person}{Jason Wu}, \bibinfo{person}{Titus Barik}, \bibinfo{person}{Jeffrey Nichols}, {and} \bibinfo{person}{Xiaoyi Zhang}.} \bibinfo{year}{2022}\natexlab{}.
\newblock \showarticletitle{Towards complete icon labeling in mobile applications}. In \bibinfo{booktitle}{\emph{Proceedings of the 2022 CHI conference on human factors in computing systems}}. \bibinfo{pages}{1--14}.
\newblock


\bibitem[Chen et~al\mbox{.}(2020b)]%
        {chen2020object}
\bibfield{author}{\bibinfo{person}{Jieshan Chen}, \bibinfo{person}{Mulong Xie}, \bibinfo{person}{Zhenchang Xing}, \bibinfo{person}{Chunyang Chen}, \bibinfo{person}{Xiwei Xu}, \bibinfo{person}{Liming Zhu}, {and} \bibinfo{person}{Guoqiang Li}.} \bibinfo{year}{2020}\natexlab{b}.
\newblock \showarticletitle{Object detection for graphical user interface: Old fashioned or deep learning or a combination?}. In \bibinfo{booktitle}{\emph{proceedings of the 28th ACM joint meeting on European Software Engineering Conference and Symposium on the Foundations of Software Engineering}}. \bibinfo{pages}{1202--1214}.
\newblock


\bibitem[Cho et~al\mbox{.}(2021)]%
        {cho2021reflect}
\bibfield{author}{\bibinfo{person}{Hyunsung Cho}, \bibinfo{person}{DaEun Choi}, \bibinfo{person}{Donghwi Kim}, \bibinfo{person}{Wan~Ju Kang}, \bibinfo{person}{Eun~Kyoung Choe}, {and} \bibinfo{person}{Sung-Ju Lee}.} \bibinfo{year}{2021}\natexlab{}.
\newblock \showarticletitle{Reflect, not regret: Understanding regretful smartphone use with app feature-level analysis}.
\newblock \bibinfo{journal}{\emph{Proceedings of the ACM on human-computer interaction}} \bibinfo{volume}{5}, \bibinfo{number}{CSCW2} (\bibinfo{year}{2021}), \bibinfo{pages}{1--36}.
\newblock


\bibitem[Datta et~al\mbox{.}(2022)]%
        {datta2022greasevision}
\bibfield{author}{\bibinfo{person}{Siddhartha Datta}, \bibinfo{person}{Konrad Kollnig}, {and} \bibinfo{person}{Nigel Shadbolt}.} \bibinfo{year}{2022}\natexlab{}.
\newblock \showarticletitle{GreaseVision: Rewriting the rules of the interface}. In \bibinfo{booktitle}{\emph{Proceedings of the First Workshop on Dynamic Adversarial Data Collection}}. \bibinfo{pages}{7--22}.
\newblock


\bibitem[Developers(2012)]%
        {monkey}
\bibfield{author}{\bibinfo{person}{Android Developers}.} \bibinfo{year}{2012}\natexlab{}.
\newblock \bibinfo{booktitle}{\emph{UI/Application Exerciser Monkey}}.
\newblock
\urldef\tempurl%
\url{https://developer.android.com/studio/test/other-testing-tools/monkey}
\showURL{%
Retrieved 12 Dec 2025 from \tempurl}


\bibitem[Di~Geronimo et~al\mbox{.}(2020)]%
        {di2020ui}
\bibfield{author}{\bibinfo{person}{Linda Di~Geronimo}, \bibinfo{person}{Larissa Braz}, \bibinfo{person}{Enrico Fregnan}, \bibinfo{person}{Fabio Palomba}, {and} \bibinfo{person}{Alberto Bacchelli}.} \bibinfo{year}{2020}\natexlab{}.
\newblock \showarticletitle{UI dark patterns and where to find them: a study on mobile applications and user perception}. In \bibinfo{booktitle}{\emph{Proceedings of the 2020 CHI conference on human factors in computing systems}}. \bibinfo{pages}{1--14}.
\newblock


\bibitem[Dong et~al\mbox{.}(2018)]%
        {dong2018frauddroid}
\bibfield{author}{\bibinfo{person}{Feng Dong}, \bibinfo{person}{Haoyu Wang}, \bibinfo{person}{Li Li}, \bibinfo{person}{Yao Guo}, \bibinfo{person}{Tegawend{\'e}~F Bissyand{\'e}}, \bibinfo{person}{Tianming Liu}, \bibinfo{person}{Guoai Xu}, {and} \bibinfo{person}{Jacques Klein}.} \bibinfo{year}{2018}\natexlab{}.
\newblock \showarticletitle{Frauddroid: Automated ad fraud detection for android apps}. In \bibinfo{booktitle}{\emph{Proceedings of the 2018 26th ACM joint meeting on European software engineering conference and symposium on the foundations of software engineering}}. \bibinfo{pages}{257--268}.
\newblock


\bibitem[Feng and Chen(2024)]%
        {feng2024prompting}
\bibfield{author}{\bibinfo{person}{Sidong Feng} {and} \bibinfo{person}{Chunyang Chen}.} \bibinfo{year}{2024}\natexlab{}.
\newblock \showarticletitle{Prompting is all you need: Automated android bug replay with large language models}. In \bibinfo{booktitle}{\emph{Proceedings of the 46th IEEE/ACM International Conference on Software Engineering}}. \bibinfo{pages}{1--13}.
\newblock


\bibitem[Feng et~al\mbox{.}(2021)]%
        {DBLP:journals/corr/abs-2105-03075}
\bibfield{author}{\bibinfo{person}{Steven~Y. Feng}, \bibinfo{person}{Varun Gangal}, \bibinfo{person}{Jason Wei}, \bibinfo{person}{Sarath Chandar}, \bibinfo{person}{Soroush Vosoughi}, \bibinfo{person}{Teruko Mitamura}, {and} \bibinfo{person}{Eduard~H. Hovy}.} \bibinfo{year}{2021}\natexlab{}.
\newblock \showarticletitle{A Survey of Data Augmentation Approaches for {NLP}}.
\newblock \bibinfo{journal}{\emph{CoRR}}  \bibinfo{volume}{abs/2105.03075} (\bibinfo{year}{2021}).
\newblock
\showeprint[arXiv]{2105.03075}
\urldef\tempurl%
\url{https://arxiv.org/abs/2105.03075}
\showURL{%
\tempurl}


\bibitem[Gak et~al\mbox{.}(2022)]%
        {gak2022distressing}
\bibfield{author}{\bibinfo{person}{Liza Gak}, \bibinfo{person}{Seyi Olojo}, {and} \bibinfo{person}{Niloufar Salehi}.} \bibinfo{year}{2022}\natexlab{}.
\newblock \showarticletitle{The distressing ads that persist: Uncovering the harms of targeted weight-loss ads among users with histories of disordered eating}.
\newblock \bibinfo{journal}{\emph{Proceedings of the ACM on Human-Computer Interaction}} \bibinfo{volume}{6}, \bibinfo{number}{CSCW2} (\bibinfo{year}{2022}), \bibinfo{pages}{1--23}.
\newblock


\bibitem[Gamma et~al\mbox{.}(1995)]%
        {gamma1995design}
\bibfield{author}{\bibinfo{person}{Erich Gamma}, \bibinfo{person}{Richard Helm}, \bibinfo{person}{Ralph Johnson}, {and} \bibinfo{person}{John Vlissides}.} \bibinfo{year}{1995}\natexlab{}.
\newblock \bibinfo{booktitle}{\emph{Design patterns: elements of reusable object-oriented software}}.
\newblock \bibinfo{publisher}{Pearson Deutschland GmbH}.
\newblock


\bibitem[Gao et~al\mbox{.}(2022)]%
        {gao2022understanding}
\bibfield{author}{\bibinfo{person}{Cuiyun Gao}, \bibinfo{person}{Jichuan Zeng}, \bibinfo{person}{David Lo}, \bibinfo{person}{Xin Xia}, \bibinfo{person}{Irwin King}, {and} \bibinfo{person}{Michael~R Lyu}.} \bibinfo{year}{2022}\natexlab{}.
\newblock \showarticletitle{Understanding in-app advertising issues based on large scale app review analysis}.
\newblock \bibinfo{journal}{\emph{Information and Software Technology}}  \bibinfo{volume}{142} (\bibinfo{year}{2022}), \bibinfo{pages}{106741}.
\newblock


\bibitem[Gao et~al\mbox{.}(2021)]%
        {gao2021users}
\bibfield{author}{\bibinfo{person}{Cuiyun Gao}, \bibinfo{person}{Jichuan Zeng}, \bibinfo{person}{Federica Sarro}, \bibinfo{person}{David Lo}, \bibinfo{person}{Irwin King}, {and} \bibinfo{person}{Michael~R Lyu}.} \bibinfo{year}{2021}\natexlab{}.
\newblock \showarticletitle{Do users care about ad’s performance costs? Exploring the effects of the performance costs of in-app ads on user experience}.
\newblock \bibinfo{journal}{\emph{Information and Software Technology}}  \bibinfo{volume}{132} (\bibinfo{year}{2021}), \bibinfo{pages}{106471}.
\newblock


\bibitem[Gray et~al\mbox{.}(2018)]%
        {gray2018dark}
\bibfield{author}{\bibinfo{person}{Colin~M Gray}, \bibinfo{person}{Yubo Kou}, \bibinfo{person}{Bryan Battles}, \bibinfo{person}{Joseph Hoggatt}, {and} \bibinfo{person}{Austin~L Toombs}.} \bibinfo{year}{2018}\natexlab{}.
\newblock \showarticletitle{The dark (patterns) side of UX design}. In \bibinfo{booktitle}{\emph{Proceedings of the 2018 CHI conference on human factors in computing systems}}. \bibinfo{pages}{1--14}.
\newblock


\bibitem[Gray et~al\mbox{.}(2025)]%
        {gray2025getting}
\bibfield{author}{\bibinfo{person}{Colin~M Gray}, \bibinfo{person}{Thomas Mildner}, {and} \bibinfo{person}{Ritika Gairola}.} \bibinfo{year}{2025}\natexlab{}.
\newblock \showarticletitle{Getting Trapped in Amazon's" Iliad Flow": A Foundation for the Temporal Analysis of Dark Patterns}. In \bibinfo{booktitle}{\emph{Proceedings of the 2025 CHI Conference on Human Factors in Computing Systems}}. \bibinfo{pages}{1--10}.
\newblock


\bibitem[Gray et~al\mbox{.}(2021)]%
        {gray2021dark}
\bibfield{author}{\bibinfo{person}{Colin~M Gray}, \bibinfo{person}{Cristiana Santos}, \bibinfo{person}{Nataliia Bielova}, \bibinfo{person}{Michael Toth}, {and} \bibinfo{person}{Damian Clifford}.} \bibinfo{year}{2021}\natexlab{}.
\newblock \showarticletitle{Dark patterns and the legal requirements of consent banners: An interaction criticism perspective}. In \bibinfo{booktitle}{\emph{Proceedings of the 2021 CHI Conference on Human Factors in Computing Systems}}. \bibinfo{pages}{1--18}.
\newblock


\bibitem[Gui et~al\mbox{.}(2017)]%
        {gui2017aspects}
\bibfield{author}{\bibinfo{person}{Jiaping Gui}, \bibinfo{person}{Meiyappan Nagappan}, {and} \bibinfo{person}{William~GJ Halfond}.} \bibinfo{year}{2017}\natexlab{}.
\newblock \showarticletitle{What aspects of mobile ads do users care about? an empirical study of mobile in-app ad reviews}.
\newblock \bibinfo{journal}{\emph{arXiv preprint arXiv:1702.07681}} (\bibinfo{year}{2017}).
\newblock


\bibitem[Gunawan et~al\mbox{.}(2021)]%
        {gunawan2021comparative}
\bibfield{author}{\bibinfo{person}{Johanna Gunawan}, \bibinfo{person}{Amogh Pradeep}, \bibinfo{person}{David Choffnes}, \bibinfo{person}{Woodrow Hartzog}, {and} \bibinfo{person}{Christo Wilson}.} \bibinfo{year}{2021}\natexlab{}.
\newblock \showarticletitle{A Comparative Study of Dark Patterns Across Web and Mobile Modalities}.
\newblock \bibinfo{journal}{\emph{Proceedings of the ACM on Human-Computer Interaction}} \bibinfo{volume}{5}, \bibinfo{number}{CSCW2} (\bibinfo{year}{2021}), \bibinfo{pages}{1--29}.
\newblock


\bibitem[Hidaka et~al\mbox{.}(2023)]%
        {hidaka2023linguistic}
\bibfield{author}{\bibinfo{person}{Shun Hidaka}, \bibinfo{person}{Sota Kobuki}, \bibinfo{person}{Mizuki Watanabe}, {and} \bibinfo{person}{Katie Seaborn}.} \bibinfo{year}{2023}\natexlab{}.
\newblock \showarticletitle{Linguistic Dead-Ends and Alphabet Soup: Finding Dark Patterns in Japanese Apps}. In \bibinfo{booktitle}{\emph{Proceedings of the 2023 CHI Conference on Human Factors in Computing Systems}}. \bibinfo{pages}{1--13}.
\newblock


\bibitem[Jaiswal et~al\mbox{.}(2020)]%
        {DBLP:journals/corr/abs-2011-00362}
\bibfield{author}{\bibinfo{person}{Ashish Jaiswal}, \bibinfo{person}{Ashwin~Ramesh Babu}, \bibinfo{person}{Mohammad~Zaki Zadeh}, \bibinfo{person}{Debapriya Banerjee}, {and} \bibinfo{person}{Fillia Makedon}.} \bibinfo{year}{2020}\natexlab{}.
\newblock \showarticletitle{A Survey on Contrastive Self-supervised Learning}.
\newblock \bibinfo{journal}{\emph{CoRR}}  \bibinfo{volume}{abs/2011.00362} (\bibinfo{year}{2020}).
\newblock
\showeprint[arXiv]{2011.00362}
\urldef\tempurl%
\url{https://arxiv.org/abs/2011.00362}
\showURL{%
\tempurl}


\bibitem[Kirkman et~al\mbox{.}(2023)]%
        {kirkman2023darkdialogs}
\bibfield{author}{\bibinfo{person}{Daniel Kirkman}, \bibinfo{person}{Kami Vaniea}, {and} \bibinfo{person}{Daniel~W Woods}.} \bibinfo{year}{2023}\natexlab{}.
\newblock \showarticletitle{DarkDialogs: Automated detection of 10 dark patterns on cookie dialogs}. In \bibinfo{booktitle}{\emph{2023 IEEE 8th European Symposium on Security and Privacy (EuroS\&P)}}. IEEE, \bibinfo{pages}{847--867}.
\newblock


\bibitem[Kollnig et~al\mbox{.}(2021)]%
        {kollnig2021want}
\bibfield{author}{\bibinfo{person}{Konrad Kollnig}, \bibinfo{person}{Siddhartha Datta}, {and} \bibinfo{person}{Max Van~Kleek}.} \bibinfo{year}{2021}\natexlab{}.
\newblock \showarticletitle{I Want My App That Way: Reclaiming Sovereignty Over Personal Devices}. In \bibinfo{booktitle}{\emph{Extended Abstracts of the 2021 CHI Conference on Human Factors in Computing Systems}}. \bibinfo{pages}{1--8}.
\newblock


\bibitem[Kowalczyk et~al\mbox{.}(2023)]%
        {kowalczyk2023understanding}
\bibfield{author}{\bibinfo{person}{Monica Kowalczyk}, \bibinfo{person}{Johanna~T Gunawan}, \bibinfo{person}{David Choffnes}, \bibinfo{person}{Daniel~J Dubois}, \bibinfo{person}{Woodrow Hartzog}, {and} \bibinfo{person}{Christo Wilson}.} \bibinfo{year}{2023}\natexlab{}.
\newblock \showarticletitle{Understanding Dark Patterns in Home IoT Devices}. In \bibinfo{booktitle}{\emph{Proceedings of the 2023 CHI Conference on Human Factors in Computing Systems}}. \bibinfo{pages}{1--27}.
\newblock


\bibitem[Lan et~al\mbox{.}(2024)]%
        {lan2024deeply}
\bibfield{author}{\bibinfo{person}{Yuanhong Lan}, \bibinfo{person}{Yifei Lu}, \bibinfo{person}{Zhong Li}, \bibinfo{person}{Minxue Pan}, \bibinfo{person}{Wenhua Yang}, \bibinfo{person}{Tian Zhang}, {and} \bibinfo{person}{Xuandong Li}.} \bibinfo{year}{2024}\natexlab{}.
\newblock \showarticletitle{Deeply reinforcing android gui testing with deep reinforcement learning}. In \bibinfo{booktitle}{\emph{Proceedings of the 46th IEEE/ACM International Conference on Software Engineering}}. \bibinfo{pages}{1--13}.
\newblock


\bibitem[Lee et~al\mbox{.}(2024)]%
        {lee2024mobilegpt}
\bibfield{author}{\bibinfo{person}{Sunjae Lee}, \bibinfo{person}{Junyoung Choi}, \bibinfo{person}{Jungjae Lee}, \bibinfo{person}{Munim~Hasan Wasi}, \bibinfo{person}{Hojun Choi}, \bibinfo{person}{Steve Ko}, \bibinfo{person}{Sangeun Oh}, {and} \bibinfo{person}{Insik Shin}.} \bibinfo{year}{2024}\natexlab{}.
\newblock \showarticletitle{Mobilegpt: Augmenting llm with human-like app memory for mobile task automation}. In \bibinfo{booktitle}{\emph{Proceedings of the 30th Annual International Conference on Mobile Computing and Networking}}. \bibinfo{pages}{1119--1133}.
\newblock


\bibitem[Li et~al\mbox{.}(2017)]%
        {li2017droidbot}
\bibfield{author}{\bibinfo{person}{Yuanchun Li}, \bibinfo{person}{Ziyue Yang}, \bibinfo{person}{Yao Guo}, {and} \bibinfo{person}{Xiangqun Chen}.} \bibinfo{year}{2017}\natexlab{}.
\newblock \showarticletitle{Droidbot: a lightweight ui-guided test input generator for android}. In \bibinfo{booktitle}{\emph{2017 IEEE/ACM 39th international conference on software engineering companion (ICSE-C)}}. IEEE, \bibinfo{pages}{23--26}.
\newblock


\bibitem[Li et~al\mbox{.}(2019)]%
        {li2019humanoid}
\bibfield{author}{\bibinfo{person}{Yuanchun Li}, \bibinfo{person}{Ziyue Yang}, \bibinfo{person}{Yao Guo}, {and} \bibinfo{person}{Xiangqun Chen}.} \bibinfo{year}{2019}\natexlab{}.
\newblock \showarticletitle{Humanoid: A deep learning-based approach to automated black-box android app testing}. In \bibinfo{booktitle}{\emph{2019 34th IEEE/ACM International Conference on Automated Software Engineering (ASE)}}. IEEE, \bibinfo{pages}{1070--1073}.
\newblock


\bibitem[Liu et~al\mbox{.}(2025)]%
        {liu2025ios}
\bibfield{author}{\bibinfo{person}{Tianming Liu}, \bibinfo{person}{Jiapeng Deng}, \bibinfo{person}{Yanjie Zhao}, \bibinfo{person}{Xiao Chen}, \bibinfo{person}{Xiaoning Du}, \bibinfo{person}{Li Li}, {and} \bibinfo{person}{Haoyu Wang}.} \bibinfo{year}{2025}\natexlab{}.
\newblock \showarticletitle{Are iOS Apps Immune to Abusive Advertising Practices?}. In \bibinfo{booktitle}{\emph{Proceedings of the 33rd ACM International Conference on the Foundations of Software Engineering}}. \bibinfo{pages}{491--502}.
\newblock


\bibitem[Liu et~al\mbox{.}(2020b)]%
        {liu2020maddroid}
\bibfield{author}{\bibinfo{person}{Tianming Liu}, \bibinfo{person}{Haoyu Wang}, \bibinfo{person}{Li Li}, \bibinfo{person}{Xiapu Luo}, \bibinfo{person}{Feng Dong}, \bibinfo{person}{Yao Guo}, \bibinfo{person}{Liu Wang}, \bibinfo{person}{Tegawend{\'e} Bissyand{\'e}}, {and} \bibinfo{person}{Jacques Klein}.} \bibinfo{year}{2020}\natexlab{b}.
\newblock \showarticletitle{Maddroid: Characterizing and detecting devious ad contents for android apps}. In \bibinfo{booktitle}{\emph{Proceedings of The Web Conference 2020}}. \bibinfo{pages}{1715--1726}.
\newblock


\bibitem[Liu et~al\mbox{.}(2024)]%
        {liu2024make}
\bibfield{author}{\bibinfo{person}{Zhe Liu}, \bibinfo{person}{Chunyang Chen}, \bibinfo{person}{Junjie Wang}, \bibinfo{person}{Mengzhuo Chen}, \bibinfo{person}{Boyu Wu}, \bibinfo{person}{Xing Che}, \bibinfo{person}{Dandan Wang}, {and} \bibinfo{person}{Qing Wang}.} \bibinfo{year}{2024}\natexlab{}.
\newblock \showarticletitle{Make llm a testing expert: Bringing human-like interaction to mobile gui testing via functionality-aware decisions}. In \bibinfo{booktitle}{\emph{Proceedings of the IEEE/ACM 46th International Conference on Software Engineering}}. \bibinfo{pages}{1--13}.
\newblock


\bibitem[Liu et~al\mbox{.}(2020a)]%
        {liu2020owl}
\bibfield{author}{\bibinfo{person}{Zhe Liu}, \bibinfo{person}{Chunyang Chen}, \bibinfo{person}{Junjie Wang}, \bibinfo{person}{Yuekai Huang}, \bibinfo{person}{Jun Hu}, {and} \bibinfo{person}{Qing Wang}.} \bibinfo{year}{2020}\natexlab{a}.
\newblock \showarticletitle{Owl eyes: Spotting ui display issues via visual understanding}. In \bibinfo{booktitle}{\emph{2020 35th IEEE/ACM International Conference on Automated Software Engineering (ASE)}}. IEEE, \bibinfo{pages}{398--409}.
\newblock


\bibitem[Lv et~al\mbox{.}(2022)]%
        {lv2022fastbot2}
\bibfield{author}{\bibinfo{person}{Zhengwei Lv}, \bibinfo{person}{Chao Peng}, \bibinfo{person}{Zhao Zhang}, \bibinfo{person}{Ting Su}, \bibinfo{person}{Kai Liu}, {and} \bibinfo{person}{Ping Yang}.} \bibinfo{year}{2022}\natexlab{}.
\newblock \showarticletitle{Fastbot2: Reusable Automated Model-based GUI Testing for Android Enhanced by Reinforcement Learning}. In \bibinfo{booktitle}{\emph{Proceedings of the 37th IEEE/ACM International Conference on Automated Software Engineering}}. \bibinfo{pages}{1--5}.
\newblock


\bibitem[Mansur et~al\mbox{.}(2023)]%
        {mansur2023aidui}
\bibfield{author}{\bibinfo{person}{SM~Hasan Mansur}, \bibinfo{person}{Sabiha Salma}, \bibinfo{person}{Damilola Awofisayo}, {and} \bibinfo{person}{Kevin Moran}.} \bibinfo{year}{2023}\natexlab{}.
\newblock \showarticletitle{Aidui: Toward automated recognition of dark patterns in user interfaces}. In \bibinfo{booktitle}{\emph{2023 IEEE/ACM 45th International Conference on Software Engineering (ICSE)}}. IEEE, \bibinfo{pages}{1958--1970}.
\newblock


\bibitem[Mathur et~al\mbox{.}(2019)]%
        {mathur2019dark}
\bibfield{author}{\bibinfo{person}{Arunesh Mathur}, \bibinfo{person}{Gunes Acar}, \bibinfo{person}{Michael~J Friedman}, \bibinfo{person}{Eli Lucherini}, \bibinfo{person}{Jonathan Mayer}, \bibinfo{person}{Marshini Chetty}, {and} \bibinfo{person}{Arvind Narayanan}.} \bibinfo{year}{2019}\natexlab{}.
\newblock \showarticletitle{Dark patterns at scale: Findings from a crawl of 11K shopping websites}.
\newblock \bibinfo{journal}{\emph{Proceedings of the ACM on Human-Computer Interaction}} \bibinfo{volume}{3}, \bibinfo{number}{CSCW} (\bibinfo{year}{2019}), \bibinfo{pages}{1--32}.
\newblock


\bibitem[Mildner et~al\mbox{.}(2023)]%
        {mildner2023engaging}
\bibfield{author}{\bibinfo{person}{Thomas Mildner}, \bibinfo{person}{Gian-Luca Savino}, \bibinfo{person}{Philip~R Doyle}, \bibinfo{person}{Benjamin~R Cowan}, {and} \bibinfo{person}{Rainer Malaka}.} \bibinfo{year}{2023}\natexlab{}.
\newblock \showarticletitle{About Engaging and Governing Strategies: A Thematic Analysis of Dark Patterns in Social Networking Services}. In \bibinfo{booktitle}{\emph{Proceedings of the 2023 CHI Conference on Human Factors in Computing Systems}}. \bibinfo{pages}{1--15}.
\newblock


\bibitem[Monge~Roffarello and De~Russis(2022)]%
        {monge2022towards}
\bibfield{author}{\bibinfo{person}{Alberto Monge~Roffarello} {and} \bibinfo{person}{Luigi De~Russis}.} \bibinfo{year}{2022}\natexlab{}.
\newblock \showarticletitle{Towards understanding the dark patterns that steal our attention}. In \bibinfo{booktitle}{\emph{Chi conference on human factors in computing systems extended abstracts}}. \bibinfo{pages}{1--7}.
\newblock


\bibitem[Monge~Roffarello et~al\mbox{.}(2023)]%
        {monge2023defining}
\bibfield{author}{\bibinfo{person}{Alberto Monge~Roffarello}, \bibinfo{person}{Kai Lukoff}, {and} \bibinfo{person}{Luigi De~Russis}.} \bibinfo{year}{2023}\natexlab{}.
\newblock \showarticletitle{Defining and Identifying Attention Capture Deceptive Designs in Digital Interfaces}. In \bibinfo{booktitle}{\emph{Proceedings of the 2023 CHI Conference on Human Factors in Computing Systems}}. \bibinfo{pages}{1--19}.
\newblock


\bibitem[Nguyen et~al\mbox{.}(2022)]%
        {nguyen2022freely}
\bibfield{author}{\bibinfo{person}{Trung~Tin Nguyen}, \bibinfo{person}{Michael Backes}, {and} \bibinfo{person}{Ben Stock}.} \bibinfo{year}{2022}\natexlab{}.
\newblock \showarticletitle{Freely Given Consent? Studying Consent Notice of Third-Party Tracking and Its Violations of GDPR in Android Apps}. In \bibinfo{booktitle}{\emph{Proceedings of the 2022 ACM SIGSAC Conference on Computer and Communications Security}}. \bibinfo{pages}{2369--2383}.
\newblock


\bibitem[Nie et~al\mbox{.}(2024)]%
        {nie2024shadows}
\bibfield{author}{\bibinfo{person}{Liming Nie}, \bibinfo{person}{Yangyang Zhao}, \bibinfo{person}{Chenglin Li}, \bibinfo{person}{Xuqiong Luo}, {and} \bibinfo{person}{Yang Liu}.} \bibinfo{year}{2024}\natexlab{}.
\newblock \showarticletitle{Shadows in the Interface: A Comprehensive Study on Dark Patterns}.
\newblock \bibinfo{journal}{\emph{Proceedings of the ACM on Software Engineering}} \bibinfo{volume}{1}, \bibinfo{number}{FSE} (\bibinfo{year}{2024}), \bibinfo{pages}{204--225}.
\newblock


\bibitem[Nouwens et~al\mbox{.}(2020)]%
        {nouwens2020dark}
\bibfield{author}{\bibinfo{person}{Midas Nouwens}, \bibinfo{person}{Ilaria Liccardi}, \bibinfo{person}{Michael Veale}, \bibinfo{person}{David Karger}, {and} \bibinfo{person}{Lalana Kagal}.} \bibinfo{year}{2020}\natexlab{}.
\newblock \showarticletitle{Dark patterns after the GDPR: Scraping consent pop-ups and demonstrating their influence}. In \bibinfo{booktitle}{\emph{Proceedings of the 2020 CHI conference on human factors in computing systems}}. \bibinfo{pages}{1--13}.
\newblock


\bibitem[OpenAI(2023)]%
        {openai2023gpt4}
\bibfield{author}{\bibinfo{person}{OpenAI}.} \bibinfo{year}{2023}\natexlab{}.
\newblock \showarticletitle{GPT-4 Technical Report}.
\newblock \bibinfo{journal}{\emph{arXiv preprint arXiv:2303.08774}} (\bibinfo{year}{2023}).
\newblock


\bibitem[Parliament(2023)]%
        {EUAIAct}
\bibfield{author}{\bibinfo{person}{European Parliament}.} \bibinfo{year}{2023}\natexlab{}.
\newblock \bibinfo{booktitle}{\emph{EU AI Act: first regulation on artificial intelligence}}.
\newblock
\urldef\tempurl%
\url{https://www.europarl.europa.eu/topics/en/article/20230601STO93804/eu-ai-act-first-regulation-on-artificial-intelligence}
\showURL{%
Retrieved 18 December 2025 from \tempurl}


\bibitem[Parliament(2025)]%
        {EUguidelines}
\bibfield{author}{\bibinfo{person}{European Parliament}.} \bibinfo{year}{2025}\natexlab{}.
\newblock \bibinfo{booktitle}{\emph{Regulating dark patterns in the EU: Towards digital fairness}}.
\newblock
\urldef\tempurl%
\url{https://www.europarl.europa.eu/RegData/etudes/ATAG/2025/767191/EPRS_ATA(2025)767191_EN.pdf}
\showURL{%
Retrieved 18 December 2025 from \tempurl}


\bibitem[Raju et~al\mbox{.}(2021)]%
        {raju2021smart}
\bibfield{author}{\bibinfo{person}{S~Hrushikesava Raju}, \bibinfo{person}{Saiyed~Faiayaz Waris}, \bibinfo{person}{S Adinarayna}, \bibinfo{person}{Vijaya~Chandra Jadala}, {and} \bibinfo{person}{G~Subba Rao}.} \bibinfo{year}{2021}\natexlab{}.
\newblock \showarticletitle{Smart dark pattern detection: Making aware of misleading patterns through the intended app}.
\newblock In \bibinfo{booktitle}{\emph{Sentimental Analysis and Deep Learning: Proceedings of ICSADL 2021}}. \bibinfo{publisher}{Springer}, \bibinfo{pages}{933--947}.
\newblock


\bibitem[Ran et~al\mbox{.}(2024)]%
        {ran2024guardian}
\bibfield{author}{\bibinfo{person}{Dezhi Ran}, \bibinfo{person}{Hao Wang}, \bibinfo{person}{Zihe Song}, \bibinfo{person}{Mengzhou Wu}, \bibinfo{person}{Yuan Cao}, \bibinfo{person}{Ying Zhang}, \bibinfo{person}{Wei Yang}, {and} \bibinfo{person}{Tao Xie}.} \bibinfo{year}{2024}\natexlab{}.
\newblock \showarticletitle{Guardian: A runtime framework for LLM-based UI exploration}. In \bibinfo{booktitle}{\emph{Proceedings of the 33rd ACM SIGSOFT International Symposium on Software Testing and Analysis}}. \bibinfo{pages}{958--970}.
\newblock


\bibitem[Robinson et~al\mbox{.}(2020)]%
        {DBLP:journals/corr/abs-2010-04592}
\bibfield{author}{\bibinfo{person}{Joshua Robinson}, \bibinfo{person}{Ching{-}Yao Chuang}, \bibinfo{person}{Suvrit Sra}, {and} \bibinfo{person}{Stefanie Jegelka}.} \bibinfo{year}{2020}\natexlab{}.
\newblock \showarticletitle{Contrastive Learning with Hard Negative Samples}.
\newblock \bibinfo{journal}{\emph{CoRR}}  \bibinfo{volume}{abs/2010.04592} (\bibinfo{year}{2020}).
\newblock
\showeprint[arXiv]{2010.04592}
\urldef\tempurl%
\url{https://arxiv.org/abs/2010.04592}
\showURL{%
\tempurl}


\bibitem[S~M Hasan~Mansur and Moran(2023)]%
        {AidUICode}
\bibfield{author}{\bibinfo{person}{Damilola~Awofisayo S~M Hasan~Mansur, Sabiha~Salma} {and} \bibinfo{person}{Kevin Moran}.} \bibinfo{year}{2023}\natexlab{}.
\newblock \bibinfo{booktitle}{\emph{Replication Package for AidUI}}.
\newblock
\urldef\tempurl%
\url{https://github.com/SageSELab/AidUI}
\showURL{%
Retrieved 18 December 2025 from \tempurl}


\bibitem[Sergeeva et~al\mbox{.}(2023)]%
        {sergeeva2023we}
\bibfield{author}{\bibinfo{person}{Anastasia Sergeeva}, \bibinfo{person}{Bj{\"o}rn Rohles}, \bibinfo{person}{Verena Distler}, {and} \bibinfo{person}{Vincent Koenig}.} \bibinfo{year}{2023}\natexlab{}.
\newblock \showarticletitle{“We Need a Big Revolution in Email Advertising”: Users’ Perception of Persuasion in Permission-based Advertising Emails}. In \bibinfo{booktitle}{\emph{Proceedings of the 2023 CHI Conference on Human Factors in Computing Systems}}. \bibinfo{pages}{1--21}.
\newblock


\bibitem[Shao et~al\mbox{.}(2018)]%
        {shao2018understanding}
\bibfield{author}{\bibinfo{person}{Rui Shao}, \bibinfo{person}{Vaibhav Rastogi}, \bibinfo{person}{Yan Chen}, \bibinfo{person}{Xiang Pan}, \bibinfo{person}{Guanyu Guo}, \bibinfo{person}{Shihong Zou}, {and} \bibinfo{person}{Ryan Riley}.} \bibinfo{year}{2018}\natexlab{}.
\newblock \showarticletitle{Understanding in-app ads and detecting hidden attacks through the mobile app-web interface}.
\newblock \bibinfo{journal}{\emph{IEEE Transactions on Mobile Computing}} \bibinfo{volume}{17}, \bibinfo{number}{11} (\bibinfo{year}{2018}), \bibinfo{pages}{2675--2688}.
\newblock


\bibitem[Shi et~al\mbox{.}(2025a)]%
        {shi202550}
\bibfield{author}{\bibinfo{person}{Zewei Shi}, \bibinfo{person}{Ruoxi Sun}, \bibinfo{person}{Jieshan Chen}, \bibinfo{person}{Jiamou Sun}, \bibinfo{person}{Minhui Xue}, \bibinfo{person}{Yansong Gao}, \bibinfo{person}{Feng Liu}, {and} \bibinfo{person}{Xingliang Yuan}.} \bibinfo{year}{2025}\natexlab{a}.
\newblock \showarticletitle{50 shades of deceptive patterns: A unified taxonomy, multimodal detection, and security implications}. In \bibinfo{booktitle}{\emph{Proceedings of the ACM on Web Conference 2025}}. \bibinfo{pages}{978--989}.
\newblock


\bibitem[Shi et~al\mbox{.}(2025b)]%
        {DPGuardCode}
\bibfield{author}{\bibinfo{person}{Zewei Shi}, \bibinfo{person}{Ruoxi Sun}, \bibinfo{person}{Jieshan Chen}, \bibinfo{person}{Jiamou Sun}, \bibinfo{person}{Minhui Xue}, \bibinfo{person}{Yansong Gao}, \bibinfo{person}{Feng Liu}, {and} \bibinfo{person}{Xingliang Yuan}.} \bibinfo{year}{2025}\natexlab{b}.
\newblock \bibinfo{booktitle}{\emph{Replication Package for DPGuard}}.
\newblock
\urldef\tempurl%
\url{https://github.com/GalaxyHBXY/DPGuard}
\showURL{%
Retrieved 18 December 2025 from \tempurl}


\bibitem[Tafradzhiyski(2025)]%
        {AustraliaAppMarket}
\bibfield{author}{\bibinfo{person}{Nayden Tafradzhiyski}.} \bibinfo{year}{2025}\natexlab{}.
\newblock \bibinfo{booktitle}{\emph{Australia App Market Statistics (2025)}}.
\newblock
\urldef\tempurl%
\url{https://www.businessofapps.com/data/australia-app-market/?utm_source=chatgpt.com}
\showURL{%
Retrieved 8 Oct 2024 from \tempurl}


\bibitem[tzutalin(2021)]%
        {labelImg}
\bibfield{author}{\bibinfo{person}{tzutalin}.} \bibinfo{year}{2021}\natexlab{}.
\newblock \bibinfo{title}{GitHub - tzutalin/labelImg}.
\newblock \bibinfo{howpublished}{\url{https://github.com/tzutalin/labelImg}}.
\newblock
\newblock
\shownote{Accessed: 24/09/2021}.


\bibitem[VirusTotal(2026)]%
        {virusTotal}
\bibfield{author}{\bibinfo{person}{VirusTotal}.} \bibinfo{year}{2026}\natexlab{}.
\newblock \bibinfo{booktitle}{\emph{VirusTotal}}.
\newblock
\urldef\tempurl%
\url{https://www.virustotal.com/gui/home/upload}
\showURL{%
Retrieved 07 April 2026 from \tempurl}


\bibitem[Vu et~al\mbox{.}(2024)]%
        {vu2024gptvoicetasker}
\bibfield{author}{\bibinfo{person}{Minh~Duc Vu}, \bibinfo{person}{Han Wang}, \bibinfo{person}{Jieshan Chen}, \bibinfo{person}{Zhuang Li}, \bibinfo{person}{Shengdong Zhao}, \bibinfo{person}{Zhenchang Xing}, {and} \bibinfo{person}{Chunyang Chen}.} \bibinfo{year}{2024}\natexlab{}.
\newblock \showarticletitle{Gptvoicetasker: Advancing multi-step mobile task efficiency through dynamic interface exploration and learning}. In \bibinfo{booktitle}{\emph{Proceedings of the 37th Annual ACM Symposium on User Interface Software and Technology}}. \bibinfo{pages}{1--17}.
\newblock


\bibitem[Vu et~al\mbox{.}(2023)]%
        {vu2023voicify}
\bibfield{author}{\bibinfo{person}{Minh~Duc Vu}, \bibinfo{person}{Han Wang}, \bibinfo{person}{Zhuang Li}, \bibinfo{person}{Gholamreza Haffari}, \bibinfo{person}{Zhenchang Xing}, {and} \bibinfo{person}{Chunyang Chen}.} \bibinfo{year}{2023}\natexlab{}.
\newblock \showarticletitle{Voicify Your UI: Towards Android App Control with Voice Commands}.
\newblock \bibinfo{journal}{\emph{Proceedings of the ACM on Interactive, Mobile, Wearable and Ubiquitous Technologies}} \bibinfo{volume}{7}, \bibinfo{number}{1} (\bibinfo{year}{2023}), \bibinfo{pages}{1--22}.
\newblock


\bibitem[Wang et~al\mbox{.}(2023)]%
        {wang2023enabling}
\bibfield{author}{\bibinfo{person}{Bryan Wang}, \bibinfo{person}{Gang Li}, {and} \bibinfo{person}{Yang Li}.} \bibinfo{year}{2023}\natexlab{}.
\newblock \showarticletitle{Enabling conversational interaction with mobile ui using large language models}. In \bibinfo{booktitle}{\emph{Proceedings of the 2023 CHI Conference on Human Factors in Computing Systems}}. \bibinfo{pages}{1--17}.
\newblock


\bibitem[Wang et~al\mbox{.}(2025)]%
        {wang2025llmdroid}
\bibfield{author}{\bibinfo{person}{Chenxu Wang}, \bibinfo{person}{Tianming Liu}, \bibinfo{person}{Yanjie Zhao}, \bibinfo{person}{Minghui Yang}, {and} \bibinfo{person}{Haoyu Wang}.} \bibinfo{year}{2025}\natexlab{}.
\newblock \showarticletitle{LLMDroid: Enhancing Automated Mobile App GUI Testing Coverage with Large Language Model Guidance}.
\newblock \bibinfo{journal}{\emph{Proceedings of the ACM on Software Engineering}} \bibinfo{volume}{2}, \bibinfo{number}{FSE} (\bibinfo{year}{2025}), \bibinfo{pages}{1001--1022}.
\newblock


\bibitem[Wen et~al\mbox{.}(2024)]%
        {wen2024autodroid}
\bibfield{author}{\bibinfo{person}{Hao Wen}, \bibinfo{person}{Yuanchun Li}, \bibinfo{person}{Guohong Liu}, \bibinfo{person}{Shanhui Zhao}, \bibinfo{person}{Tao Yu}, \bibinfo{person}{Toby Jia-Jun Li}, \bibinfo{person}{Shiqi Jiang}, \bibinfo{person}{Yunhao Liu}, \bibinfo{person}{Yaqin Zhang}, {and} \bibinfo{person}{Yunxin Liu}.} \bibinfo{year}{2024}\natexlab{}.
\newblock \showarticletitle{Autodroid: Llm-powered task automation in android}. In \bibinfo{booktitle}{\emph{Proceedings of the 30th Annual International Conference on Mobile Computing and Networking}}. \bibinfo{pages}{543--557}.
\newblock


\bibitem[Wodinsky(2022)]%
        {EpicFine}
\bibfield{author}{\bibinfo{person}{Shoshana Wodinsky}.} \bibinfo{year}{2022}\natexlab{}.
\newblock \bibinfo{booktitle}{\emph{The ‘dark patterns’ used by Epic Games that led to the largest FTC penalties ever}}.
\newblock
\urldef\tempurl%
\url{https://www.marketwatch.com/story/the-dark-patterns-in-fortnite-that-led-to-the-largest-ftc-penalties-ever-11671488228}
\showURL{%
Retrieved 8 Oct 2024 from \tempurl}


\bibitem[Yang et~al\mbox{.}(2023)]%
        {yang2023imagedataaugmentationdeep}
\bibfield{author}{\bibinfo{person}{Suorong Yang}, \bibinfo{person}{Weikang Xiao}, \bibinfo{person}{Mengchen Zhang}, \bibinfo{person}{Suhan Guo}, \bibinfo{person}{Jian Zhao}, {and} \bibinfo{person}{Furao Shen}.} \bibinfo{year}{2023}\natexlab{}.
\newblock \bibinfo{title}{Image Data Augmentation for Deep Learning: A Survey}.
\newblock
\showeprint[arxiv]{2204.08610}~[cs.CV]
\urldef\tempurl%
\url{https://arxiv.org/abs/2204.08610}
\showURL{%
\tempurl}


\bibitem[Yoon et~al\mbox{.}(2024)]%
        {yoon2024intent}
\bibfield{author}{\bibinfo{person}{Juyeon Yoon}, \bibinfo{person}{Robert Feldt}, {and} \bibinfo{person}{Shin Yoo}.} \bibinfo{year}{2024}\natexlab{}.
\newblock \showarticletitle{Intent-driven mobile gui testing with autonomous large language model agents}. In \bibinfo{booktitle}{\emph{2024 IEEE Conference on Software Testing, Verification and Validation (ICST)}}. IEEE, \bibinfo{pages}{129--139}.
\newblock


\bibitem[Yuan et~al\mbox{.}(2024)]%
        {yuan2024designrepair}
\bibfield{author}{\bibinfo{person}{Mingyue Yuan}, \bibinfo{person}{Jieshan Chen}, \bibinfo{person}{Zhenchang Xing}, \bibinfo{person}{Aaron Quigley}, \bibinfo{person}{Yuyu Luo}, \bibinfo{person}{Tianqi Luo}, \bibinfo{person}{Gelareh Mohammadi}, \bibinfo{person}{Qinghua Lu}, {and} \bibinfo{person}{Liming Zhu}.} \bibinfo{year}{2024}\natexlab{}.
\newblock \showarticletitle{Designrepair: Dual-stream design guideline-aware frontend repair with large language models}.
\newblock \bibinfo{journal}{\emph{arXiv preprint arXiv:2411.01606}} (\bibinfo{year}{2024}).
\newblock


\bibitem[Yue et~al\mbox{.}(2024)]%
        {yue2024darkfleece}
\bibfield{author}{\bibinfo{person}{Chang Yue}, \bibinfo{person}{Chen Zhong}, \bibinfo{person}{Kai Chen}, \bibinfo{person}{Zhiyu Zhang}, {and} \bibinfo{person}{Yeonjoon Lee}.} \bibinfo{year}{2024}\natexlab{}.
\newblock \showarticletitle{$\{$DARKFLEECE$\}$: Probing the Dark Side of Android Subscription Apps}. In \bibinfo{booktitle}{\emph{33rd USENIX Security Symposium (USENIX Security 24)}}. \bibinfo{pages}{1543--1560}.
\newblock


\bibitem[Zhang et~al\mbox{.}(2025)]%
        {zhang2025appagent}
\bibfield{author}{\bibinfo{person}{Chi Zhang}, \bibinfo{person}{Zhao Yang}, \bibinfo{person}{Jiaxuan Liu}, \bibinfo{person}{Yanda Li}, \bibinfo{person}{Yucheng Han}, \bibinfo{person}{Xin Chen}, \bibinfo{person}{Zebiao Huang}, \bibinfo{person}{Bin Fu}, {and} \bibinfo{person}{Gang Yu}.} \bibinfo{year}{2025}\natexlab{}.
\newblock \showarticletitle{Appagent: Multimodal agents as smartphone users}. In \bibinfo{booktitle}{\emph{Proceedings of the 2025 CHI Conference on Human Factors in Computing Systems}}. \bibinfo{pages}{1--20}.
\newblock


\bibitem[Zhang et~al\mbox{.}(2023)]%
        {zhang2023automated}
\bibfield{author}{\bibinfo{person}{Yuxin Zhang}, \bibinfo{person}{Sen Chen}, \bibinfo{person}{Lingling Fan}, \bibinfo{person}{Chunyang Chen}, {and} \bibinfo{person}{Xiaohong Li}.} \bibinfo{year}{2023}\natexlab{}.
\newblock \showarticletitle{Automated and context-aware repair of color-related accessibility issues for android apps}. In \bibinfo{booktitle}{\emph{Proceedings of the 31st ACM Joint European Software Engineering Conference and Symposium on the Foundations of Software Engineering}}. \bibinfo{pages}{1255--1267}.
\newblock


\end{thebibliography}

\appendix

\section{Taxonomy}
\revision{Table~\ref{tab:full_taxonomy} shows the detailed definitions of each deceptive pattern types and their detection types (i.e. \revision{intra-page} or \revision{inter-page}).}

\begin{table*}[ht]
\caption{Revised Taxonomy of Deceptive Patterns from Chen et al.~\cite{chen2023unveiling}. From left to right: category, type, sub-type, description, and detection types.}
\vspace{-6pt}
\centering
\resizebox{1\textwidth}{!}{
\begin{tabular}{p{0.25\linewidth} p{0.18\linewidth} p{0.22\linewidth} p{0.45\linewidth} p{0.12\linewidth}}
\toprule
\textbf{Category} & \textbf{Type} & \textbf{Sub-Type} & \textbf{Description} & \textbf{Detection Types} \\
\midrule
\textbf{NAGGING* (NG*)} &  &  & A pop-up window unexpectedly and repeatedly appears and interrupts user tasks. Examples like pop up ads, pop up to rate the app, and pop up to upgrade to premium version. & Inter-page \\

\midrule
\multirow[t]{3}{*}{\begin{minipage}[t]{\linewidth}\textbf{OBSTRUCTION (OB)} \newline Makes a task more difficult or time-consuming to discourage certain behaviors.  \end{minipage}} 
& \textbf{Roach Motel} &  & Easy to opt-in but hard/impossible to opt out. & Inter-page\\
& \textbf{Intermediate \newline Currency} &  & Disconnects real cost by forcing virtual currency use. &  Inter-page \\
& \textbf{Price Comparison Prevention} &  & Hinders direct comparison with other markets. & Intra-page \\

\midrule
\multirow[t]{4}{*}{\begin{minipage}[t]{\linewidth}\textbf{SNEAKING (SN)} \newline Hide, disguise or delay information, preventing informed user choices. \end{minipage}} 
& \textbf{Bait and Switch} &  & A normal action leads to an unexpected outcome. & Inter-page \\
& \textbf{Forced Continuity} &  & Users continue to be charged after trial ends. & Intra-page\\
& \textbf{Hidden Costs} &  & Costs are disclosed only at the final step. & Inter-page \\
& \textbf{Sneak into Basket} &  & Adds items not explicitly chosen by users. & Inter-page \\
\midrule

\multirow[t]{4}{*}{\begin{minipage}[t]{\linewidth}\textbf{FORCED ACTION (FA)} \newline Users are forced to perform some actions to get rewards, unlock features or achieve some tasks. \end{minipage}}
& \textbf{Social Pyramid} &  & Requires social sharing/invites to unlock features. & Intra-page \\
& \textbf{Privacy Zuckering} &  & Forces disclosure of personal data for access. & Intra-page\\
& \textbf{Gamification} &  & Repeats unrelated actions to earn rewards. & Intra-page \\
& \textbf{General Types} &  & Coercive access through ads or payment. & Inter-page \\
\midrule

\multirow[t]{7}{*}{\begin{minipage}[t]{\linewidth}\textbf{INTERFACE \newline INTERFERENCE (II)} \newline Privilege some options over others to confuse and hide the information from users. \end{minipage}}
& \textbf{Preselection} &  & Options are preselected to bias user choice. & Intra-page \\
& \textbf{Hidden Information} &  & Relevant options/info are not readily accessible. & Intra-page \\
&  &  & Absence of explicit consent control. & Intra-page\\
& \textbf{Aesthetic Manipulation (AM)} 
& \textit{Toying with Emotion} & Uses language/color/style to pressure users. & Intra-page \\
&  & \textit{False Hierarchy} & One option made more salient than peers. & Intra-page \\
&  & \textit{Disguised Ad} & Ads presented as native content. & Inter-page \\
&  & \textit{Tricked Questions} & Confusing/negated wording to elicit consent. & Intra-page \\
&  & \textit{General Types} & Generic visual nudges that bias actions. & Intra-page\\
\bottomrule

\end{tabular}}
\label{tab:full_taxonomy}
\end{table*}

\section{Prompts for Task-Oriented App Exploration}
\begin{figure}
    \centering
    \includegraphics[width=1.0\linewidth]{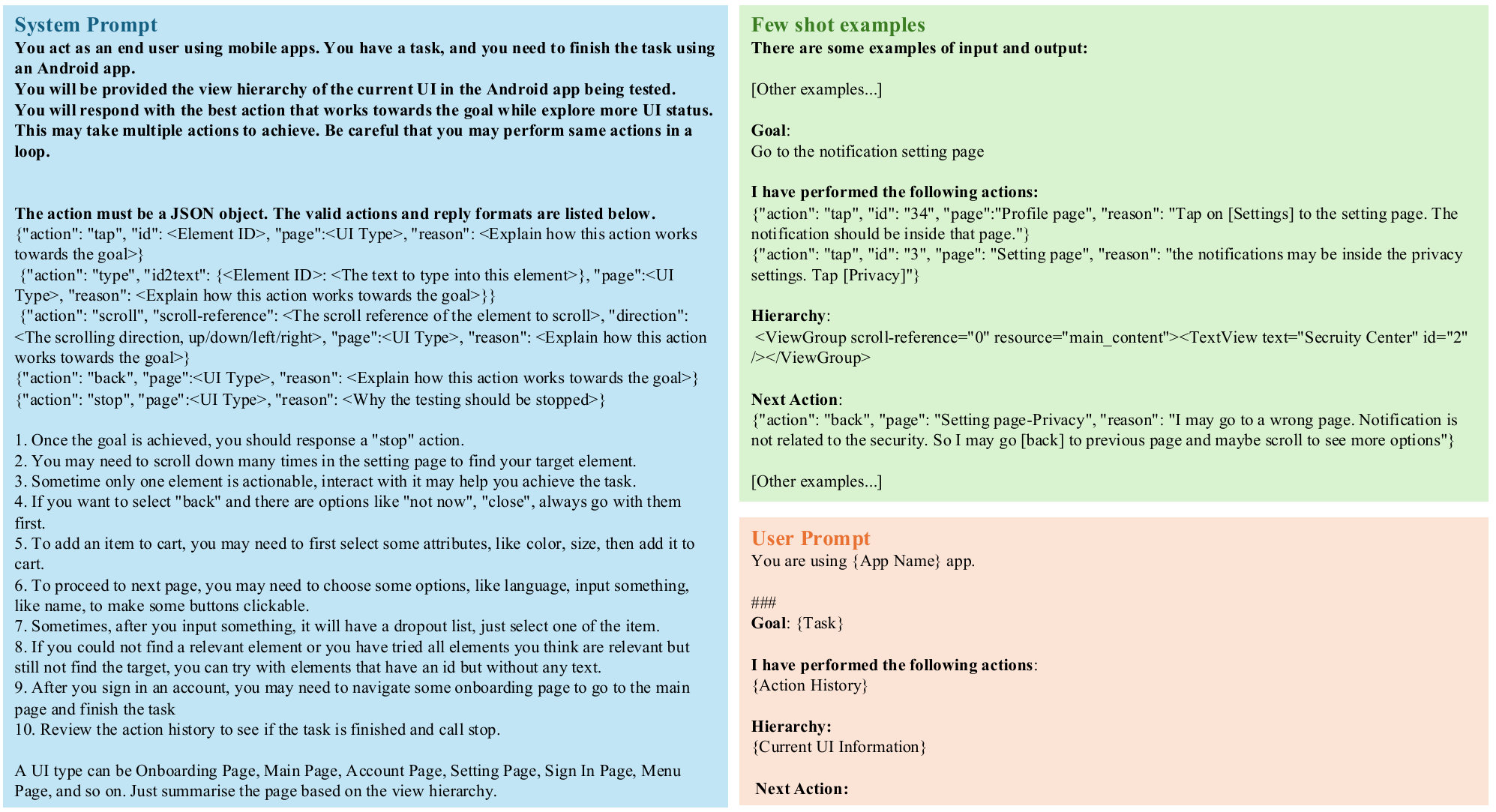}
    \caption{Prompts for Task-Oriented App Exploration.}
    \label{fig:prompt}
\end{figure}

\revision{Figure~\ref{fig:prompt} shows our prompts used for task-oriented app exploration.}

\section{Illustration of Inference Process}
\label{sec:inference_process}
\begin{figure}
    \centering
    \includegraphics[width=1.0\linewidth]{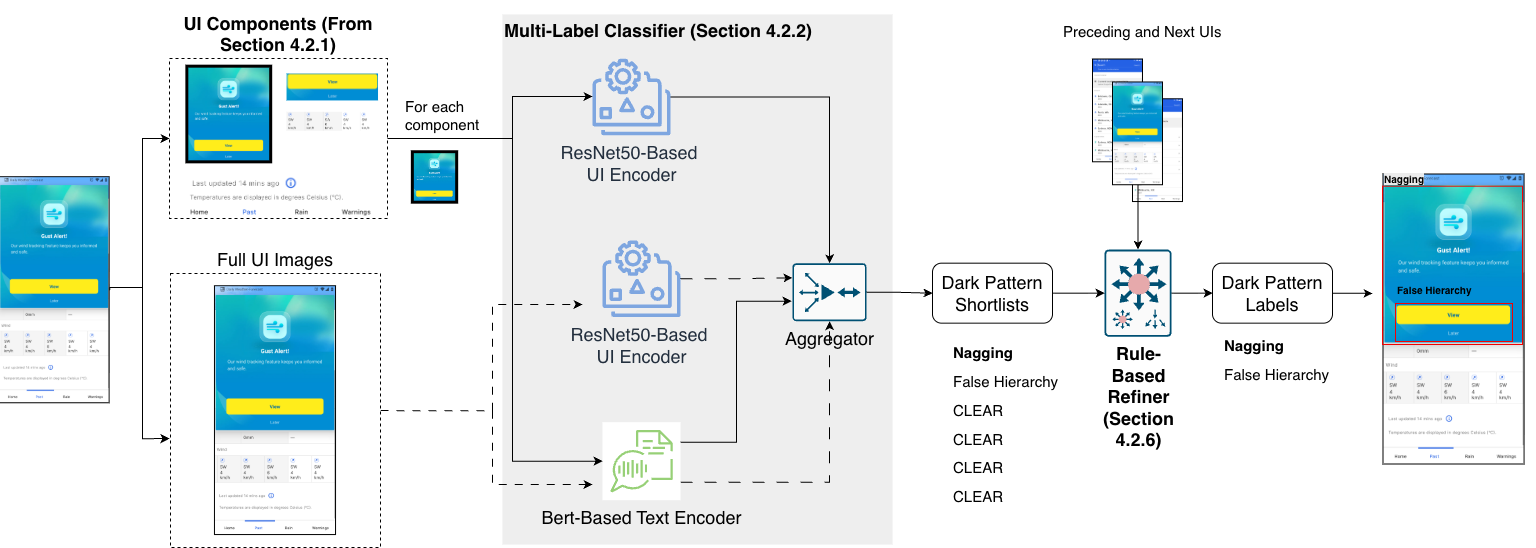}
    \caption{\revision{Illustration of the inference process. Given a UI under test, we first group UI elements into semantically distinct components (Section~\ref{sec:ui_extraction_grouping}). For each component, a multi-label classifier encodes the component image, the full UI image, and associated text to generate initial deceptive pattern predictions, which are then refined using rule-based reasoning over surrounding UIs. This process is applied to all components to obtain final detections, which are localized to UI regions using component bounding boxes. In this example, six components are extracted, yielding initial predictions of Nagging and False Hierarchy, which remain unchanged after refinement and are aggregated as the final UI-level detections.}}
    \label{fig:inference}
\end{figure}

Figure~\ref{fig:inference} shows an illustration of the inference process.

\section{Results and Analysis on ContextDP-mobile and UIGuard-Rico datasets}
\label{sec:baseline_datasets_analysis}

\begin{table}
    \centering
    \caption{Overall Classification Performance in UIGuard-Rico and ContextDP-mobile datasets.}    
    \resizebox{1\columnwidth}{!}{%
        \begin{tabular}{l c
                | c c >{\columncolor{f1gray}}c 
                | c c >{\columncolor{f1gray}}c
                | c c >{\columncolor{f1gray}}c
                | c c >{\columncolor{f1gray}}c
                | c c >{\columncolor{f1gray}}c
                | c c >{\columncolor{f1gray}}c
                }
        \toprule
         &  &  \multicolumn{3}{c|}{\textbf{AidUI}} & \multicolumn{3}{c|}{\textbf{UIGuard}} & \multicolumn{3}{c|}{\textbf{DPGuard (GPT-4o)}} & \multicolumn{3}{c|}{\textbf{DPGuard (GPT-5.2)}} & \multicolumn{3}{c|}{\textbf{DPGuard (Gemini3-pro)}} & \multicolumn{3}{c}{\textbf{\tool{}}} \\
         \midrule
         \textbf{Datasets} & \textbf{DP/non-DP}  & \textbf{Pre.} & \textbf{Rec.} & \textbf{F1} 
         & \textbf{Pre.} & \textbf{Rec.} & \textbf{F1}  
         & \textbf{Pre.} & \textbf{Rec.} & \textbf{F1}  
         & \textbf{Pre.} & \textbf{Rec.} & \textbf{F1} 
         & \textbf{Pre.} & \textbf{Rec.} & \textbf{F1} 
         & \textbf{Pre.} & \textbf{Rec.} & \textbf{F1} \\
         \midrule
         ContextDP-mobile   & DP      
         & 0.91 & 0.69 & 0.79 
         & 0.93 & 0.58 & 0.71 
         & 0.96 & 0.78 & \textbf{0.86} 
         & 0.93 & 0.84 & \textbf{0.88} 
         & 0.95 & 0.74 & 0.83
         & 0.92 & 0.79 & \textbf{0.85} 
         \\
         ContextDP-mobile   & non-DP  
         & 0.63 & 0.89 & 0.74 
         & 0.57 & 0.93 & 0.70 
         & 0.72 & 0.95 & 0.82 
         & 0.77 & 0.89 & 0.83 
         & 0.68 & 0.94 & 0.79
         & 0.95 & 0.94 & \textbf{0.94} 
         \\
         \midrule
         UIGuard-Rico       & DP      
         & 0.26 & 0.26 & 0.26 
         & 0.85 & 0.82 & \textbf{0.84} 
         & 0.72 & 0.77 & 0.74 
         & 0.73 & 0.80 & 0.76 
         & 0.76 & 0.59 & 0.67
         & 0.92 & 0.81 & \textbf{0.86} 
         \\
         UIGuard-Rico       & non-DP  
         & 0.80 & 0.80 & 0.80 
         & 0.95 & 0.96 & \textbf{0.96} 
         & 0.94 & 0.92 & \textbf{0.93} 
         & 0.94 & 0.92 & \textbf{0.93} 
         & 0.90 & 0.95 & 0.92
         & 0.93 & 0.95 & \textbf{0.94} 
         \\
         \bottomrule
        \end{tabular}
    }
    \label{tab:classification_on_baseline_datasets}
\end{table}

\begin{table*}
    \centering
    \caption{Performance of baselines and \tool{} in the ContextDP-mobile-dark dataset.}    
    \resizebox{1\columnwidth}{!}{%
        \begin{tabular}{l c
                | c c >{\columncolor{f1gray}}c 
                | c c >{\columncolor{f1gray}}c
                |c c >{\columncolor{f1gray}}c
                |c c >{\columncolor{f1gray}}c
                |c c >{\columncolor{f1gray}}c
                |c c >{\columncolor{f1gray}}c
                }
        \toprule
         & &  \multicolumn{3}{c|}{\textbf{AidUI}} & \multicolumn{3}{c|}{\textbf{UIGuard}} & \multicolumn{3}{c|}{\textbf{DPGuard (GPT-4o)}} & \multicolumn{3}{c|}{\textbf{DPGuard (GPT-5.2)}} & \multicolumn{3}{c|}{\textbf{DPGuard (Gemini3-pro)}} & \multicolumn{3}{c}{\texttt{\textbf{\tool{}}}} \\
         \midrule
         \textbf{DP Category} & \textbf{\# of Instance} 
         & \textbf{Pre.} & \textbf{Rec.} & \textbf{F1} 
         & \textbf{Pre.} & \textbf{Rec.} & \textbf{F1}  
         & \textbf{Pre.} & \textbf{Rec.} & \textbf{F1}  
         & \textbf{Pre.} & \textbf{Rec.} & \textbf{F1}  
         & \textbf{Pre.} & \textbf{Rec.} & \textbf{F1} 
         & \textbf{Pre.} & \textbf{Rec.} & \textbf{F1} \\
        \midrule
        Nagging              & 9   
        & 0.66 & 0.55 & 0.60
        & 0.84 & 0.23 & 0.36
        & 0.65 & 0.22 & 0.33
        & 0.76 & 0.35 & 0.48
        & 0.84 & 0.59 & 0.69 
        & 0.93 & 0.90 & \textbf{0.91} \\
        ForcedAction-General & 20  
        & 1.00 & 0.20 & 0.33
        & 0.55 & 0.30 & 0.39
        & 0.42 & 0.65 & 0.51
        & 0.61 & 0.55 & 0.58
        & 0.67 & 0.70 & 0.68 
        & 0.74 & 0.70 & \textbf{0.72} \\
        Preselection         & 104 
        & 0.11 & 0.08 & 0.09
        & 0.76 & 0.45 & 0.57
        & 0.95 & 0.57 & 0.71
        & 0.95 & 0.66 & 0.78
        & 0.94 & 0.70 & 0.80 
        & 0.81 & 0.83 & \textbf{0.82} \\
        False Hierarchy      & 59  
        & 1.00 & 0.07 & 0.13
        & 0.79 & 0.32 & 0.46
        & 0.82 & 0.86 & 0.84
        & 0.67 & 0.95 & 0.78
        & 0.68 & 0.92 & 0.78 
        & 0.93 & 0.86 & \textbf{0.89} \\
        Disguised Ads        & 45  
        & 0.53 & 0.36 & 0.43
        & 0.54 & 0.16 & 0.24
        & 0.51 & 0.82 & 0.63
        & 0.51 & 0.89 & 0.65
        & 0.64 & 0.56 & 0.60 
        & 0.91 & 0.65 & \textbf{0.76} \\
        \midrule
        \textbf{Micro Avg.}  & 339 
        & 0.44 & 0.26 & 0.33
        & 0.74 & 0.31 & 0.44
        & 0.70 & 0.56 & 0.62
        & 0.70 & 0.65 & 0.68
        & 0.78 & 0.69 & 0.73
        & 0.92 & 0.88 & \textbf{0.90} \\
        \textbf{Macro Avg.}  & 339 
        & 0.66 & 0.25 & 0.32
        & 0.69 & 0.29 & 0.40
        & 0.67 & 0.62 & 0.60
        & 0.70 & 0.68 & 0.65
        & 0.75 & 0.69 & 0.71
        & 0.88 & 0.81 & \textbf{0.84} \\
        \bottomrule
        \end{tabular}
    }
    \label{tab:rq3_detector_contextdp}
\end{table*}

\begin{table*}
    \centering
    \caption{Performance of baselines and \tool{} in the UIGuard-Rico dataset.}    
    \resizebox{1\columnwidth}{!}{%
        \begin{tabular}{l c
                | c c >{\columncolor{f1gray}}c 
                | c c >{\columncolor{f1gray}}c
                |c c >{\columncolor{f1gray}}c
                |c c >{\columncolor{f1gray}}c
                |c c >{\columncolor{f1gray}}c
                |c c >{\columncolor{f1gray}}c
                }
        \toprule
         & &  \multicolumn{3}{c|}{\textbf{AidUI}} & \multicolumn{3}{c|}{\textbf{UIGuard}} & \multicolumn{3}{c|}{\textbf{DPGuard (GPT-4o)}} & \multicolumn{3}{c|}{\textbf{DPGuard (GPT-5.2)}} & \multicolumn{3}{c|}{\textbf{DPGuard (Gemini3-pro)}} & \multicolumn{3}{c}{\texttt{\textbf{\tool{}}}} \\
         \midrule
         \textbf{DP Category} & \textbf{\# of Instance} 
         & \textbf{Pre.} & \textbf{Rec.} & \textbf{F1} 
         & \textbf{Pre.} & \textbf{Rec.} & \textbf{F1}  
         & \textbf{Pre.} & \textbf{Rec.} & \textbf{F1}  
         & \textbf{Pre.} & \textbf{Rec.} & \textbf{F1} 
         & \textbf{Pre.} & \textbf{Rec.} & \textbf{F1} 
         & \textbf{Pre.} & \textbf{Rec.} & \textbf{F1} \\
        \midrule
        Nagging & 188
        & 0.18 & 0.20 & 0.19
        & 0.90 & 0.73 & 0.81
        & 0.24 & 0.13 & 0.17
        & 0.46 & 0.27 & 0.34
        & 0.75 & 0.80 & 0.77 
        & 0.89 & 0.82 & \textbf{0.86} \\
        Forced Continuity & 1
        & - & - & -
        & 1.00 & 1.00 & \textbf{1.00}
        & 0.08 & 1.00 & 0.14
        & 0.11 & 1.00 & 0.20
        & 0.25 & 1.00 & 0.40 
        & 1.00 & 1.00 & \textbf{1.00} \\
        Social Pyramid & 16
        & - & - & -
        & 0.88 & 0.88 & \textbf{0.88}
        & 0.27 & 0.50 & 0.35
        & 0.50 & 0.44 & 0.47
        & 0.89 & 0.50 & 0.64 
        & 0.59 & 0.92 & 0.72 \\
        Privacy Zuckering & 117
        & - & - & -
        & 0.96 & 0.67 & 0.79
        & 0.71 & 0.38 & 0.49
        & 0.52 & 0.26 & 0.34
        & 0.79 & 0.62 & 0.70 
        & 1.00 & 0.70 & \textbf{0.83} \\
        ForcedAction-General & 103
        & 0.00 & 0.00 & 0.00
        & 0.97 & 0.89 & \textbf{0.93}
        & 0.66 & 0.43 & 0.52
        & 0.83 & 0.44 & 0.57
        & 0.78 & 0.66 & 0.72 
        & 0.71 & 0.84 & 0.77 \\
        Preselection & 232
        & 0.04 & 0.02 & 0.02
        & 0.97 & 0.72 & \textbf{0.83}
        & 0.81 & 0.36 & 0.50
        & 0.89 & 0.43 & 0.58
        & 0.83 & 0.44 & 0.57
        & 0.91 & 0.75 & 0.82 \\
        False Hierarchy & 273
        & 0.75 & 0.01 & 0.02
        & 0.92 & 0.45 & 0.60
        & 0.66 & 0.43 & 0.52
        & 0.50 & 0.50 & 0.50
        & 0.56 & 0.55 & 0.55 
        & 0.90 & 0.80 & \textbf{0.84} \\
        Disguised Ads & 46
        & 0.10 & 0.15 & 0.12
        & 0.43 & 0.50 & 0.46
        & 0.03 & 0.57 & 0.06
        & 0.03 & 0.61 & 0.07
        & 0.08 & 0.54 & 0.14
        & 0.37 & 0.74 & \textbf{0.50} \\
        Interface Interference - General & 684
        & - & - & -
        & 1.00 & 0.94 & \textbf{0.97}
        & 0.79 & 0.38 & 0.51
        & 0.73 & 0.43 & 0.54
        & 0.83 & 0.37 & 0.51  
        & 0.90 & 0.80 & 0.85 \\
        \midrule
        \textbf{Micro Avg.} & 1,660
        & 0.13 & 0.06 & 0.08
        & 0.95 & 0.77 & 0.85
        & 0.36 & 0.36 & 0.36
        & 0.37 & 0.41 & 0.39
        & 0.60 & 0.50 & 0.54
        & 0.95 & 0.89 & \textbf{0.92} \\
        \textbf{Macro Avg.} & 1,660
        & 0.21 & 0.08 & 0.07
        & 0.89 & 0.75 & \textbf{0.81}
        & 0.47 & 0.46 & 0.36
        & 0.51 & 0.48 & 0.40
        & 0.64 & 0.61 & 0.56
        & 0.82 & 0.83 & \textbf{0.81} \\
        \bottomrule
        \end{tabular}
    }
    \label{tab:rq3_detector_baselines_uiguard_rico}
\end{table*}

\subsection{Results for ContextDP-mobile}
\revision{
Table~\ref{tab:classification_on_baseline_datasets} and Table~\ref{tab:rq3_detector_contextdp} report the classification and detection performance on the ContextDP-mobile dataset for all baselines and our \tool{}. Overall, all methods achieve reasonable performance in distinguishing deceptive from non-deceptive UIs. Rule-based methods reach F1 scores between 0.71 and 0.79 in the classification task, while all other methods achieve F1 scores above 0.8. The overall performance trends largely mirror those observed on \revision{AppRay-Tainted-UIs }: rule-based methods perform the worst, MLLM-based methods improve upon them due to stronger reasoning capabilities, and \tool{} consistently achieves the best results. We also observe that rule-based methods perform noticeably better on ContextDP-mobile than on \revision{AppRay-Tainted-UIs }, which is expected given that their detection rules are developed and tuned based on similar datasets.
}

\begin{figure}
    \centering
    \includegraphics[width=1.0\linewidth]{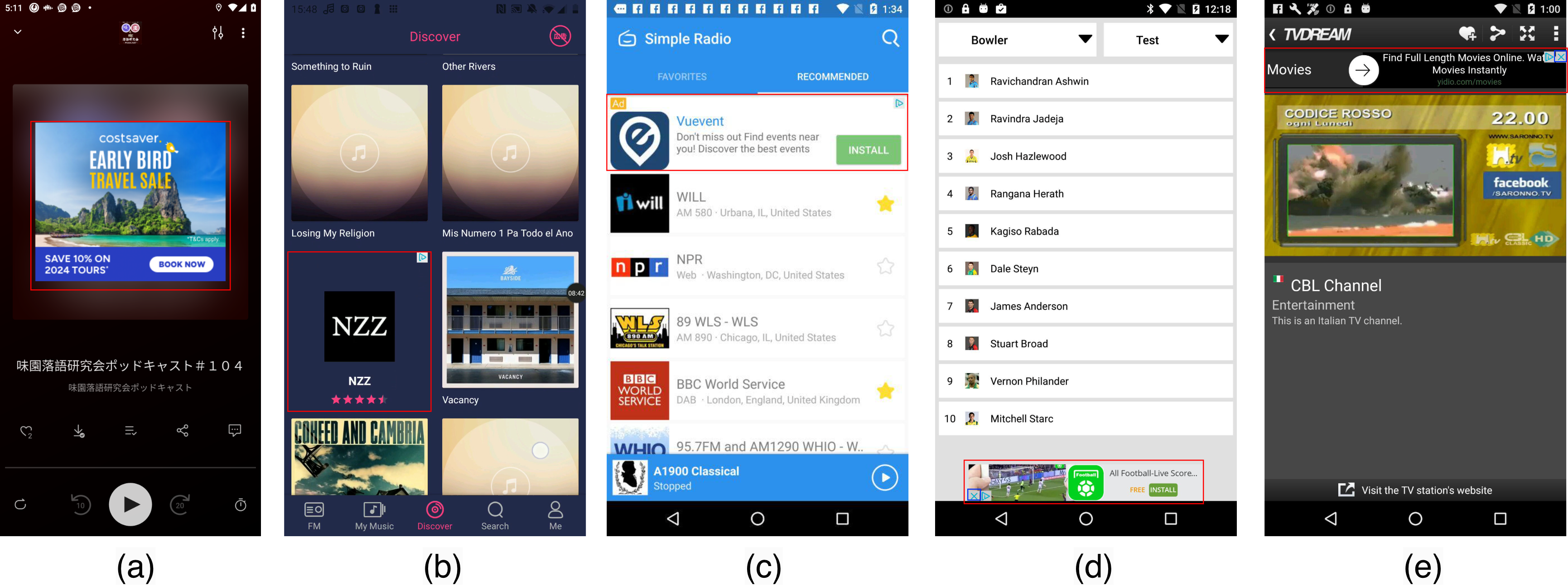}
    \caption{\revision{Examples of Disguised Advertisement Deceptive Patterns. Red boxes indicate the locations of deceptive patterns in the user interface. (a) Advertisements presented without explicit disclosure cues. (b–c) Advertisements that are visually similar to surrounding content, marked only by a small “ad” icon in the top-right corner. (d–e) Typical bottom- or top-placed disguised advertisements, featuring an ad icon and a small close icon.}}
    \label{fig:disguisedAds}
\end{figure}

\revision{
Among MLLM-based methods, a notable difference emerges for DPGuard (Gemini-3-Pro) on the Disguised Ads category. While Gemini-3-Pro performs substantially worse than other MLLMs on \revision{AppRay-Tainted-UIs }, its performance on ContextDP-mobile becomes comparable. This change can be attributed to differences in dataset characteristics. ContextDP-mobile contains a large number of disguised advertisements with relatively explicit visual cues, such as clear “Ad” icons or textual labels combined with close buttons (see Figure~\ref{fig:disguisedAds}(c)). In contrast, \revision{AppRay-Tainted-UIs } includes more diverse and challenging cases, including advertisements without close buttons and interaction-dependent disguised ads, which are more difficult for general-purpose models to identify reliably (see Figure~\ref{fig:disguisedAds}(a)(b)).
}

\revision{
These results suggest that Gemini-3-Pro is more effective at detecting visually explicit and structurally simple disguised ads, but struggles when confronted with more diverse and implicit implementations. As a result, its performance improves on the more homogeneous ContextDP-mobile dataset, while degrading on the more challenging and diverse \revision{AppRay-Tainted-UIs } benchmark. This observation is consistent with the overall trends observed across datasets and further highlights the importance of evaluating disguised ads under diverse UI implementations.
}

\subsection{Results for UIGuard-Rico}
\revision{
Table~\ref{tab:classification_on_baseline_datasets} and Table~\ref{tab:rq3_detector_baselines_uiguard_rico} report the classification and detection performance on the UIGuard-Rico dataset for all baselines and our \tool{}. Overall, the relative performance trends across different methods are consistent with observations on the other datasets.
}

\revision{
The performance differences on UIGuard-Rico primarily reflect the dataset’s high degree of homogeneity. Compared to \revision{AppRay-Tainted-UIs }, UIGuard-Rico contains highly repetitive interface structures, visual cues, and linguistic patterns, which strongly favor rule-based and template-driven approaches. As shown in Table~\ref{tab:rq3_detector_baselines_uiguard_rico}, rule-based methods achieve particularly strong performance on categories with stable and explicit cues, such as Interface Interference - General and Preselection.
}

\revision{
However, this same homogeneity also exposes limitations for MLLM-based methods. Rather than benefiting uniformly from repeated patterns, MLLMs exhibit higher variance when confronted with visually and structurally similar instances. For example, in the Interface Interference - General category (Figure~\ref{fig:disguisedAds}(d)(e) - blue boxes, small close buttons), instances that are visually indistinguishable to human observers are not always detected consistently by MLLMs. 
}

\revision{
Furthermore, the two datasets differ substantially in terms of sample distribution for certain deceptive patterns. For Forced Continuity, UIGuard-Rico contains only a single annotated instance, whereas AppRay includes a much larger and more diverse set. Similarly, Social Pyramid is sparsely represented in both datasets, with fewer than 15 instances. Such data sparsity leads to high performance variance and highlights the need for richer and more representative benchmarks.
}

\revision{
Overall, the results on UIGuard-Rico highlight how dataset design systematically emphasizes different detection capabilities. High visual regularity and conservative annotations benefit rule-based approaches, whereas MLLM-based methods are more sensitive to annotation boundaries, limited sample sizes, and implicit cues. The observed performance differences therefore reflect dataset-induced emphasis rather than inherent instability of the models.
However, across these dataset-specific variances, \tool{} maintains relatively stable performance, reflecting its ability to balance visual cue detection, semantic reasoning, and interaction-aware analysis under different dataset characteristics.
}

\begin{table*}[t]
\centering
\caption{Detailed Performance for each ablation of AppRay}
\label{tab:encoder_ablation}
\resizebox{0.8\textwidth}{!}{
\begin{tabular}{l|ccc|ccc|ccc}
\toprule
\multirow{2}{*}{\textbf{DP Category}} 
& \multicolumn{3}{c|}{\textbf{Text+DA+NS+CW}}
& \multicolumn{3}{c|}{\textbf{Image+DA+NS+CW}}
& \multicolumn{3}{c}{\textbf{Text+Image}} \\
\cline{2-10}
& P & R & F1 & P & R & F1 & P & R & F1 \\
\midrule
Nagging & 0.712 & 0.806 & 0.755 & 0.875 & 0.704 & 0.780 & 0.884 & 0.633 & 0.759 \\
Bait And Switch & 0.737 & 0.764 & 0.750 & 0.603 & 0.553 & 0.574 & 0.821 & 0.454 & 0.588 \\
Forced Continuity & 0.719 & 0.561 & 0.629 & 0.903 & 0.766 & 0.829 & 0.754 & 0.717 & 0.735 \\
Roach Motel & 0.926 & 0.294 & 0.447 & 0.935 & 0.103 & 0.185 & 0.580 & 0.525 & 0.551 \\
Intermediate Currency & 0.990 & 0.338 & 0.504 & 0.910 & 0.815 & 0.860 & 0.661 & 0.860 & 0.747 \\
Social Pyramid & 0.938 & 0.970 & 0.953 & 0.949 & 0.118 & 0.210 & 0.900 & 0.880 & 0.890 \\
Privacy Zuckering & 0.806 & 0.940 & 0.868 & 0.794 & 0.557 & 0.654 & 0.844 & 0.607 & 0.707 \\
Gamification & 0.916 & 0.888 & 0.902 & 0.886 & 0.486 & 0.628 & 0.745 & 0.881 & 0.807 \\
ForcedAction-General & 0.951 & 0.634 & 0.761 & 0.771 & 0.934 & 0.845 & 0.820 & 0.782 & 0.801 \\
Preselection & 0.914 & 0.909 & 0.911 & 0.783 & 0.870 & 0.824 & 0.911 & 0.885 & 0.898 \\
Hidden Information & 0.963 & 0.932 & 0.947 & 0.792 & 0.942 & 0.861 & 0.923 & 0.894 & 0.908 \\
Toying with Emotion & 0.874 & 0.814 & 0.843 & 0.763 & 0.684 & 0.721 & 0.742 & 0.725 & 0.733 \\
False Hierarchy & 0.777 & 0.842 & 0.808 & 0.781 & 0.622 & 0.692 & 0.721 & 0.680 & 0.701 \\
Disguised Ads & 0.940 & 0.934 & 0.937 & 0.704 & 0.922 & 0.807 & 0.790 & 0.900 & 0.841 \\
Tricked Questions & 0.736 & 0.685 & 0.709 & 0.911 & 0.114 & 0.202 & 0.766 & 0.826 & 0.795 \\
Interface Interference-General & 0.800 & 0.650 & 0.717 & 0.810 & 0.822 & 0.816 & 0.724 & 0.799 & 0.761 \\
\midrule
Micro Avg. & 0.837 & 0.805 & 0.821 & 0.773 & 0.845 & 0.800 & 0.867 & 0.835 & 0.851 \\
Macro Avg. & 0.879 & 0.729 & 0.758 & 0.840 & 0.649 & 0.675 & 0.797 & 0.707 & 0.722 \\
\hline
\hline
\multirow{2}{*}{\textbf{DP Category}} 
& \multicolumn{3}{c|}{\textbf{Text+Image+DA}}
& \multicolumn{3}{c|}{\textbf{Text+Image+DA+NS}}
& \multicolumn{3}{c}{\textbf{Text+Image+DA+NS+CW}} \\
\cline{2-10}
& P & R & F1 & P & R & F1 & P & R & F1 \\
\midrule
Nagging & 0.888 & 0.681 & 0.770 & 0.941 & 0.711 & 0.810 & 0.952 & 0.767 & 0.849 \\
Bait And Switch & 0.744 & 0.758 & 0.751 & 0.792 & 0.703 & 0.745 & 0.801 & 0.733 & 0.766 \\
Forced Continuity & 0.756 & 0.747 & 0.751 & 0.812 & 0.723 & 0.765 & 0.821 & 0.752 & 0.785 \\
Roach Motel & 0.615 & 0.640 & 0.627 & 0.703 & 0.548 & 0.616 & 0.603 & 0.447 & 0.514 \\
Intermediate Currency & 0.675 & 0.892 & 0.770 & 0.713 & 0.832 & 0.768 & 0.731 & 0.852 & 0.787 \\
Social Pyramid & 0.912 & 0.905 & 0.908 & 0.932 & 0.863 & 0.896 & 1.000 & 1.000 & 1.000 \\
Privacy Zuckering & 0.848 & 0.646 & 0.733 & 0.903 & 0.619 & 0.735 & 0.868 & 0.618 & 0.722 \\
Gamification & 0.760 & 0.910 & 0.828 & 0.823 & 0.878 & 0.860 & 0.802 & 0.995 & 0.888 \\
ForcedAction-General & 0.826 & 0.807 & 0.816 & 0.881 & 0.803 & 0.840 & 0.873 & 0.892 & 0.883 \\
Preselection & 0.917 & 0.896 & 0.906 & 0.943 & 0.901 & 0.922 & 0.962 & 0.941 & 0.951 \\
Hidden Information & 0.927 & 0.913 & 0.920 & 0.951 & 0.931 & 0.941 & 0.967 & 0.951 & 0.959 \\
Toying with Emotion & 0.778 & 0.852 & 0.813 & 0.828 & 0.854 & 0.841 & 0.846 & 0.867 & 0.857 \\
False Hierarchy & 0.726 & 0.716 & 0.721 & 0.781 & 0.658 & 0.714 & 0.802 & 0.701 & 0.748 \\
Disguised Ads & 0.812 & 0.922 & 0.863 & 0.884 & 0.879 & 0.881 & 0.834 & 1.000 & 0.910 \\
Tricked Questions & 0.835 & 0.560 & 0.671 & 0.872 & 0.507 & 0.641 & 0.823 & 0.565 & 0.671 \\
Interface Interference-General & 0.738 & 0.832 & 0.781 & 0.814 & 0.772 & 0.792 & 0.766 & 0.906 & 0.830 \\
\midrule
Micro Avg. & 0.897 & 0.859 & 0.877 & 0.906 & 0.881 & 0.893 & 0.903 & 0.855 & 0.878 \\
Macro Avg. & 0.813 & 0.752 & 0.767 & 0.874 & 0.734 & 0.770 & 0.841 & 0.812 & 0.820 \\
\bottomrule
\end{tabular}}
\label{tab:detailed_ablation}
\end{table*}

\end{document}